\documentclass[fleqn,usenatbib]{mnras}

\usepackage{newtxtext,newtxmath}
\usepackage[T1]{fontenc}
\usepackage{ae,aecompl}

\usepackage{graphicx}	% Including figure files
\usepackage{amsmath}	% Advanced maths commands
\usepackage{amssymb}	% Extra maths symbols

\title[Spectro-photometric decomposition of galaxies]{Spectro-photometric decomposition of galaxy structural components}

\author[J. M\'endez-Abreu et al.]{
J. M\'endez-Abreu,$^{1,2}$\thanks{E-mail: jairomendezabreu@gmail.com} S. F. S\'anchez,$^{3}$ A. de Lorenzo-C\'aceres$^{1,2}$
\\
$^{1}$Instituto de Astrof\'isica de Canarias, Calle V\'ia L\'actea s/n, E-38205 La Laguna, Tenerife, Spain\\
$^{2}$Departamento de Astrof\'isica, Universidad de La Laguna, E-38200 La Laguna, Tenerife, Spain\\
$^{3}$Instituto de Astronom\'ia, Universidad Nacional Aut\'onoma de M\'exico, A.P. 70-264, 04510 M\'exico, D.F., M\'exico.\\
}

\date{Accepted XXX. Received YYY; in original form ZZZ}

\pubyear{2018}

\begin{document}
\label{firstpage}
\pagerange{\pageref{firstpage}--\pageref{lastpage}}
\maketitle

\begin{abstract}
Galaxies are complex systems made up of different structural components such as bulges, discs, and bars. Understanding galaxy evolution requires unveiling, independently, their history of stellar mass and metallicity assembly. We introduce {\sc c2d}, a new algorithm to perform spectro-photometric multi-component decompositions of integral field spectroscopy (IFS) datacubes. The galaxy surface-brightness distribution at each wavelength (quasi-monochromatic image) is fitted using GASP2D, a 2D photometric decomposition code. As a result, {\sc c2d} provides both a characteristic one-dimensional spectra and a full datacube with all the spatial and spectral information for every component included in the fit. We show the basic steps of the {\sc c2d} spectro-photometric fitting procedure, tests on mock datacubes demonstrating its reliability, and a  first application of {\sc c2d} to a sample of three early-type galaxies (ETGs) observed within the CALIFA survey. The resulting datacubes from {\sc c2d} are processed through the {\sc Pipe3D} pipeline obtaining both the stellar populations and ionised gas properties of bulges and discs. This paper presents an overview of the potential of {\sc c2d}+{\sc Pipe3D} to unveil the formation and evolution of galaxies.
\end{abstract}

\begin{keywords}
galaxies: bulge - galaxies: disc - galaxies: evolution - galaxies: formation - galaxies: structure - galaxies: photometry
\end{keywords}

%%%%%%%%%%%%%%%%%%%%%%%%%%%%%%%%%%%%%%%%%%%%%%%%%%

%%%%%%%%%%%%%%%%% BODY OF PAPER %%%%%%%%%%%%%%%%%%

%---------------------------------------------------
\section{Introduction}
\label{sec:intro}

Understanding the history of the mass assembly in galaxies is fundamental to shed light on their formation and evolution. The galaxy mass growth depends on the equilibrium between the physical processes that trigger and shutdown/quench their star formation (SF). However, galaxies are complex systems made up of several stellar structures such as bulges, discs, and bars, and their relative contribution can modify this fragile balance. It is therefore mandatory to understand the mass growth of the different galactic structures independently in order to provide a general view of galaxy evolution.

At high redshift, gas-rich galaxy discs are highly turbulent with SF occurring in massive clumps, giving rise to the observed irregular morphologies \citep{abraham96, vandenbergh96,elmegreen07,hinojosagoni16}. The subsequent coalescence of these clumps has been proposed as a mechanism to form current-day bulges, which in turn change again the morphology of galaxies \citep[e.g.,][]{noguchi99,bournaud07,bournaud16}. Mergers of galaxies are also common in the early Universe and have been proposed to drive the morphological  evolution of galaxies, as well as the SF quenching \citep[e.g.,][]{hopkins09}. However, numerical simulations demonstrated that mergers alone cannot reproduce the time evolution of the star-forming vs. passive galaxy population \citep{dimatteo08}. Other processes have been suggested to quench SF: i) cold gas stripping in a cluster environment \citep{tonnesenbryant09,feldmann10}, ii) heating by virial shocks of gas infalling in a massive dark matter halo \citep{birnboimdekel03,keres09}, iii) energy feedback from an active galactic nucleus \citep[AGN;][]{hopkins05, cattaneo09}, and iv) disc stabilization against SF due to the presence of a dominant bulge \citep{martig09}. All these physical processes produce distinctive patterns in the relation between galaxy morphology and star formation history (SFH). Every galaxy structure keeps a unique memory of their past evolution. For instance, bulges, with their location at the center of the potential well, are insensitive to environmental harassment contrary to outer discs that are more prone to these effects. In addition, bulges and discs contribute differently to galaxy evolution depending on the galaxy/halo mass, environment, and redshift. Thus, the chronology of the galaxy mass assembly is imprinted in the time evolution of the galaxy structures and their stellar populations \citep{perezgonzalez08,perez13,ibarramedel16}.

Many studies have tried to separate the light contribution from the bulge and disc to better understand their formation, hence the evolution of the galaxy. Most of these studies are based on photometric decomposition techniques applied to broad-band galaxy images. Nowadays, two-dimensional (2D) photometric decomposition codes are widespread (e.g., GIM2D, \citealt{simard02}; GALFIT, \citealt{peng02}; BUDDA, \citealt{desouza04}; GASP2D, \citealt{mendezabreu08a}; IMFIT, \citealt{erwin15}; PHI, \citealt{argyle18}) and they have been applied to both detailed multi-component analysis of relatively small galaxy samples \citep[e.g.,][]{laurikainen05,gadotti09,mendezabreu17,delorenzocaceres19} and structurally simple photometric decomposition of large galaxy samples \citep[e.g.][]{simard11,meert16,dimauro18}. These studies have highlighted that the formation of galaxy substructures is a complex process, and that the spatial distribution of the different stellar populations gives us clues on how they have been redistributed throughout the galaxy during mergers (classical bulges and inside-out disc formation at $z > 1$) and secular processes (disc-like bulges and outside-in disc evolution at $z < 1$). 

Similar decomposition techniques can be applied on spectroscopic data. In fact, the idea of applying standard photometric decomposition techniques (used in broad-band photometry) to quasi-monochromatic images extracted from spectroscopy is not new. Using long-slit spectroscopy, \citet{jahnke02} used this technique to decouple the spectra of both the nucleus and the host galaxy in a sample of quasars. \citet{johnston12} used a similar methodology to produce separated long-slit spectra of bulges and discs in a sample of S0 galaxies \citep[see also][]{johnston14}.

Like in the case of the photometric decompositions, the extension of these spectro-photometric techniques to the more general 2D case arrived concomitantly with integral field spectrographs (IFS). \citet{wisotzki03} applied 2D modelling techniques for successfully deblending quasar lenses. \citet{sanchez04} used this technique to separate both the nuclear and host galaxy of the central region of 3C 120 \citep[see also][]{jahnke04,garcialorenzo05}, and \citet{sanchez07} applied it to extract the deblended spectra of the galaxies in the core of Abell~2218. The advent of new IFS surveys with the necessary combination of depth, angular resolution, and area to enable detailed studies of galaxy morphologies provides a fertile ground for the further development of these spectro-photometric methodologies which are able to fully exploit the data. \citet{husemann13} improved the previous methodologies to separate the quasar emission from the host galaxies developing {\sc QDEBLEND$^{\rm 3D}$}, a new tool especially designed to work with IFS. These techniques were also applied to the case of stellar spectroscopy in crowded fields \citep{kamann13}. An example of their application to the bulge/disc spectro-photometric decomposition of galaxies is BUDDI \citep{johnston17}, which uses GALFITM \citep{haussler13} to decompose the light distribution of galaxies in an IFS datacube into two components (a bulge and a disc), providing both the integrated spectra and the 2D datacube for each component. This work represents the extension of their previous work using long-slit spectra \citep{johnston12,johnston14} to  the IFS case. Other spectro-photometric methods, including the possibly different kinematic distribution of the galaxy components, have also been developed \citep{coccato11,tabor17,catalantorrecilla17,coccato18}.

In this paper we present {\sc c2d}, a new code to separate the spectral information of multiple structural components in an IFS datacube using 2D photometric decompositions of their surface-brightness distributions. We apply {\sc c2d} to IFS galaxy datacubes observed within the Calar Alto Legacy Integral Field Area \citep[CALIFA;][]{sanchez12}. Galaxies were modeled assuming they are composed by an inner bulge and an outer disc, as already demonstrated using Sloan Digital Sky Survey (SDSS) broad-band imaging by \citet{mendezabreu17}. The resulting, independent, datacubes for each component are analysed using the extensively tested {\sc Pipe3D} pipeline \citep{sanchez16}, providing spatially resolved information on the stellar population and ionised gas properties of bulges and discs separately. The wealth of information given by the combined application of {\sc c2d}+{\sc Pipe3D} to the CALIFA database is overwhelming, and the analysis of the whole CALIFA sample will be presented in forthcoming papers.

Here, we show a pilot analysis on a sample of three early type galaxies (ETGs) observed with CALIFA and studied in detail in \citet[][Sect.~\ref{sec:CALIFA}]{gomes16}. They are ideal showcases since they are structurally simple (bulge and disc), but with clear signs of SF in their outer regions (not usual in ETGs). The {\sc c2d} algorithm is presented in Sect.~\ref{sec:decomposition}. A synthetic description of the {\sc Pipe3D} capabilities is depicted in Sect.~\ref{sec:pipe3D}. A number of tests to demonstrate the robustness and accuracy of {\sc c2d} are presented in Sect.~\ref{sec:idealmock} for mock datacubes. The stellar population and ionised-gas properties of the bulges and discs in our three sample galaxies are detailed in Sect.~\ref{sec:results}. The interpretation of our results in terms of the different formation and evolution scenarios is provided in Sect.~\ref{sec:discussion}. The conclusions are described in Sect.~\ref{sec:conclusions}. Throughout the paper we assume a flat cosmology with $\Omega_{\rm m}$ = 0.3, $\Omega_{\rm\lambda}$ = 0.7, and a Hubble constant $H_0$ = 70 km s$^{-1}$ Mpc$^{-1}$.

%---------------------------------------------------
%---------------------------------------------------
\section{Sample and CALIFA observations}
\label{sec:CALIFA}
The CALIFA survey obtained a statistically well-defined sample of 667 galaxies in the local universe (0.005 $< z <$ 0.03). Observations were performed with the PMAS/PPAK integral field spectrophotometer mounted on the Calar Alto 3.5m telescope. Galaxies were observed using two different setups. The V500 grating has a nominal resolution of $R$ = 850 at 5000 \AA\ and covers from 3745 to 7300 \AA. This grating is particularly suitable for stellar population studies and it has been extensively used within the CALIFA collaboration. The second setup is based on the V1200 grating with better spectral resolution $R$ = 1650 at 4500 \AA, but covering a shorter wavelength range from 3700\AA\ to 5000 \AA. This setup is particularly suited for spatially resolved stellar kinematic studies.

Recently, \citet{mendezabreu17} carried out a multi-component, multi-band photometric decomposition of the CALIFA data release 3 \citep[DR3;][]{sanchez16} galaxy sample using images from the SDSS-DR7 \citep{abazajian09}. From this analysis, we obtained that ~170 galaxies covering a wide range of stellar masses ($10^9 < M_{\star}/M_{\sun} < 10^{11}$) and Hubble types (S0-Sd) are well described with a bulge and disc component. In \citet{mendezabreu18} we carefully analysed the early-type galaxies of this sample. We applied a combination of a logical filtering and the BIC statistical method to separate, on a photometric basis, elliptical from lenticular galaxies. Our results showed the intrinsic difficulty of separating these two classes when no other structures, such as bars, are present in the galaxies. In fact, we classified 26, 34, and 21 galaxies as ellipticals, lenticulars, and unknown (not possible to classify), respectively.

%--------------------------------------------------------
\begin{figure*}
\begin{center}
\includegraphics[bb=150 0 450 850,angle=90,width=\textwidth]{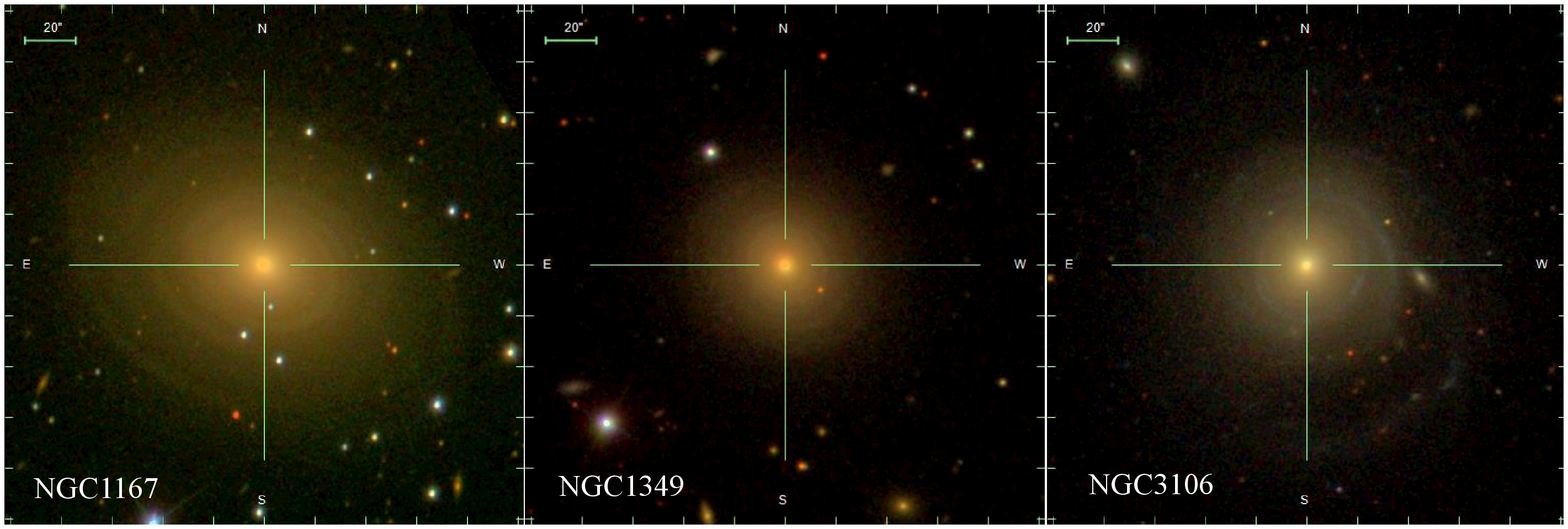}
\caption{SDSS composite-color images of the three galaxies in our sample.}
\label{fig:galaxies}
\end{center}
\end{figure*}
%--------------------------------------------------------

The scope of this paper is to show the potential of our new spectro-photometric technique to unveil the formation of the different structures shaping the galaxies. Therefore, we pick three early-type galaxies observed with CALIFA and analysed in detail in \citet{gomes16}. These are  NGC1167, NGC1349, and NGC3106 (Fig.\ref{fig:galaxies}). Early-type galaxies are defined as those without clear signs of SF, generally associated to a lack of cold gas. However, \citet{gomes16} found clear signs of SF following a spiral-like shape in the outer parts of these galaxies. The morphological analysis performed in \citet{mendezabreu18} revealed that NGC1349 and NGC3106 can be  photometrically classified as lenticulars, therefore hosting a bulge and a disc. However, NGC1167 was classified as unknown. In this paper, we used the combination of {\sc c2d}+{\sc Pipe3D}, applied to the CALIFA IFS data, to provide further constraints on their formation and evolution,  and possibly to dissentangle the actual number of structural components in each galaxy. Table~\ref{tab:sample} shows the main properties of the sample. A complete analysis of the 170 galaxies composed by a bulge and a disc in the CALIFA survey will be carried out in a forthcoming paper.

%---------------------------------------------------------------------------
\begin{table}
 \centering
  \caption{Main properties of the galaxy sample}
  \label{tab:sample}
  \begin{tabular}{lcccc}
  \hline
Name  & Morph. type & Distance & Mass & Model\\
 (1)  & (2) & (3) & (4) & (5) \\
\hline
\hline
NGC\,1167 & SA0$^-$  & 66.2 & 11.3 & Unknown\\
NGC\,1349 & S0       & 87.7 & 10.8 &BD\\
NGC\,3106 &  S0      & 90.1 & 11.0 &BD\\
\hline
\end{tabular}
\begin{minipage}{8cm}
(1) Galaxy name; (2) morphological type from NED, (3) luminosity distance (in Mpc) as extracted from \citet{gomes16}, (4) logarithm of the stellar mass in solar units from \citet{walcher14}, (5) number of components from the photometric decomposition of \citet{mendezabreu17}: Unknown shows that either a single S\'ersic (B) or a S\'ersic+Exponential (BD) model returns statistically compatible results; BD shows that a S\'ersic+Exponential is the best representation of the galaxy.
\end{minipage}\end{table}
%-----------------------------------------------------------------------

%---------------------------------------------------
%---------------------------------------------------
\section{{\sc c2d}. A new spectro-photometric decomposition algorithm}
\label{sec:decomposition}

\begin{figure*}
\begin{center}
\includegraphics[bb= 100 50 500 810,width=10cm]{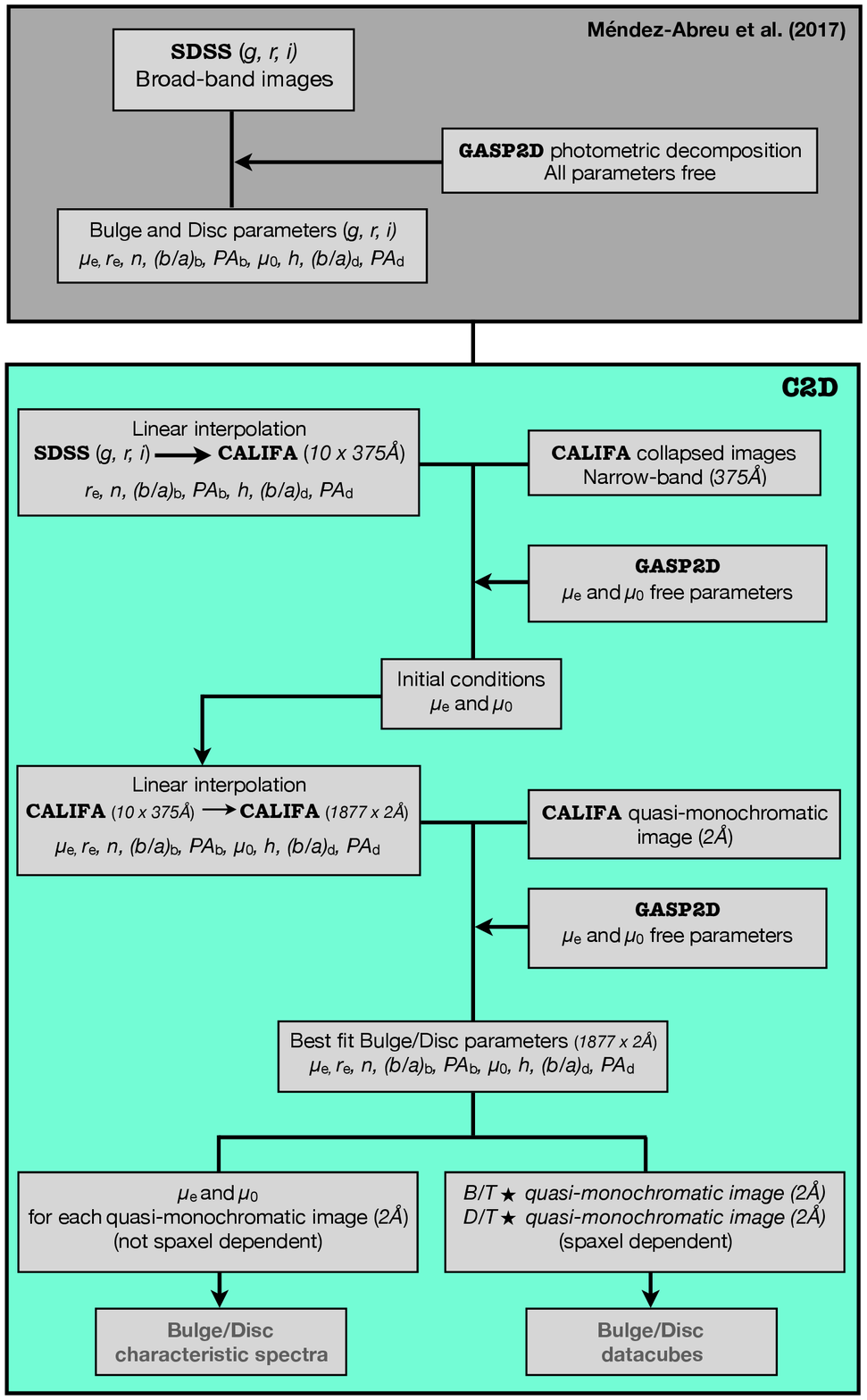}
\caption{Schematic of the {\sc c2d} code. The upper panel describes the 2D photometric decomposition carried out using SDSS images, and used as an input to {\sc c2d} (see \citealt{mendezabreu17}). The lower panel shows the different steps followed by {\sc c2d} to provide the two outputs: the characteristic 1D bulge/disc spectra and the 2D datacubes.}
\label{fig:scheme}
\end{center}
\end{figure*}

IFS provides the best of two worlds: photometry and spectroscopy. Each datacube can be considered as a set of 2D galaxy images obtained at different wavelengths. Therefore one could easily anticipate that standard 2D photometric decomposition techniques, such as those used with broad-band imaging, can be applied to each quasi-monochromatic 2D image of the datacube, disregarding the kinematic distribution within the field-of-view (FoV). This simple idea has not been extensively tested yet, but some attempts in small galaxy samples have been done in \citet{johnston17},  with the main difference that they obliterate the stellar kinematics. {\sc c2d} is based on this concept. The detailed description of the code (see Fig.~\ref{fig:scheme} for a sketch) and its application to the CALIFA datacubes are presented in the following sections.

%--------------------------------------------------------
%--------------------------------------------------------

%---------------------------------------------------
\subsection{GASP2D. The photometric decomposition engine.}
\label{sec:GASP2D}

The structure of the {\sc c2d} code is divided in a number of steps necessary to prepare the datacubes and produce the output results. However, its main engine consists on performing robust and accurate 2D photometric decompositions of large set of images (see Fig.~\ref{fig:scheme}). This task is carried out in {\sc c2d}  by the GASP2D algorithm \citep[see][for a detailed description]{mendezabreu08a,mendezabreu14}. GASP2D have been tested in a wide range of extragalactic studies, from the relatively simple brightest cluster galaxies \citep{mendezabreu12}, galaxies hosting AGNs at their centers \citep{benitez13}, or more complex systems with multiple structural components such as bulges, truncated discs, double bars, barlenses, and inner discs \citep{delorenzocaceres19b}. The code works by fitting a set of analytical models to the galaxy image in a pixel-by-pixel basis. The minimisation is performed using a Levenberg-Marquardt algorithm and the galaxy model is convolved with the data point spread function (PSF) at each iteration before comparing with the real observations.

GASP2D is a highly flexible algorithm with a variety of analytical functions available to represent the different galactic structures. In this particular application, we used a S\'ersic function \citep{sersic68} to describe the light of the bulge component:

%-------------------------------------------------------------------- 
\begin{equation} 
I_{\rm b}(r_{\rm b})=I_{\rm e}10^{-b_n\left[\left(\frac{r_{\rm b}}{r_{\rm e}} 
\right)^{\frac{1}{n}}-1\right]}, 
\label{eqn:bulge_surfbright} 
\end{equation} 
%-------------------------------------------------------------------- 
% 
where $r_{\rm  b}$ is the radius measured in the reference  system of the bulge. $r_{\rm e}$, $I_{\rm e}$, and $n$ are the effective (or half-light) radius, the surface brightness at $r_{\rm e}$, and the S\'ersic index describing the curvature of the profile, respectively, and $b_n \simeq 0.868\,n-0.142$ \citep{caon93}.

The surface  brightness distribution of a galaxy disc was described as a single exponential \citep{freeman70}

%-------------------------------------------------------------------- 
\begin{equation} 
I_{\rm d}(r_{\rm d})=I_{\rm 0}\, e^{-\frac{r_{\rm d}}{h}} ,
\label{eqn:disk} 
\end{equation} 
%-------------------------------------------------------------------- 
% 
where I$_{\rm 0}$ and $h$ represents the central intensity and scale-length of the disc.

Both components were built into two independent reference frames centred at the galaxy center ($x_0$, $y_0$), but with different axis ratios ($(b/a)_{\rm b}$,$(b/a)_{\rm d}$) and position angles ($PA_{\rm b}$,$PA_{\rm d}$).

%---------------------------------------------------
\subsection{CALIFA specific pre-processing analysis}
\label{sec:preprocessing}

In general, we would like to apply a standard 2D bulge/disc photometric decomposition to each quasi-monochromatic image of the corresponding datacube. This would imply fitting 11 free parameters for each image: five for the bulge ($\mu_{\rm e}$, $r_{\rm e}$, $n$, $(b/a)_{\rm b}$, and $PA_{\rm b}$), four for the disc ($\mu_{\rm 0}$, $h$, $(b/a)_{\rm d}$, and $PA_{\rm d}$), and the galaxy centre. Besides the amount of computing power necessary to carry out this task, the CALIFA datacubes present another drawback that prevent us from using this brute force computation. First, the signal-to-noise (S/N) of every quasi-monochromatic image might not be sufficient for a robust photometric decomposition. In general, the central regions of the galaxy do not represent a problem, but the outer disc parameters can sometimes be difficult to constrain. It is important to take into account that the fitting is performed in the reconstructed datacubes without any  kind of spatial binning. Second, the spatial resolution of CALIFA is only enough to resolve the largest bulges in the survey. This problem was highlighted in \citet{mendezabreu18} and it implies that a photometric decomposition performed directly on the CALIFA datacubes, even if collapsed to increase the S/N, might not be reliable for a significant fraction of the CALIFA sample. We should note that the situation would be worst for data extracted from other IFS surveys like MaNGA and SAMI, due to their slightly worse physical spatial resolution and lower S/N at the same galactocentric distances \citep{sanchez17}. On the other hand, more recent datasets like the one provided by the AMUSING survey using MUSE observations \citep{galbany16,galbany18}, would be better placed for a direct photometric analysis on the datacubes.

To overcome these problems we adopted an alternative strategy. We used the information from our previous SDSS multi-band photometric decompositions of the CALIFA sample \citep{mendezabreu17}. SDSS imaging provides the right combination of photometric depth (to reach the outer disc with enough S/N) and image quality (the typical CALIFA PSF is $\sim$ 2 times worst than in SDSS). The suitability of SDSS imaging to perform careful photometric decompositions have been demonstrated in several works in the literature \citep{gadotti09,mendezabreu17, costantin17a}. Therefore, for each galaxy we have  a robust photometric decomposition in the standard $g$, $r$, and $i$ SDSS bands available, each image centred at 4770\AA, 6231\AA, and 7625\AA, respectively. The structural parameters of the different galaxy components are known to smoothly vary with wavelength \citep[e.g.,][]{vika13}. In fact, our derived values for the different structural parameters of both the bulge and disc are not the same in the three bands. Thus, we performed a linear fit to the wavelength dependence of each parameter in order to assign a given value to each quasi-monochromatic image. The structural parameters are then fixed during the fit to the datacube according to this variation. Only the respective intensities of the bulge and the disc ($\mu_{\rm e}$ and $\mu_{\rm 0}$) are left free to vary during the fit. We explored alternative parameterisations of the wavelength dependence of the different structural parameters, in particular constant variation and second order polynomial. However, the first revealed to provide a poor representation of the galaxies in the datacubes, as seen in the residuals, whereas the latter allowed some non-physical solutions (i.e., negative values of the scale-length of the components) due to the fact that only three observational datapoints were available. Our assumptions are somehow similar to those presented in \citet{johnston17}. They also used a linear interpolation of the structural parameters over their wavelength range, but their bulge and disc intensities are allowed to vary smoothly according to Chebyshev polynomials. On the other hand, they considered broad-band data obtained from collapsing the spectroscopic datacube, i.e., at the same spatial resolution and not from SDSS or other imaging survey. The input photometric parameter for our sample galaxies obtained from the SDSS imaging are shown in Table~\ref{tab:phot}. 

%--------------------------------------------------------------------------------------------------------------
\begin{table*}
 \centering
  \caption{Broad-band photometric properties of the galaxy sample}
  \label{tab:phot}
  \begin{tabular}{lcccccccccc}
  \hline
Name  & Band & $r_{\rm e}$ & $n$ & $(b/a)_{\rm b}$ & $PA_{\rm b}$ & $h$ & $(b/a)_{\rm d}$ & $PA_{\rm d}$ & $B/T$ & $r_{\rm e,gal}$ \\
 (1)  & (2) & (3) & (4) & (5) & (6) & (7) & (8) & (9) & (10) & (11)\\
\hline
\hline
NGC\,1167 & g & 7.6$\pm$0.6  & 2.21$\pm$0.08 & 0.91$\pm$0.01 & 69.4$\pm$1.6 & 24.5$\pm$0.7 & 0.787$\pm$0.008 & 73.0$\pm$0.6 & 0.31 & --\\
NGC\,1167 & r & 9.1$\pm$0.7  & 2.66$\pm$0.09 & 0.88$\pm$0.01 & 64.4$\pm$1.6 & 25.8$\pm$0.7 & 0.836$\pm$0.008 & 72.3$\pm$0.6 & 0.34 & 24.9\\
NGC\,1167 & i & 11.4$\pm$0.8 & 3.21$\pm$0.11 & 0.90$\pm$0.01 & 70.9$\pm$1.6 & 27.5$\pm$0.8 & 0.757$\pm$0.008 & 70.2$\pm$0.6 & 0.40 & --\\
NGC\,1349 & g & 2.2$\pm$0.3  & 1.54$\pm$0.09 & 0.95$\pm$0.02 & 98.0$\pm$3.1 & 12.4$\pm$0.8 & 0.88$\pm$0.01   & 97.8$\pm$1.1 & 0.16 & --\\ 
NGC\,1349 & r & 3.6$\pm$0.4  & 2.98$\pm$0.18 & 0.95$\pm$0.01 & 98.0$\pm$3.1 & 13.6$\pm$0.9 & 0.87$\pm$0.01   & 97.7$\pm$1.1 & 0.26 & 17.0\\ 
NGC\,1349 & i & 3.2$\pm$0.4  & 2.45$\pm$0.15 & 0.95$\pm$0.02 & 98.0$\pm$3.1 & 13.6$\pm$0.9 & 0.88$\pm$0.01   & 97.7$\pm$1.1 & 0.24 & --\\
NGC\,3106 & g & 5.9$\pm$0.4  & 5.17$\pm$0.18 & 0.95$\pm$0.01 & 142.1$\pm$1.6& 18.9$\pm$0.5 & 0.89$\pm$0.01   & 133.3$\pm$0.7& 0.33 & --\\
NGC\,3106 & r & 4.8$\pm$0.4  & 3.86$\pm$0.13 & 0.96$\pm$0.01 & 149.3$\pm$1.6& 18.1$\pm$0.5 & 0.901$\pm$0.008 & 133.1$\pm$0.6& 0.34 & 21.4\\
NGC\,3106 & i & 6.1$\pm$0.5  & 5.23$\pm$0.18 & 0.96$\pm$0.01 & 144.0$\pm$1.6& 18.1$\pm$0.5 & 0.903$\pm$0.008 & 135.3$\pm$0.6& 0.37 & --\\
\hline
\end{tabular}
\begin{minipage}{16.5cm}
(1) Galaxy name; (2) SDSS photometric band, (3) bulge effective radius in arcsec, (4), (5), and (6) S\'ersic index, semi-axis ratio, and position angle of the bulge, (7) disc scale-lenght in arcsec, (8) and (9) semi-axis ratio and position angle of the disc, (10) bulge-to-total luminosity ratio, (11) galaxy effective radius in arcsec. (3) to (10) values have been taken from \citet{mendezabreu17}. (11) values taken from \citet{walcher14}.
\end{minipage}\end{table*}
%--------------------------------------------------------------------------------------------------------------

%---------------------------------------------------
\subsection{Application of {\sc c2d} to the datacubes}
\label{sec:c2d}

The pre-processing stage described in Sect.~\ref{sec:preprocessing} was necessary due to the technical characteristics of the CALIFA datacubes. In this section we will describe the general case where every quasi-monochromatic image of the datacube have enough S/N and spatial resolution to perform a standard 2D photometric decomposition with all the structural parameters free to vary. An example of this analysis with {\sc c2d}, using a single S\'ersic component, was presented in \citet{galbany18} using observations obtained with MUSE at the VLT.

The first step is to collapse the datacube into {\it narrow-band-like} images to increase the S/N. The CALIFA datacube was collapsed into 10 different narrow band images of $\sim$375 \AA\, each. The photometric decomposition using GASP2D is then carried out over these images. In the case of CALIFA datacubes, the structural parameters of the bulge and disc are kept fixed to the linearly interpolated values in the corresponding wavelength bins. Only the intensities are then fitted. The best fitted intensities are then linearly interpolated to the full wavelength grid to be used as initial conditions for the final fit. In a general case, all parameters would have been let free during the fit to the {\it narrow-band} images and the linear interpolation to get initial values to the final fit would have applied to all parameters.

The next step is to perform the photometric decomposition to each quasi-monochromatic image of the datacube. In a general case, the values for each parameter derived from the linear interpolation of the narrow band images will be used a initial conditions. Then, the fit is carried out in a standard way with all parameters allowed to vary. In our specific case, we kept fixed the structural parameters to the values obtained from the linear fit to the SDSS broad-band images, and only the intensities are free to vary  (see Sect.~\ref{sec:preprocessing}).

%---------------------------------------------------
\subsection{{\sc c2d} data-products}

Our analysis of the CALIFA datacubes provides the characteristic spectra of both bulge and disc as a 0th order data-product: the  fitted values of the intensities for each quasi-monochromatic image returns directly these characteristic spectra. In addition, {\sc c2d} computes an independent datacube (with spatial and spectral information) for each component. We derived, for each quasi-monochromatic image, the relative contribution of both the bulge (bulge-to-total, $B/T$) and the disc (disc-to-total, $D/T=1-B/T$) component for each spaxel. Each fraction is then multiplied by the observed CALIFA datacube in that spaxel and wavelength. The $B/T$ (and $D/T$) fraction is spatially dependent, thus, we can compute a different bulge and disc quasi-monochromatic image. Rearranging the quasi-monochromatic images as a function of the wavelength produces the independent bulge and disc datacubes that we analyse using the {\sc Pipe3D} algorithm in the next sections.

%---------------------------------------------------
%---------------------------------------------------
\section{{\sc Pipe3D}. Derivation of the stellar and ionised-gas properties.}
\label{sec:pipe3D}

The analysis of both the bulge and disc datacubes, obtained from {\sc c2d}, was carried out using the {\sc Pipe3D} pipeline \citep{sanchez16b,sanchez16a}. This analysis uses an improved version of the FIT3D fitting tool, developed to explore the properties of the stellar populations and ionized gas from IFS data. {\sc Pipe3D} has been extensively tested on the datasets extracted from different IFS galaxy surveys, including CALIFA, MaNGA, SAMI, and AMUSING \citep{sanchez15,sanchez16a,sanchezmenguiano18,sanchez19}. Thus, since the outputs of {\sc c2d} are produced in the CALIFA format, the combination of {\sc c2d}+{\sc Pipe3D} represents the best strategy to derive the stellar and ionised-gas properties of our bulges and discs. We refer the reader to the presentation paper of {\sc Pipe3D} for a detailed description of the code and its application to CALIFA data \citep{sanchez16a}. For the sake of clarity, we describe here its main characteristics and data-products. 

{\sc Pipe3D} makes use of the GSD156 library of simple stellar populations \citep[SSPs;][]{cidfernandes13}. It comprises 156 templates spanning 39 stellar ages (from 1 Myr to 14.1 Gyr), and 4 metallicities ($Z/Z_{\sun}$=0.2 dex, 0.4 dex, 1 dex, and 1.5 dex). These templates have been largely used within the CALIFA collaboration \citep[e.g.,][]{perez13,gonzalezdelgado14a}, and analyses from other IFS surveys \citep{ibarramedel16,sanchez18}. The datacubes are first binned to increase the S/N of the spectra. The binning strategy is described in \citet{sanchez16b} and it is designed to minimise the impact on the original shape of the galaxy. Two criteria are followed during the binning process: i) a foreseen goal of S/N=50 per \AA \, is set for the binned spectra (allowing for a derivation of the stellar population properties with uncertainties $\sim$10-15\%, \citealt{sanchez16b}), and ii) a maximum difference in the flux intensity between adjacent spaxels of 15\%. Once the spatial binning/segmentation is performed, the spectra from the spaxels of each bin are co-added prior to any further analysis.

The stellar population fitting procedure comprises two steps: first, the stellar kinematic parameters (velocity and velocity dispersion) are derived together with the average dust attenuation affecting the stellar populations ($A_{\rm V, ssp}$). Second, a multi-SSP linear fitting is performed, using the GSD156 library, and adopting the stellar kinematics and dust attenuation derived in the first step. This second step is repeated in a Monte-Carlo fashion including the errors of the original spectrum. The best coefficients of the linear fit and their errors are then obtained by marginalising over the remaining parameters derived for the stellar populations. At the end of this analysis we have a model of the stellar populations for each spectrum in each spectral bin.

Finally, we estimate the stellar population model for each spaxel by re-scaling the best model within each spatial bin to the continuum flux intensity in the corresponding spaxel \citep{sanchez16a}. This model is used to derive the average stellar properties at each spaxel, including the stellar mass density, light- and mass-weighted stellar age and metallicity, and the average dust attenuation. Moreover, the same parameters are derived as a function of look-back time thus providing the star formation and chemical enrichment histories of the galaxy at different locations.

The stellar population model spectrum for each spaxel is then subtracted from the original cube to produce a gas-pure cube with only the ionized-gas emission lines. Individual emission line fluxes are measured spaxel by spaxel using two different procedures (1) fitting a single Gaussian function for each emission line and spectrum, and (2) by a non-parametric weighted momentum analysis.

%---------------------------------------------------
%---------------------------------------------------
\section{Test with idealised mock datacubes}
\label{sec:idealmock}

A set of reliability tests were performed on {\sc c2d} using idealised datacubes with the same observational characteristics of the CALIFA datacubes. The general procedure used to build the simulated datacubes is explained in Appendix B of \citet{mendezabreu18}. For the sake of completeness we will describe here their main characteristics.

Mock datacubes were built following the same technical characteristics of real CALIFA V500 data in terms of spatial and spectral resolution. Models were created with a 1 arcsec spaxel resolution in a frame of 77$\times$77 arcsec$^2$ field-of-view (FoV), covering the wavelength range 3745\AA-7300\AA\ in a grid of 2\AA/px and with an instrumental spectral resolution of 6\AA. All datacubes were created from scratch including kinematic, photometric, and stellar population properties similar to those of the CALIFA galaxies. 

The stellar kinematic properties of the mock sample were assumed to be the same for the whole set of simulations. The velocity field was modeled following the parameterisation of \citet{salucci07}, with a maximum rotational velocity $v_{\rm max}=100, 250$ km/s and a typical spatial scale of $r_{\rm v}=5, 10$ arcsec for the bulge and the disc, respectively. The velocity dispersion distribution was modeled using an exponential profile with a maximum velocity dispersion $\sigma_{\rm max}=250, 100$ km/s and a scale-lenght $r_{\sigma}=10, 10$ arcsec for the bulge and the disc, respectively. These values are consistent with those measured for the CALIFA sample of ETGs described in \citet{mendezabreu18}.

%--------------------------------------------------------
\begin{figure*}
\begin{center}
\includegraphics[bb=54 141 394 707,width=12cm]{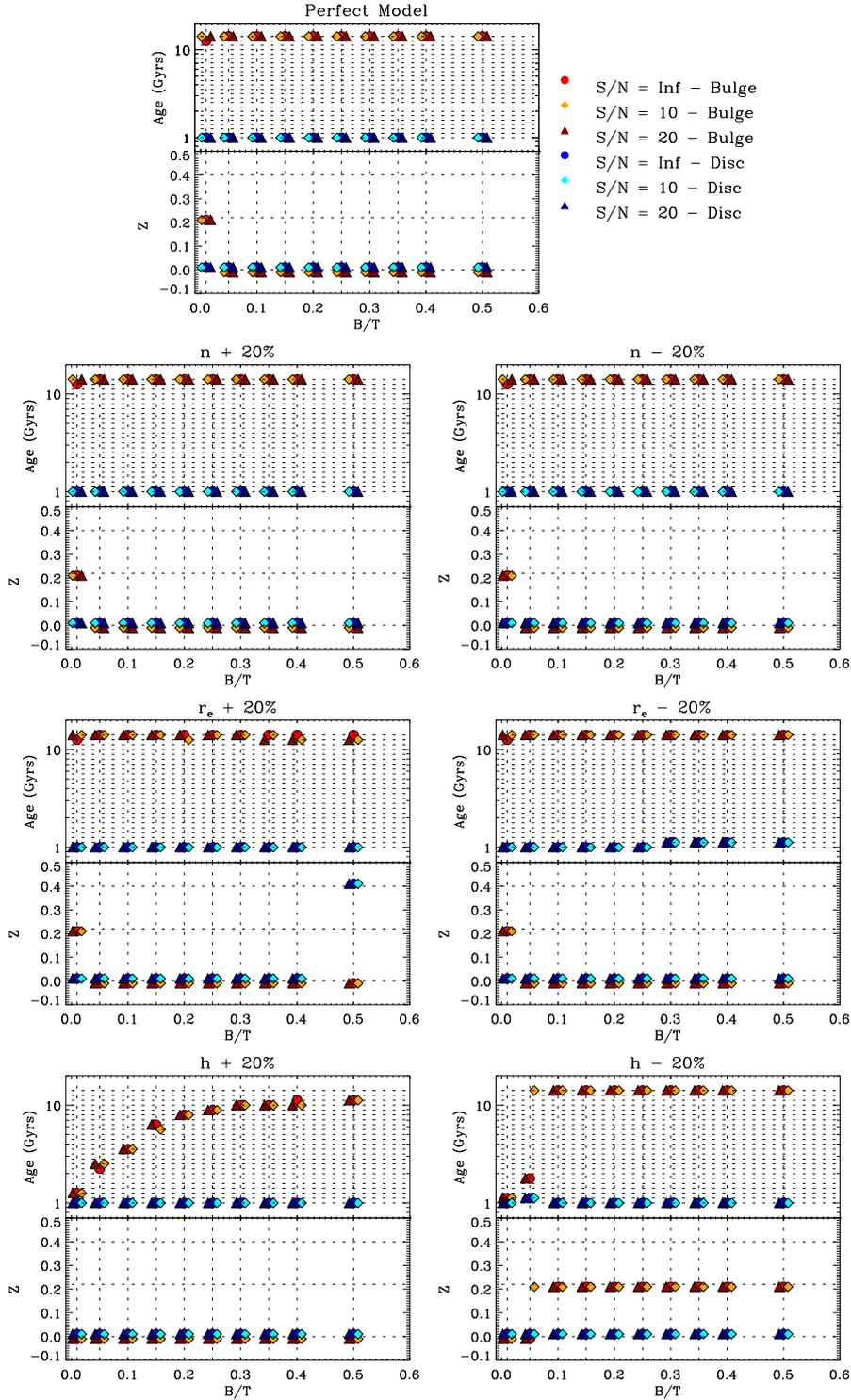}
\caption{Distribution of ages and metallicities for  the bulge and disc recovered using {\sc c2d} as a function of the $B/T$ luminosity ratio of the model. The bulge component was simulated using a SSP model with age 14 Gyr and metallicity $Z$=0 dex. The disc component was represented with a SSP with age 1 Gyr and metallicity $Z$=0 dex. Slight offsets with respect to the $Z$=0 dex metallicity are used in the figure for the sake of separating bulge and disc measurements. The full description of kinematic and photometric properties of the mock datacubes is detailed in the text. Results for datacubes with different S/N ratios are represented with the symbols and colors shown in the legend. The top panel shows the results from {\sc c2d} using the same structural parameters in {\sc c2d} as those used to create the models. The second row from the top shows the results when we imposed a plus (left) or minus (right) 20\% difference in the S\'ersic index ($n$) value. The third and fourth rows from the top are analogous, but for the bulge effective radius ($r_{\rm e}$) and disc scale-length ($h$), respectively. Dashed lines show the whole grid of MILES SSP spectra used in the simulations.}
\label{fig:ideal}
\end{center}
\end{figure*}
%--------------------------------------------------------

The photometric properties of the mock galaxies were assumed to be the sum of a central bulge modeled with a S\'ersic distribution and an outer disc modeled with an exponential profile. We used values of $n=2$, $r_{\rm e}=4$ arcsec, and $h=10$ arcsec for the S\'ersic index, bulge effective radius, and disc scale-length, respectively. These are the typical values for simple bulge and disc galaxies found in \citet{mendezabreu17}. The relative flux between both components was allowed to vary to check the robustness of {\sc c2d} against different $B/T$ luminosity ratios. We explore ten different values of $B/T$=0.01, 0.05, 0.1, 0.15, 0.2, 0.25, 0.3, 0.35, 0.4, 0.5. All models were created at an inclination of 45 degrees and convolved with a Gaussian PSF with the typical full width at half maximum (FWHM) of CALIFA observations \citep[FWHM=2.5 arcsec,][]{garciabenito15}. Random Gaussian noise was added to mimic CALIFA observations with S/N=$\infty$ (no noise), 20, and 10.

The stellar population properties were included as a SSP model for each component. SSPs were obtained from the MILES model library \citep{vazdekis15} restricted to the spectral range and downgraded to the instrumental dispersion of the CALIFA data\footnote{http://www.iac.es/proyecto/miles/pages/stellar-libraries/miles-library.php}. We used single age and single metallicity solar-scaled models with values of 14 Gyrs and $Z$=0 dex for the bulge and 1 Gyr and $Z$=0 dex for the disc.

The mock datacubes, including the previous kinematic, photometric, stellar population, and observational characteristics, were then analysed using {\sc c2d} as if they were real galaxies. Using our {\it a priori} knowledge of the input photometric properties of the simulated galaxies, we can test the accuracy of the recovered stellar population properties against both different S/N ratios and possible errors in the photometric parameters. For the latter, we run {\sc c2d} with values of $n$, $r_{\rm e}$, and $h$ a 20\% upper/lower than the actual values. It is worth noting that we did not include any spatial variation of the stellar populations. The characteristic output spectra obtained from {\sc c2d} for both the bulge and disc were then compared with the full set of models (all ages and metallicities) provided in the MILES library using a simple $\chi^2$ test. The best fitted model parameters were finally compared with the input ones. In total we created 70 different datacubes and performed 250 different fits with {\sc c2d}.

Fig.~\ref{fig:ideal} shows a summary of the results for some of these tests. The top panel represents the perfect case where the input photometric parameters for both the bulge and disc are the same as those used to create the models. {\sc c2d} is able to recover the right values for the ages (14 Gyr for the bulge and 1 Gyr for the disc) for all values of $B/T$ ($<$30$\%$). Something similar occurs for the metallicities ($Z$=0 dex for both components) except for the bulge models with the lowest values of $B/T$=0.01. We therefore consider a limit of $B/T$=0.05 as our lower limit to recover the separated stellar populations of bulges and discs with our technique. Regarding the errors induced by a wrong determination of either the bulge or disc structural parameters, we found that neither an underestimation or overestimation of a 20\% in the S\'ersic index of the bulge biases our results for galaxies with $B/T \geq 0.05$. On the other hand, possible errors in the derived spatial scales of the bulge and disc are more problematic. In particular we found a strong dependence of our results with the scale-length of the disc ($h$). Using a disc scale-length 20\% larger than the real one might cause a catastrophic error on the ages of the stellar populations of the bulge, specially for low $B/T$ values. Using a disc scale-length 20\% smaller than the real one produces a smaller impact in the derived ages of the bulge, but it modifies their metallicities. Despite this caveat, the typical error on the disc scale-lengths for bulge-to-disc galaxies is $\sim10\%$ \citep{mendezabreu17} so this effect should be less important for real galaxies given their estimated errors. We also found some slight biases when the perfect bulge effective radius is not used. This happens specially for larger $B/T$ values, but the differences are well within the typical errors of a stellar population analysis \citep[e.g., 0.1-0.15 dex in the mean ages;][]{cidfernandes14,sanchez16a}.

\begin{figure*}
\begin{center}
\includegraphics[bb=54 141 394 566,height=11cm]{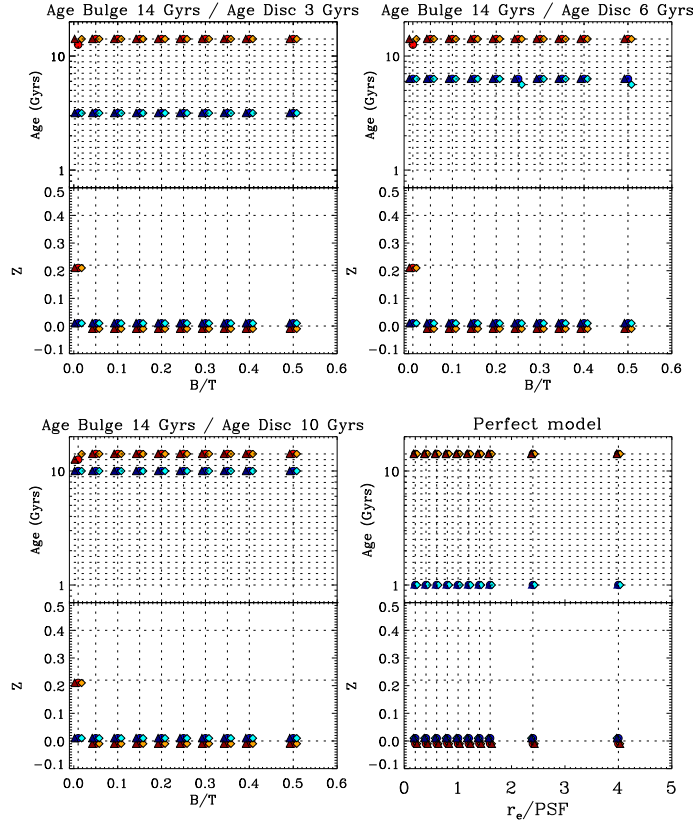}
\caption{Distribution of ages and metallicities for both the bulge and disc recovered using {\sc c2d} as a function of the $B/T$ and $r_e/PSF$ ratios of the model. Symbols, lines, and colors are the same as in Fig.~\ref{fig:ideal}. Models using a disc SSP with $Z$=0 dex and ages of 3 Gyrs, 6 Gyrs, and 10 Gyrs are represented in the top-left, top-right, and bottom-left panels, respectively. The bottom-right panel represents models created with different bulge effective radius, thus producing different $r_e/PSF$ ratios. All models were fitted with {\sc c2d} assuming a perfect knowledge of the photometric parameters.  }
\label{fig:appendix_ideal}
\end{center}
\end{figure*}

 We also analysed a more complex set of idealised mock simulations including both a different set of ages for the disc ($A_{\rm d}$=3, 6, and 10 Gyr) and a different $r_e/PSF$ ratio. Fig.~\ref{fig:appendix_ideal} shows the result of these tests. All fits were performed assuming a perfect knowledge of the photometric model and the comparison with the MILES spectral library was carried out as explained before. It is clear that {\sc c2d} is able to robustly separate the stellar populations of both the bulge and the disc even when they are similar in ages and metallicities. In fact, even in the case of a model with a bulge represented by a SSP of 14 Gyrs and a disc with a SSP of 10 Gyrs, the results are quite accurate for $B/T > 0.05$. The bottom right panel of Fig.~\ref{fig:appendix_ideal} shows the case of a model with $B/T=0.3$ and stellar populations for the bulge and disc of 14 Gyrs ($Z$=0 dex) and 1 Gyr ($Z$=0 dex), respectively. The fits were also carried out using different values of the bulge effective radius. Therefore, the results are represented as a function of the $r_e/PSF$ where the PSF FWHM was fixed to 2.5 arcsecs.

%--------------------------------------------------------
%--------------------------------------------------------

%---------------------------------------------------
%---------------------------------------------------
\section{Analysis of real galaxies from the CALIFA survey}
\label{sec:results}

The aim of this section is to show the potential of the new methodology presented in this paper. We apply the {\sc c2d} algorithm to the CALIFA observations of a sample of 3 ETGs (NGC1167, NGC1349, and NGC3106) analysed in \citet{gomes16}. These galaxies were classified as i+ based on the properties of their warm interstellar medium \citep{gomes16b}, i.e., they are ETGs  characterized by a nearly constant EW (H$\alpha$) $\leq$ 3 \AA\, throughout their extra-nuclear component, with a steep EW (H$\alpha$) increase in their outermost periphery.

We derived the independent bulge and disc datacubes applying the {\sc c2d} code to the CALIFA observations (Sect.~\ref{sec:decomposition}). Then, they were analysed using the {\sc Pipe3D} pipeline (Sect.~\ref{sec:pipe3D}). We finally obtained the spatially resolved stellar population and ionised-gas properties for both components. The results of this analysis are presented in this section.

%----------------------------------------------
\subsection{Stellar population properties of bulges and discs}
\label{sec:SFH}

\begin{figure}
\begin{center}
\includegraphics[width=0.49\textwidth]{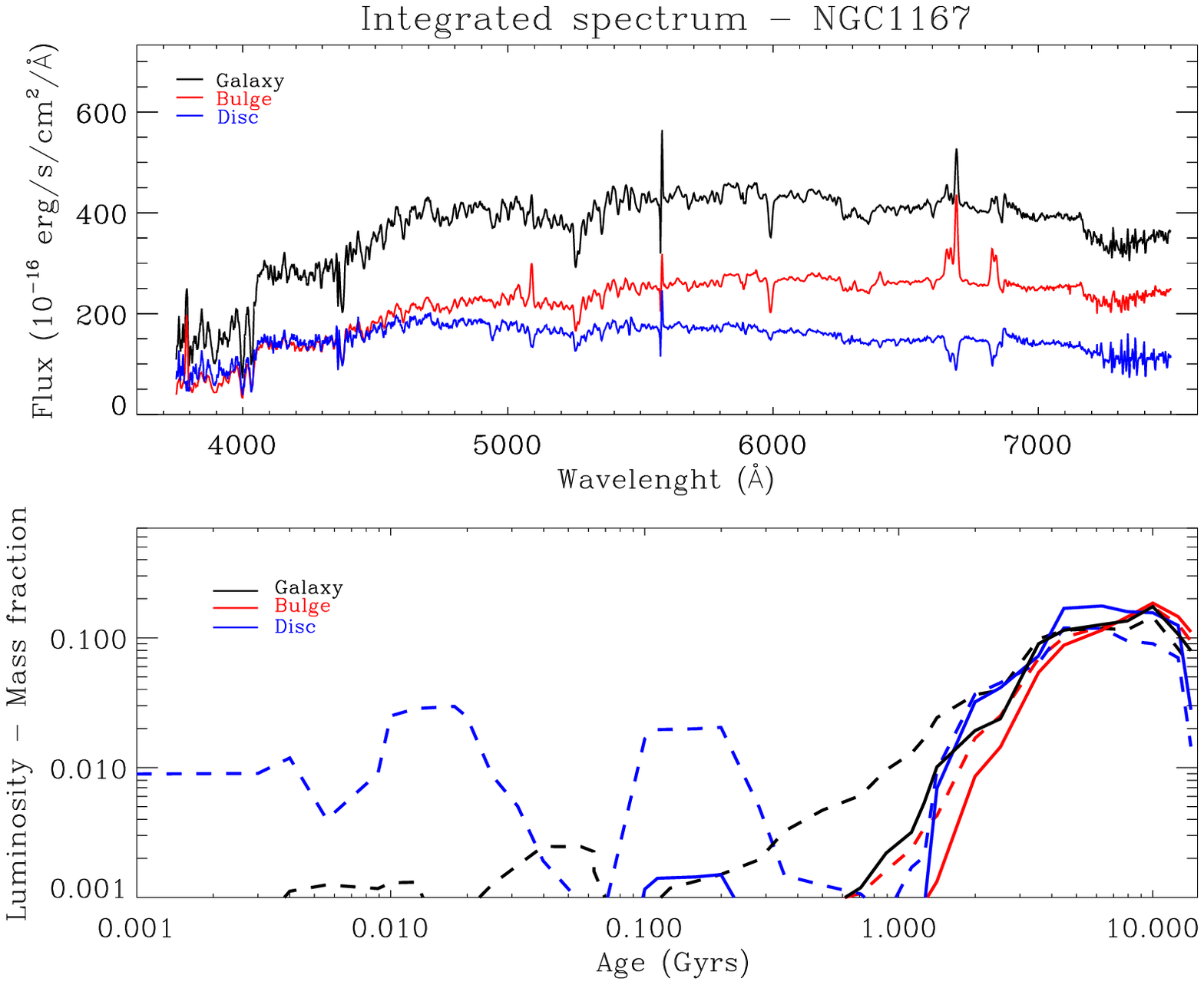}

\vspace{0.5cm}
\includegraphics[width=0.49\textwidth]{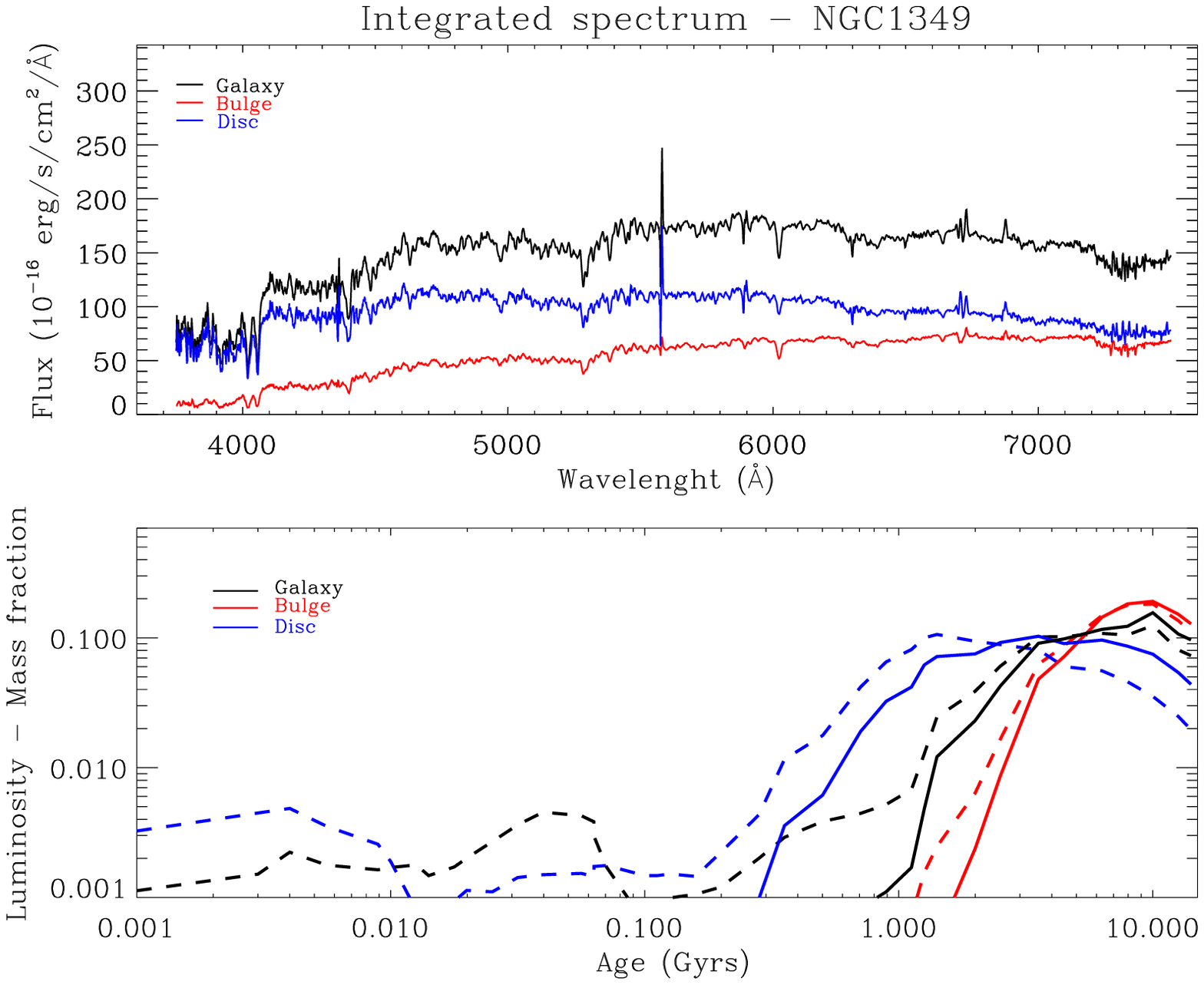}

\vspace{0.5cm}
\includegraphics[width=0.49\textwidth]{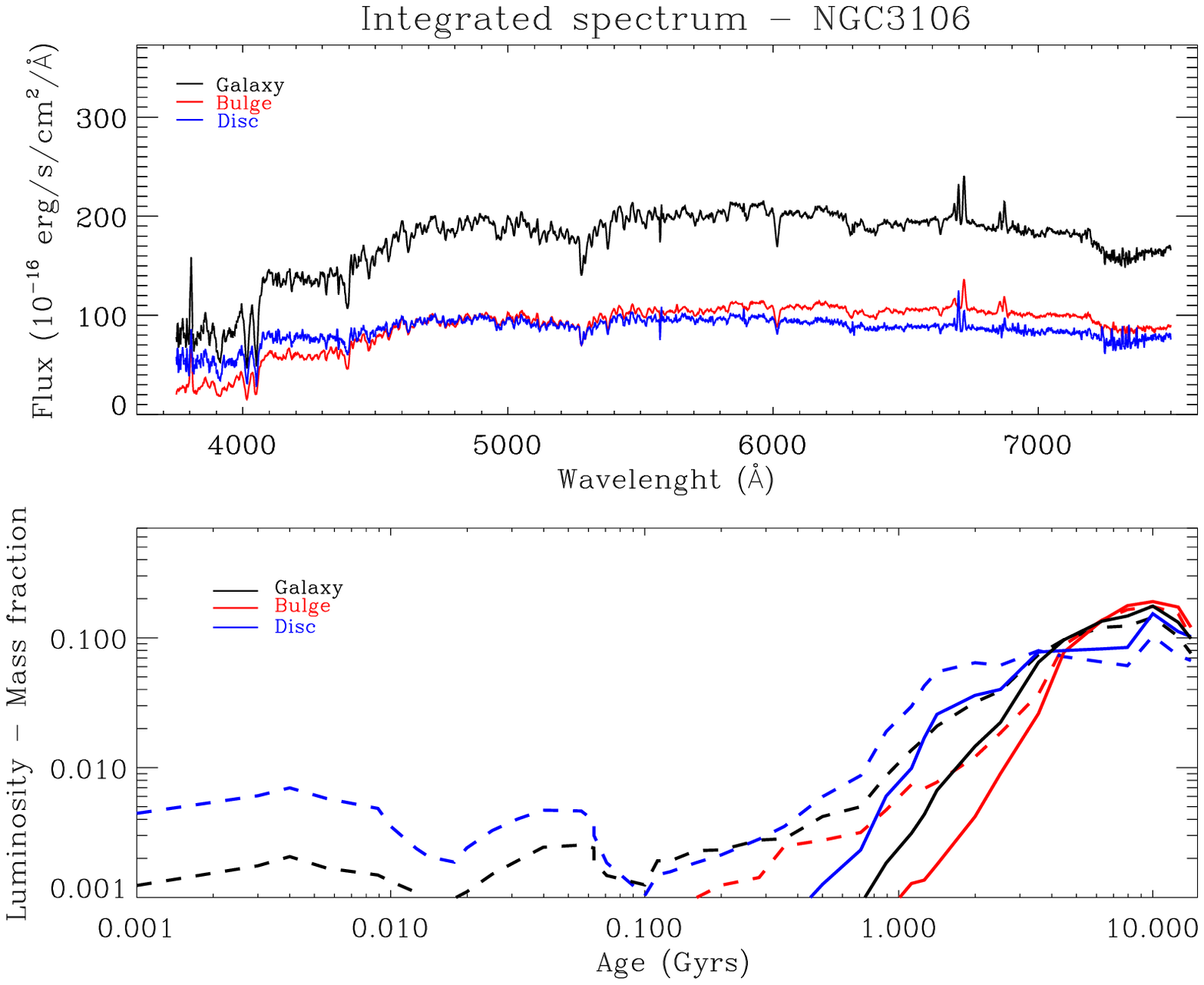}
\caption{Upper panels. Integrated spectrum over one effective radius of the galaxy ($r_{\rm e, gal}$) for the whole galaxy (black), bulge (red), and disc (blue). Bottom panels. Luminosity (dashed lines) and mass (solid lines) fraction of stars contributing to a given stellar population age. Different colors as in the upper panels. Fractions are also computed within 1 $r_{\rm e, gal}$.}
\label{fig:integratedSFH}
\end{center}
\end{figure}

Figure~\ref{fig:integratedSFH} shows the spectra (upper panels) and the fraction of light/mass (bottom panels) contributed by stars of different ages (marginalised over all possible metallicities) for the galaxy, bulge, and disc. All these quantities are integrated within an ellipse with semi-major axis of one effective radius of the galaxy ($r_{\rm e, gal}$, see Table~\ref{tab:phot}) and oriented with the ellipticity and position angle of the outer disc. The mass fraction distribution as a function of the age of the stellar population is frequently referred to as the SFH. In purity, the SFH would be the distribution of the star formation rate (SFR) along cosmological times, and it therefore would be $\frac{\Delta M}{\Delta t}$ at each time. However, one can be easily transformed into the other, as shown in \citet{sanchez19}. We used the effective radius of the galaxy in $r-$band as computed by  \citet{walcher14}. This definition allowed us to compare the derived SFH between different structures within a galaxy and also amongst different galaxies.

All bulges in our sample show a very similar SFH profile. They are old structures with a typical exponentially declining SFH. The SFH of discs is significantly different from that of the bulges. They show a more extended SFH towards younger ages. We quantify this different behaviour in Table~\ref{tab:massfrac}. It shows the stellar mass fraction of young (Age < 1 Gyr), intermediate (1 < Age (Gyr) < 6) , and old stars (Age > 6 Gyr) building up our bulges and discs within 1 $r_{\rm e, gal}$. More than 90\% of the stars in our bulges are older than 6 Gyr with only a tiny fraction formed at ages between 1 < Age (Gyr) < 6. On the contrary, galaxy discs show a significant fraction of stars ($>20\%$) formed at these intermediate ages, with a smaller fraction of old stars.

%--------------------------------------------------------
%--------------------------------------------------------

%---------------------------------------------------------------------------
\begin{table}
 \centering
  \caption{Mass fractions in bulges and discs within 1 $r_{\rm e,gal}$}
  \label{tab:massfrac}
  \begin{tabular}{lcccc}
  \hline
Name  & Structure & Young & Intermediate & Old \\
 (1)  & (2) & (3) & (4) & (5)\\
\hline
\hline
NGC\,1167 & Bulge & 0.00 & 0.07 & 0.93 \\
NGC\,1167 & Disc  & 0.01 & 0.21 & 0.78 \\
NGC\,1349 & Bulge & 0.00 & 0.04 & 0.96 \\
NGC\,1349 & Disc  & 0.03 & 0.59 & 0.38 \\
NGC\,3106 & Bulge & 0.00 & 0.05 & 0.95 \\
NGC\,3106 & Disc  & 0.01 & 0.22 & 0.77 \\
\hline
\end{tabular}
\begin{minipage}{8cm}
(1) Galaxy name; (2) galaxy structure; (3), (4), (5) mass fractions in stars younger than 1 Gyr, between 1 < Age (Gyr) < 6, and older than 6 Gyr, respectively. 
\end{minipage}\end{table}
%-----------------------------------------------------------------------

%--------------------------------------------------------
\begin{figure*}
\begin{center}
\includegraphics[width=8cm]{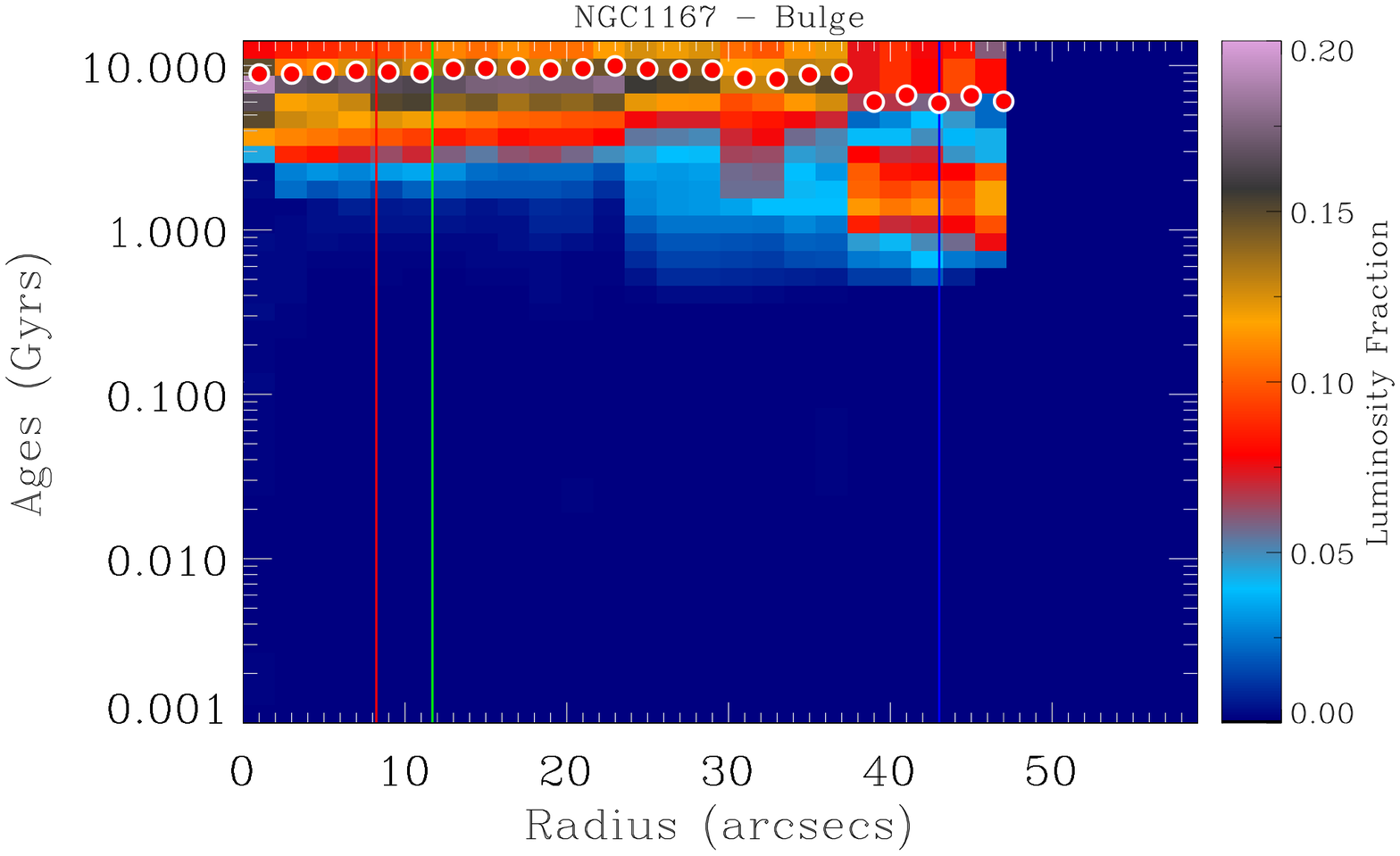}
\includegraphics[width=8cm]{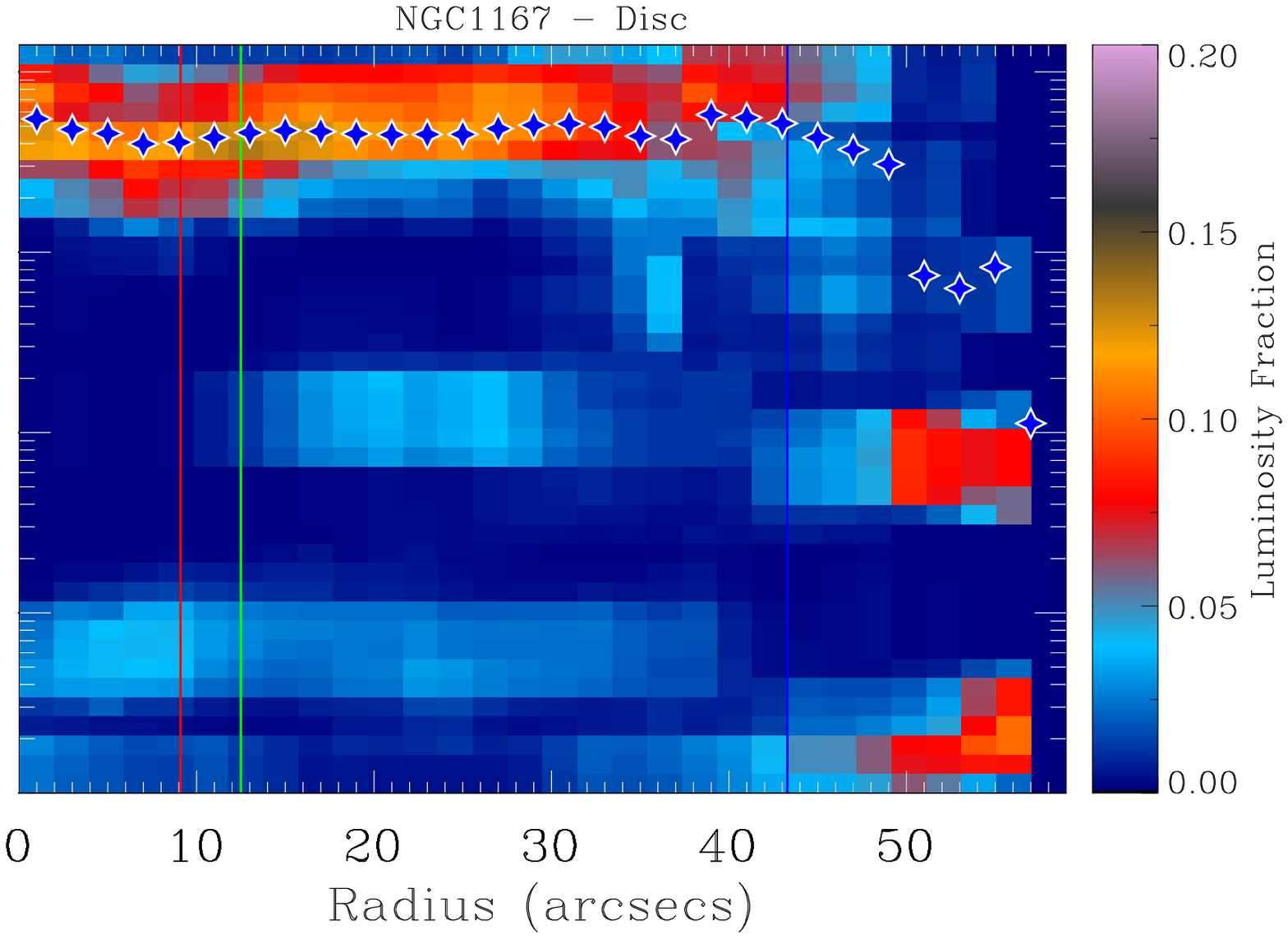}
\includegraphics[width=8cm]{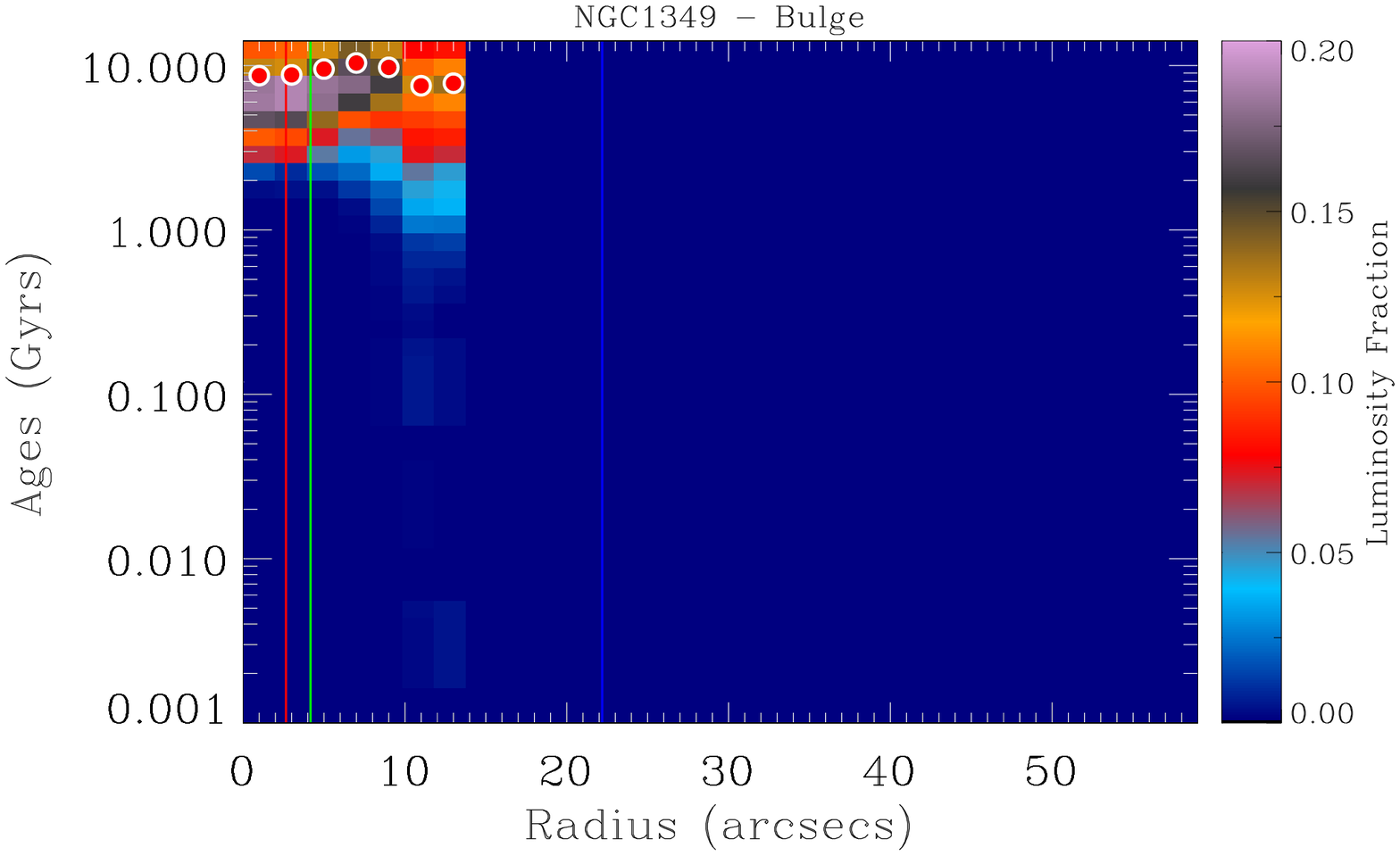}
\includegraphics[width=8cm]{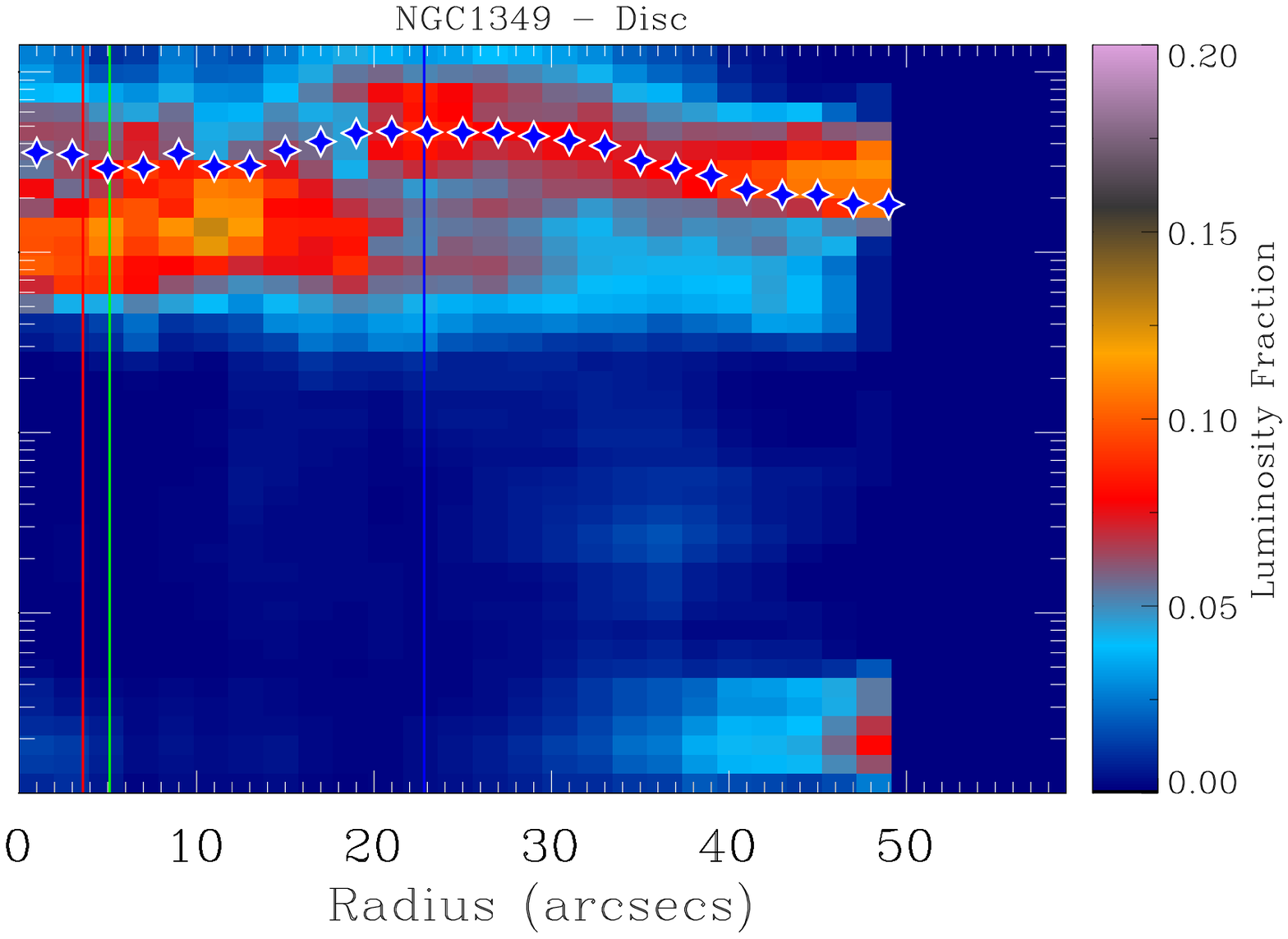}
\includegraphics[width=8cm]{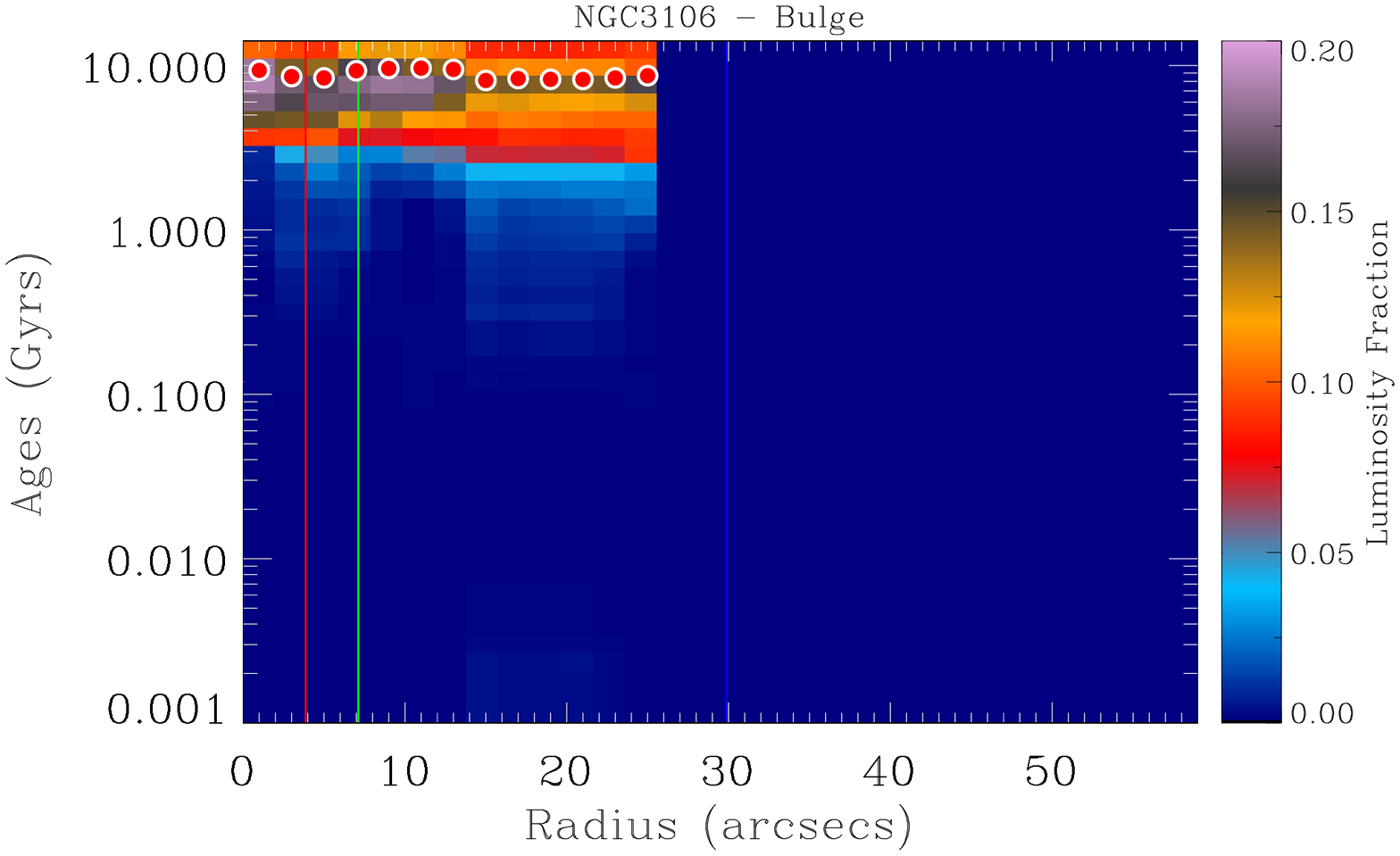}
\includegraphics[width=8cm]{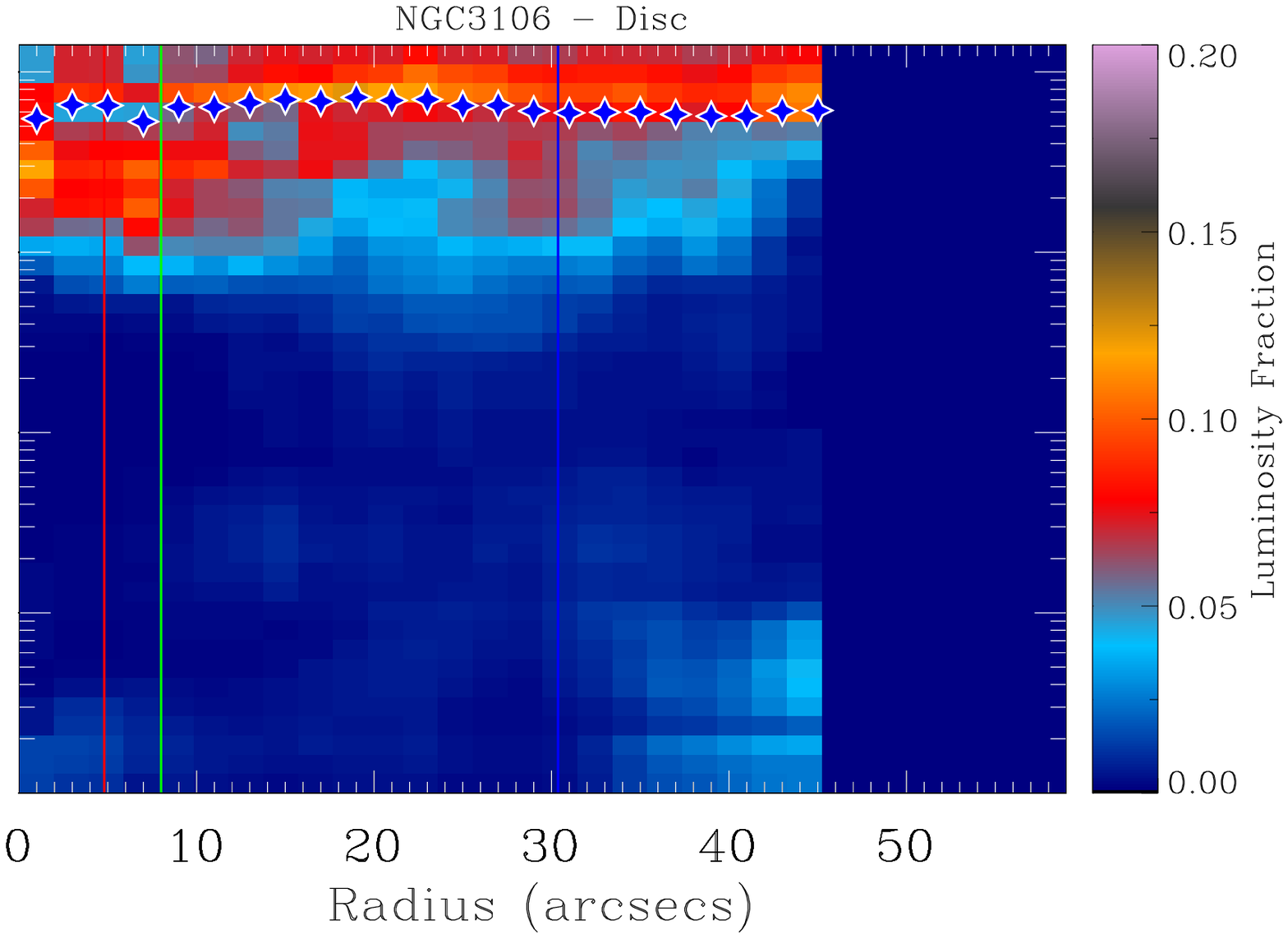}
\caption{Luminosity fraction produced by stars of different ages in elliptically averaged radii with ellipticity and position angle corresponding to either the bulges (left panels) and discs (right panels). The fractions are normalised within each radial bin. The red dots and blue stars show the luminosity weighted ages at the corresponding radii fro the bulge and the disc, respectively. The red, blue, and green vertical lines indicate the effective radius of the bulge, disc, and the radius where the surface brightness of the bulge and  disc are the same.}
\label{fig:SFH}
\end{center}
\end{figure*}
%--------------------------------------------------------

A more complete view to understand the history of mass assembly of bulges and discs can be obtained from Fig.~\ref{fig:SFH}. It shows the radial distribution of the SFH (in luminosity fraction) for the bulges and discs in our sample. The luminosity fraction was integrated in elliptical annuli of 2 arcsec width following the ellipticity and position angle of either the bulge or the disc in the $r-$band (see Table~\ref{tab:phot}). After that, it was normalised to produce an integral of unity in each radial bin. The red circles and blue stars show the luminosity weighted ages of the stellar population at each radius for bulges and discs, respectively. Fig.~\ref{fig:SFH} shows the luminosity fraction, instead of the mass fraction, to highlight the differences in the young stellar populations of both components.

%--------------------------------------------------------
\begin{figure}
\begin{center}
\includegraphics[width=0.45\textwidth]{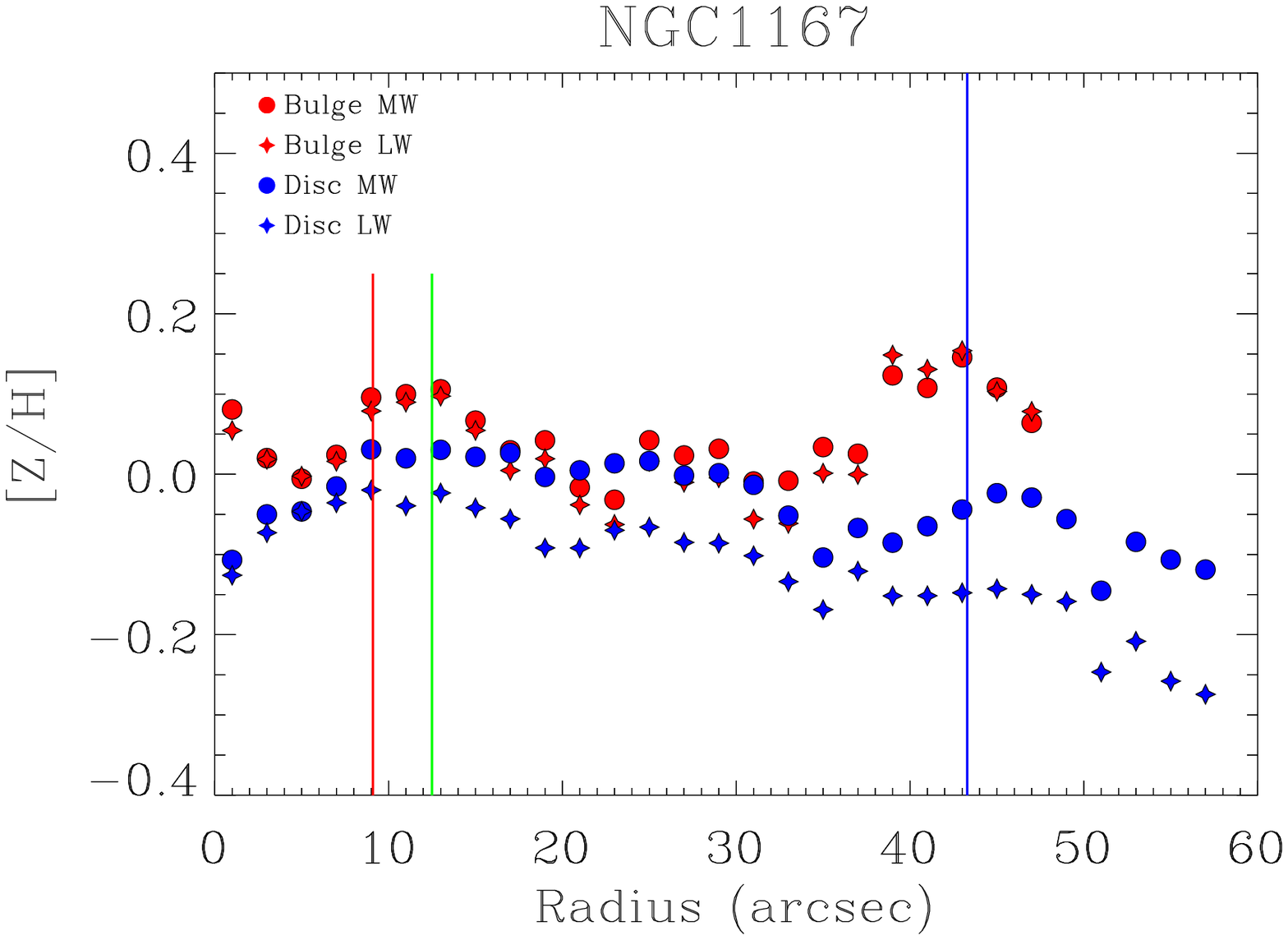}

\vspace{0.5cm}
\includegraphics[width=0.45\textwidth]{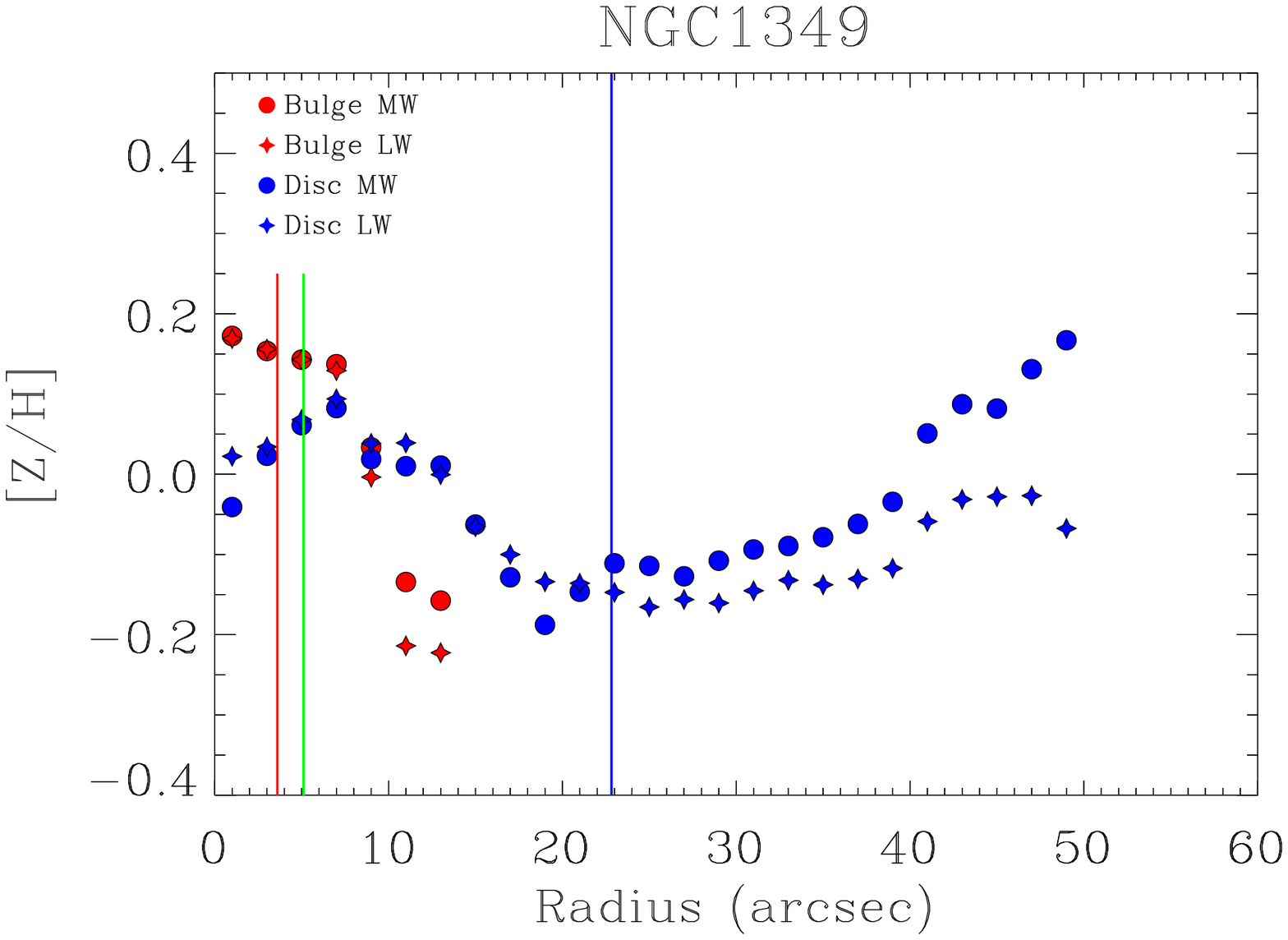}

\vspace{0.5cm}
\includegraphics[width=0.45\textwidth]{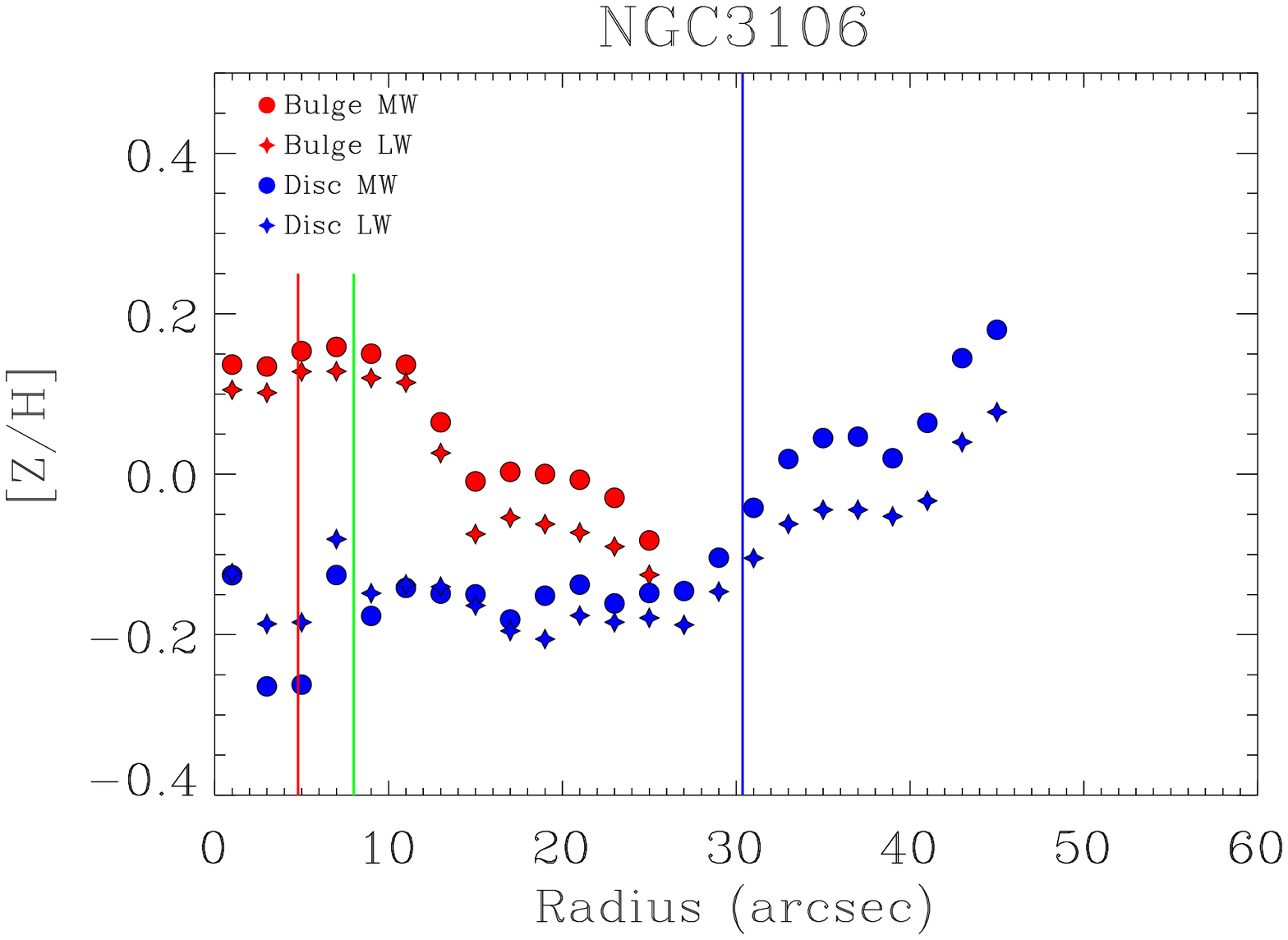}
\caption{Radial profiles for the stellar metallicity of the bulge (red symbols) and the disc (blue symbols). Both the luminosity weighted (LW) and mass weighted (MW) quantities are shown. The red, blue, and green vertical lines indicate the effective radius of the bulge, disc, and the radius where the surface brightness of the bulge and disc are the same. }
\label{fig:metallicity}
\end{center}
\end{figure}
%--------------------------------------------------------

Bulges show a simple radial distribution of the luminosity fraction. Their SFH (luminosity fraction) distribution at all radii is dominated by old stellar population with mean ages $\sim$9 Gyr for all three galaxies. On the other hand, galaxy discs show a different pattern. They have a similar SFH to the bulges (even if with younger ages, $\sim$5 Gyr) at low radii, but Fig.~\ref{fig:SFH} shows evidence of multiple bursts of SF at larger radii (generally beyond the effective radius of the disc). These bursty events are stronger in NGC1167, where they appear at two different ages ($\sim$3 Myr and $\sim$70 Myr). Such SF events represent a small contribution to the total galaxy mass, but they confirm the presence of gas able to form stars in the outer regions of these ETGs, in agreement with \citet{gomes16}.
 
Figure~\ref{fig:metallicity} shows the luminosity weighted (LW) and mass weighted (MW) radial profiles of the stellar metallicity in our bulges and discs. Two different trends can be seen for the three galaxies under study. Both the bulge and disc of NGC1167 show a nearly constant stellar metallicity radial profile over the whole extension of the galaxy, i.e., even beyond the effective radius of the disc (blue lines in Figure~\ref{fig:metallicity}). This trend is also independent of the metallicity being  light or mass weighted. We note that NGC1167 was photometrically classified as unknown, i.e., the presence of a disc cannot be assured statistically, and therefore this result might be interpreted as a lack of such a disc. On the contrary, the bulges of both NGC1349 and NGC3106 are more metal-rich at their centers showing a steep decline at larger radii while their discs show an increase in the metallicity at large radii (beyond the effective radius of disc), but they are constant or decreasing for the bulge and disc in the more central regions.

%----------------------------------------------
\subsection{Ionised-gas properties of bulges and discs}
\label{sec:gas}

\begin{figure}
\begin{center}
\includegraphics[width=0.45\textwidth]{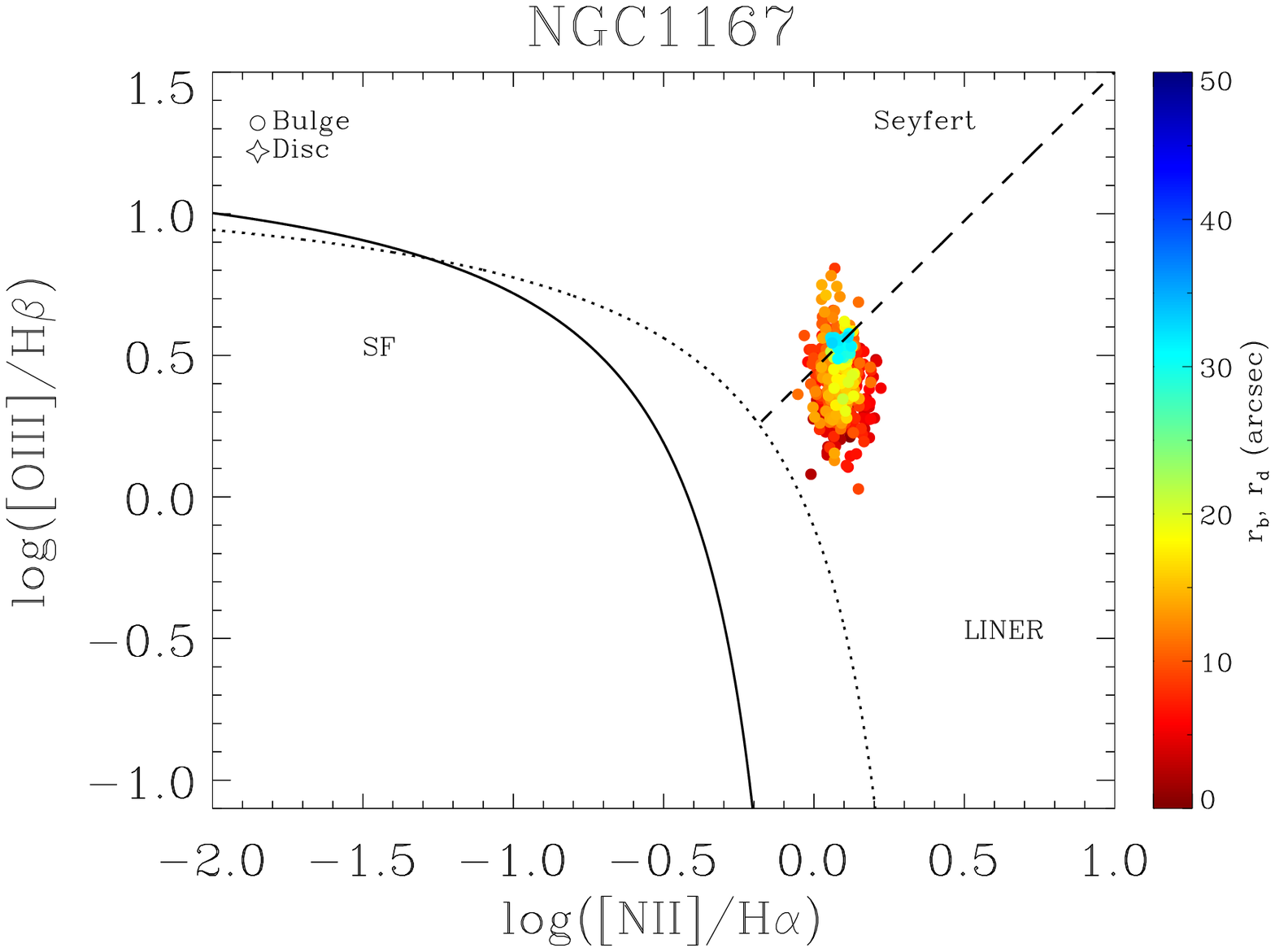}

\vspace{0.5cm}

\includegraphics[width=0.45\textwidth]{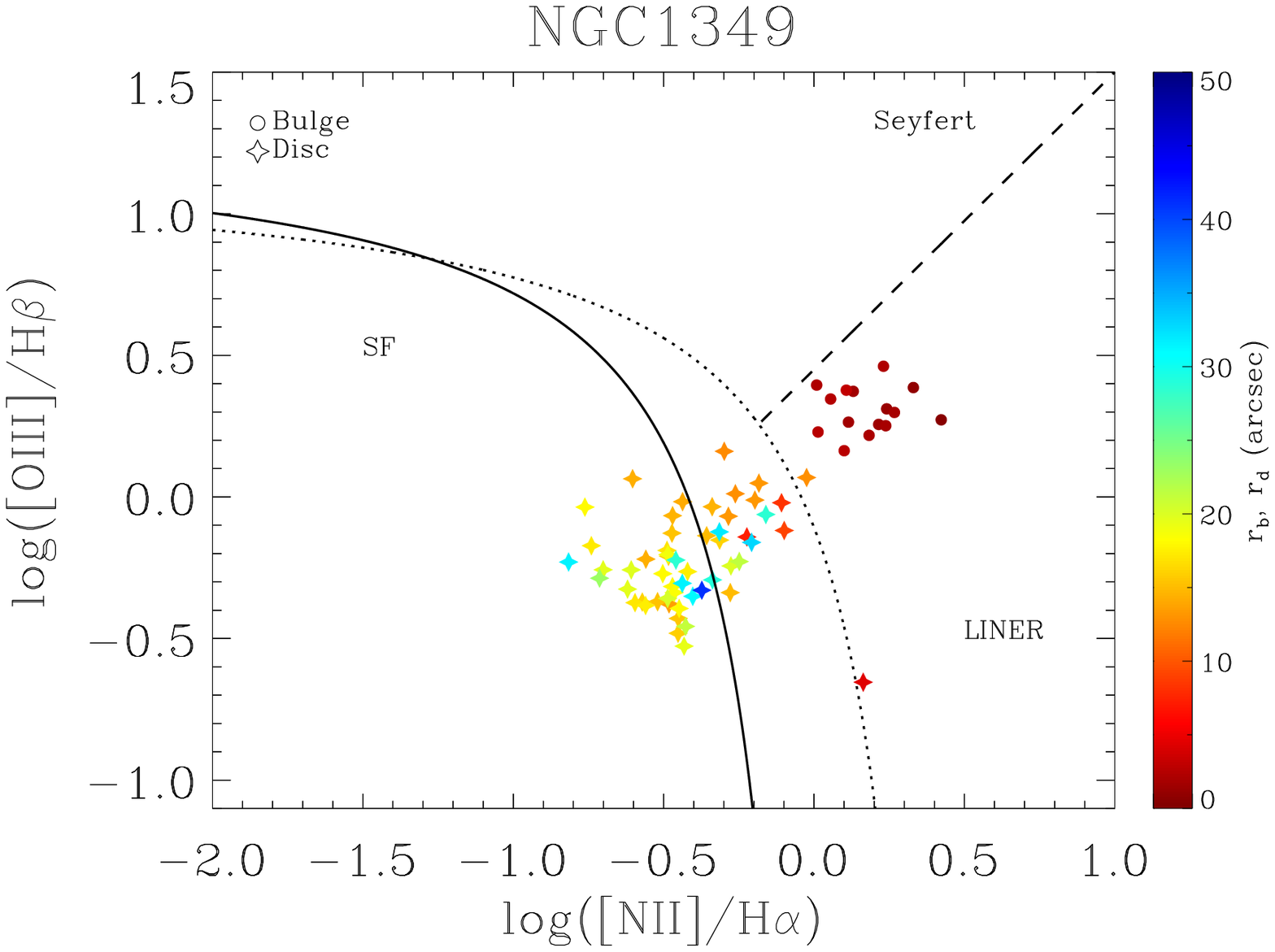}

\vspace{0.5cm}
\includegraphics[width=0.45\textwidth]{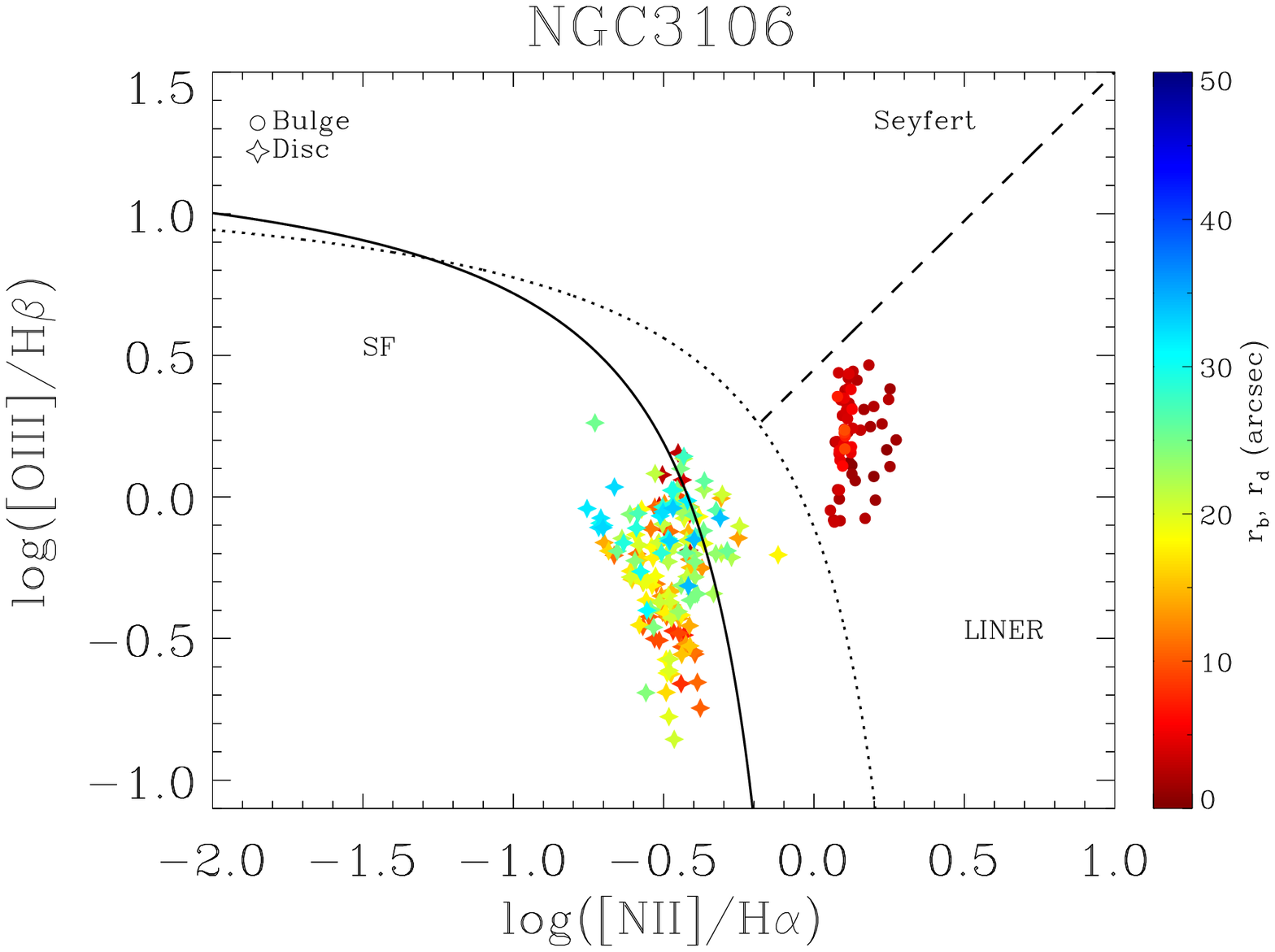}
\caption{BPT diagram showing the [NII]/H$\alpha$ vs. [OIII]/H$\beta$ strong lines diagnostic to separate regions dominated by photo-ionisation due to star formation from those ionised by an AGN/LINER-like source. Solid and dotted lines separate star-forming from AGN ionisation following \citet{kewley01,kewley06}. The dashed line marks the division between Seyfert and LINERs following \citet{schawinski07}. Circles and stars  represent spaxels from the bulge and disc datacubes, respectively. The different colors show the galacto-centric distance for each spaxel. Only spaxels with significant emission over the continuum ($> 1\sigma$) in the four emission lines are represented.}
\label{fig:BPT}
\end{center}
\end{figure}

\begin{figure}
\begin{center}
\includegraphics[width=0.45\textwidth]{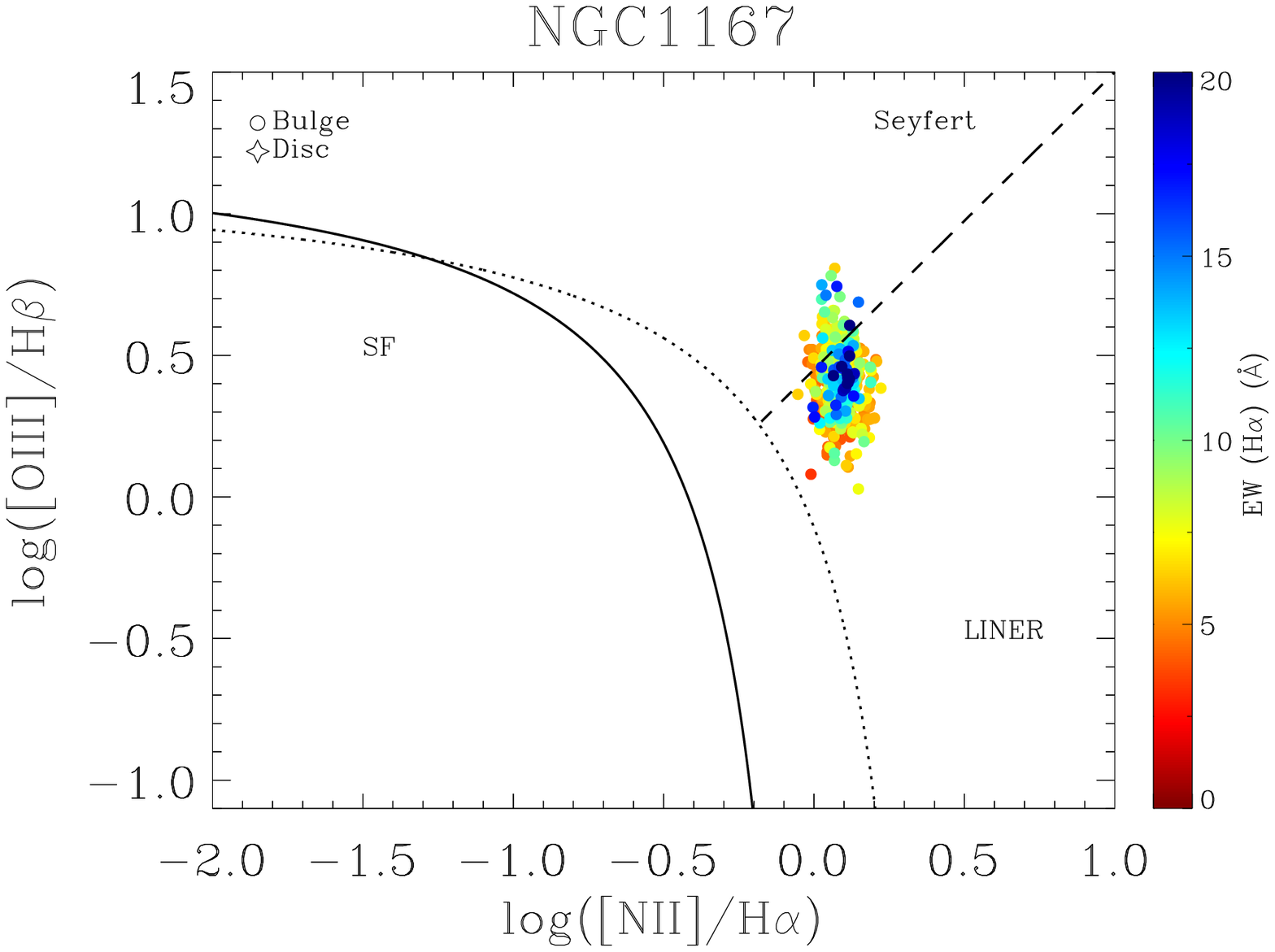}

\vspace{0.5cm}

\includegraphics[width=0.45\textwidth]{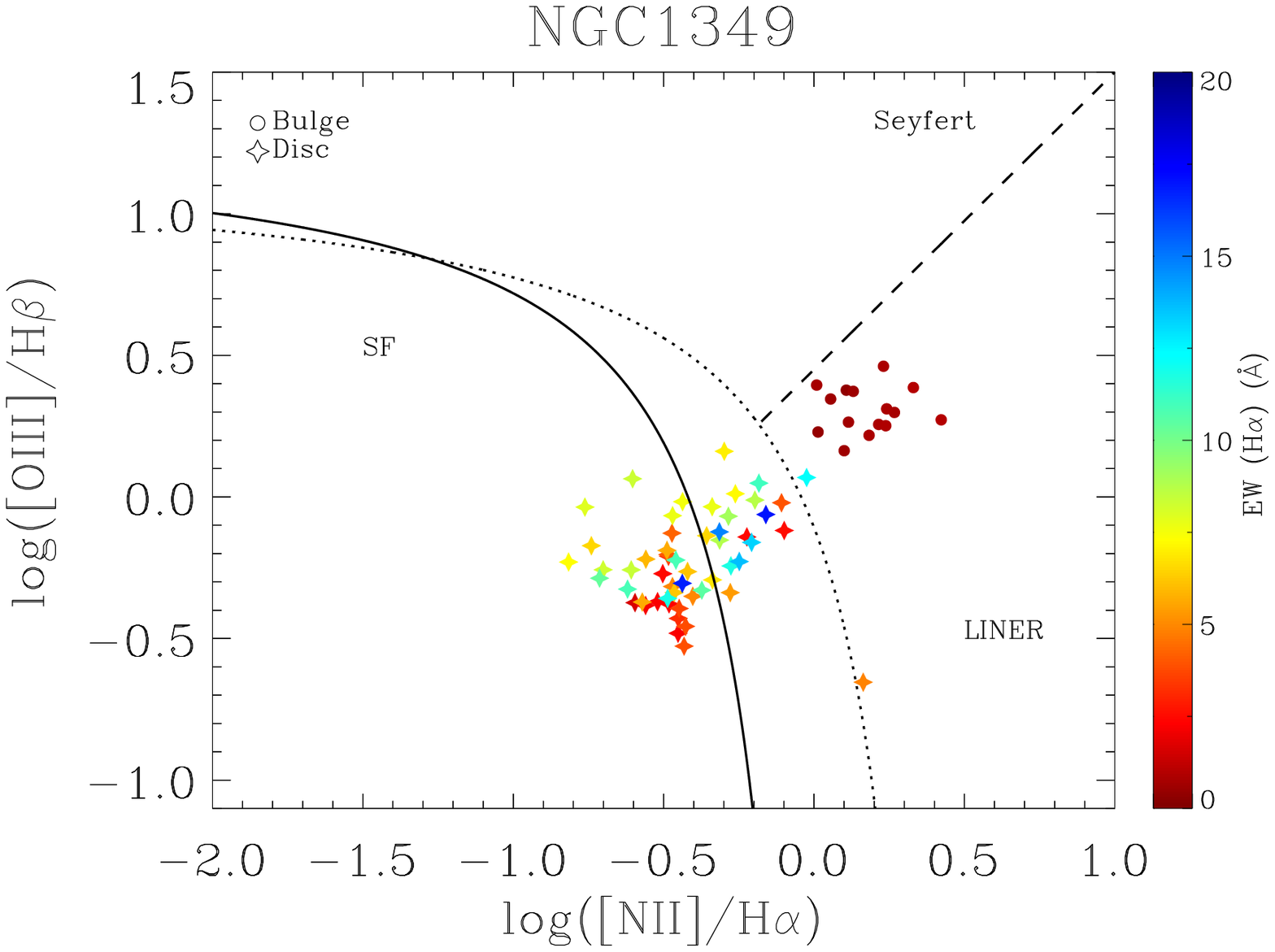}

\vspace{0.5cm}
\includegraphics[width=0.45\textwidth]{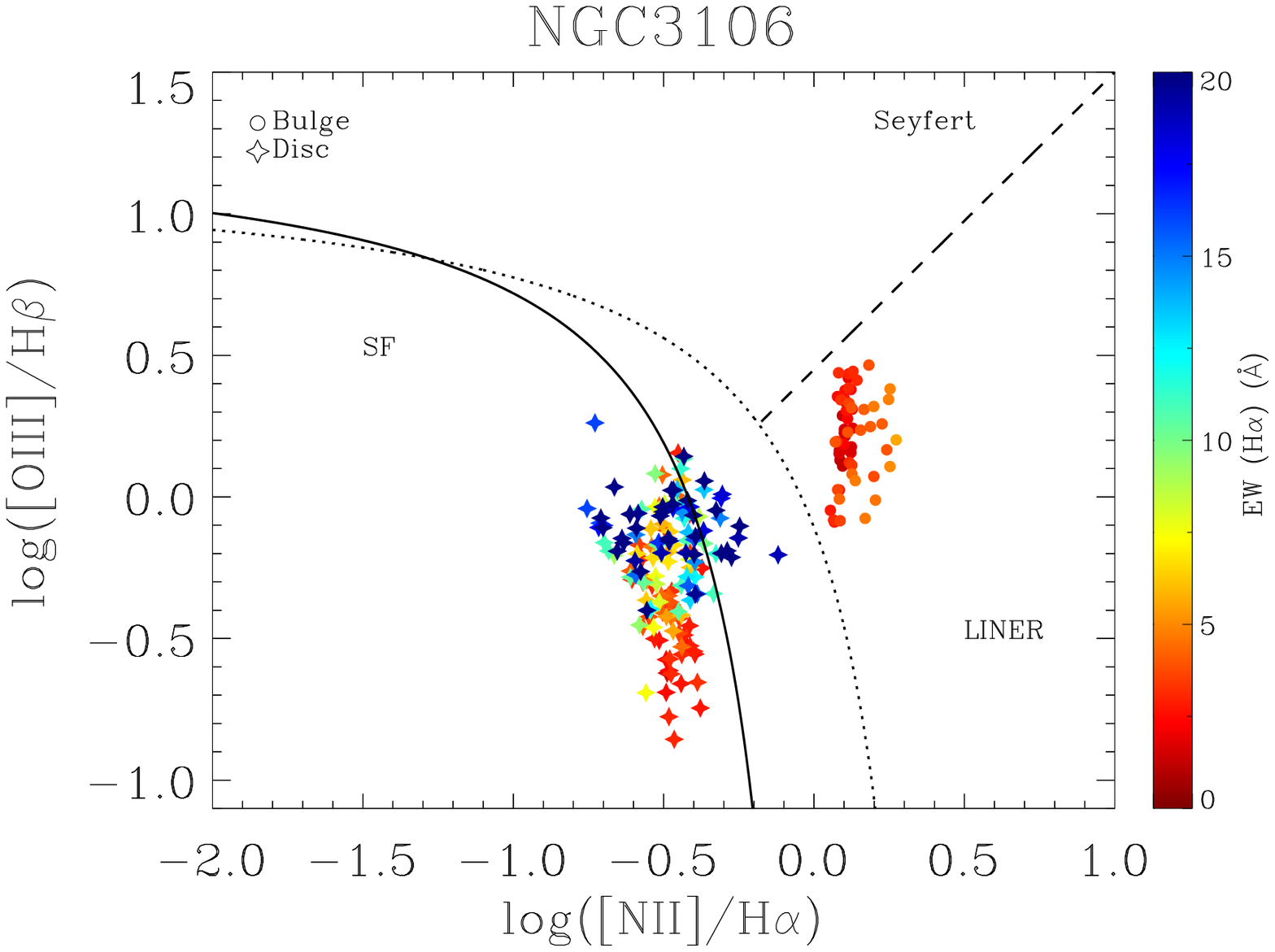}
\caption{Same as Fig~\ref{fig:BPT}, but the datapoints are colour-coded depending on the EW (H$\alpha$).}
\label{fig:BPT_EW}
\end{center}
\end{figure}

\begin{figure*}
\begin{center}
\includegraphics[bb=70 120 318 700,width=0.33\textwidth]{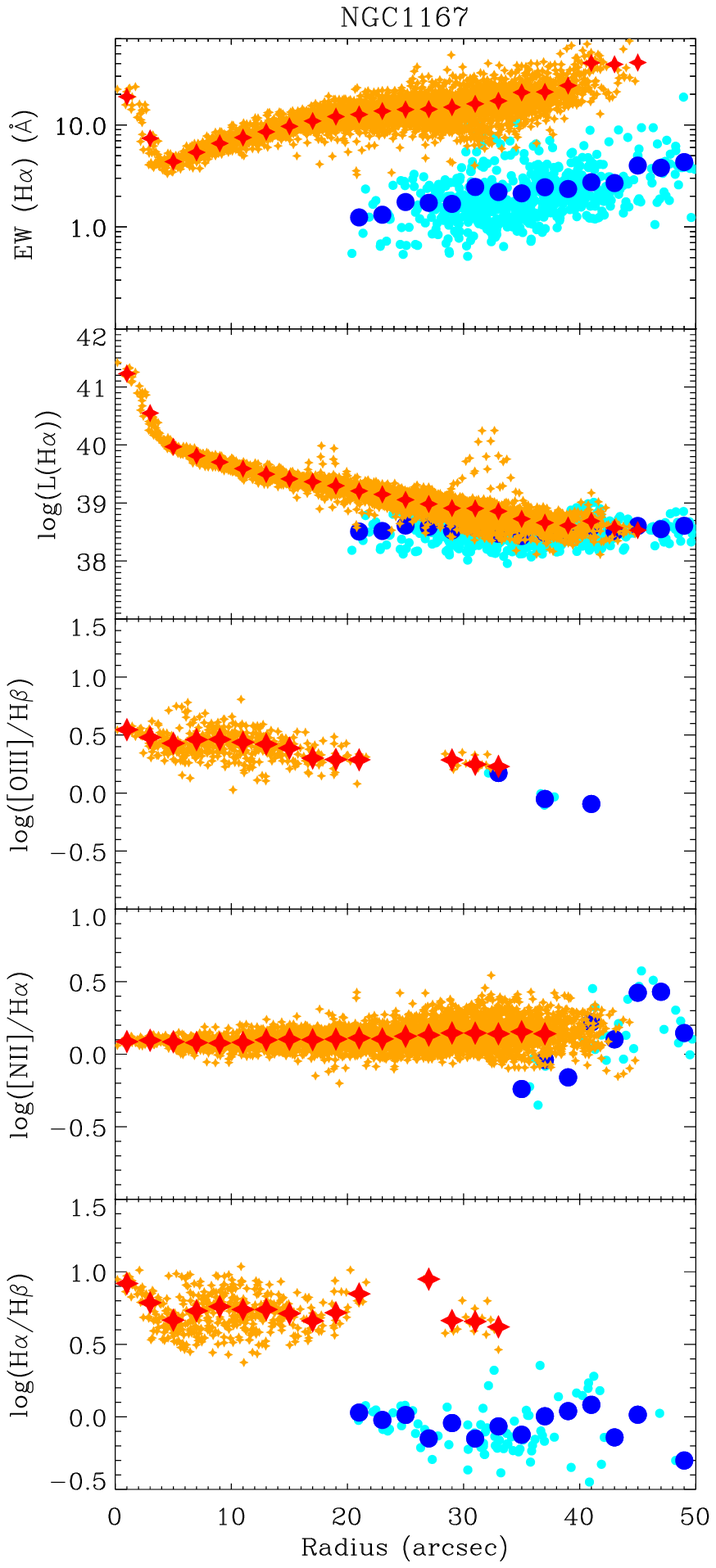}
\includegraphics[bb=70 120 318 700,width=0.33\textwidth]{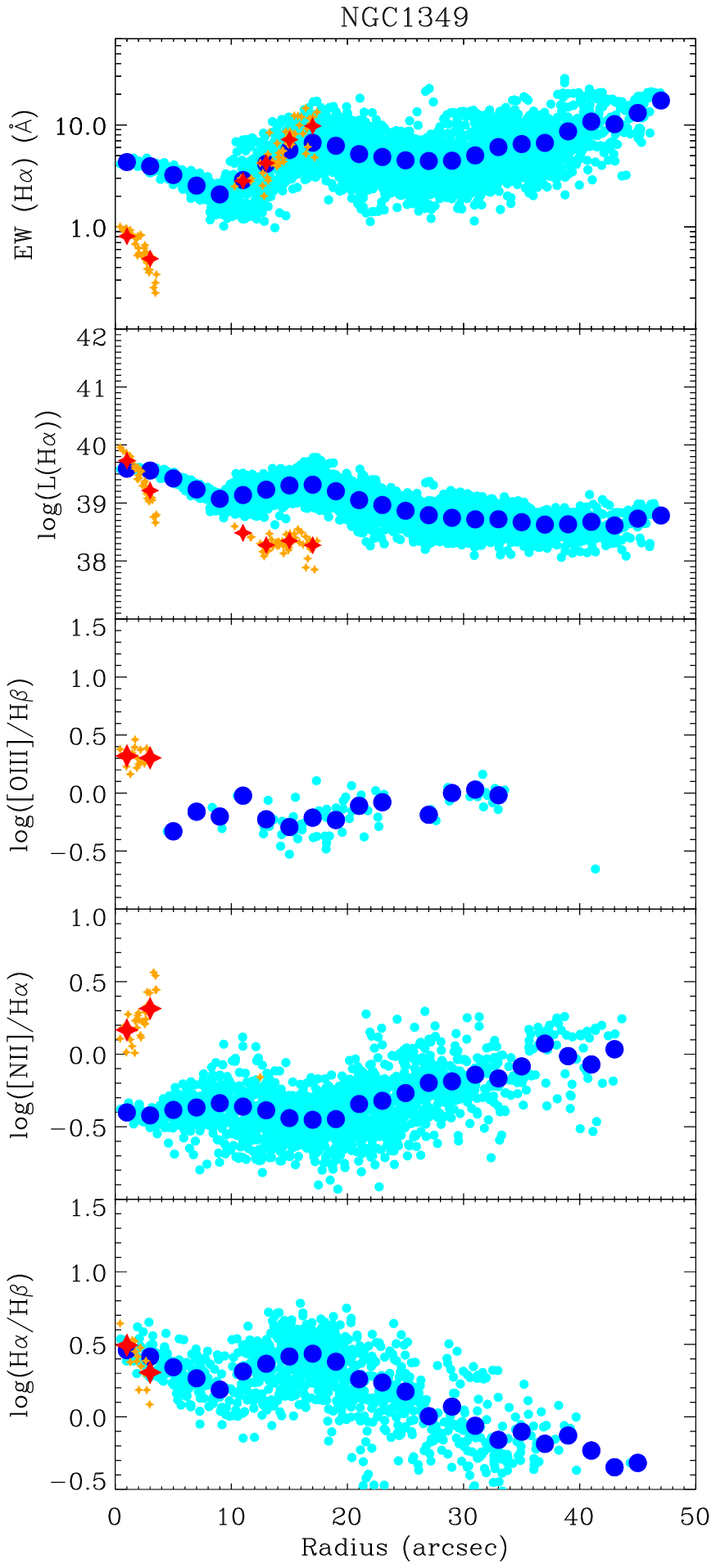}
\includegraphics[bb=70 120 318 700,width=0.33\textwidth]{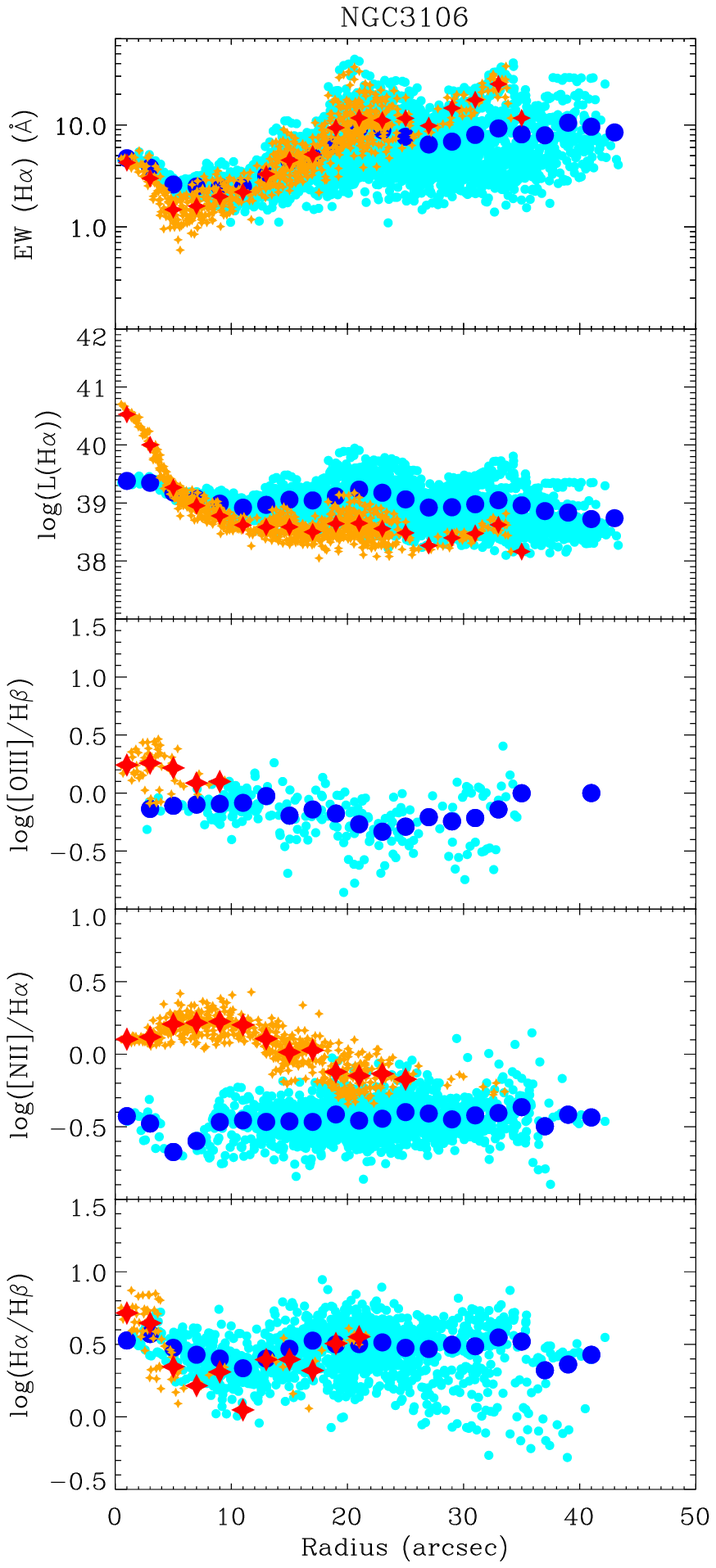}
\caption{From top to bottom, radial profiles for the bulge and disc of the H$\alpha$ equivalent width, H$\alpha$ luminosity, [OIII]/H$\beta$, [NII]/H$\alpha$, and H$\alpha$/H$\beta$ line ratios. Orange stars and cyan circles represent the measurements for individual spaxels in the bulge and disc datacubes, respectively. Red stars and blue circles show the radially averaged values in elliptical annuli oriented with the ellipticity and position angle of the bulge and disc, respectively.}
\label{fig:emission}
\end{center}
\end{figure*}

The analysis performed on the individual bulge and disc datacubes using {\sc Pipe3D} also provides us with information on the ionised-gas properties. Fig.~\ref{fig:BPT} shows the classical BPT diagram \citep{baldwin81} to identify the structures and regions within the galaxy ionised by different mechanisms such as recent star formation or AGNs. We represented each spaxel with significant emission over the continuum ($> 1\sigma$) in the individual bulge (circles) and disc (stars) datacubes. The radial distance to the galaxy center is indicated with different colors. From the BPT diagram we first notice that bulge spaxels always lie on the Seyfert/LINER region of the diagram, independently of their galacto-centric radii. In addition, only the bulge of NGC1167 contain spaxels in the Seyfert region, whereas for NGC1349 and NGC3106 we only found spaxels in the LINER regime. In order to further explore the ionisation mechanism present in our bulges and discs we represent in Fig.~\ref{fig:BPT_EW} the same BPT diagram as a function of the EW H$\alpha$ (see also the top panel in Fig.~\ref{fig:emission}). We found that the bulge spaxels in NGC1167 also show high values of the EW H$\alpha$, therefore confirming the Seyfert nature of the ionisation. However, this is not the case for the bulges of NGC1349 and NGC3106, where their ionisation might come from hot low-mass evolved stars \citep[e.g.,][]{lacerda18}. The same classification was found by \citet{gomes16} attending to the BPT values obtained from the galaxy spectra in the central regions of the galaxy. It is worth noting that, despite the central spaxels are generally dominated by the bulge light, SF in the discs can also be found at the very center of some galaxies (e.g., NGC3106). Similarly, our results suggest that star-formation is only occurring in the disc component of our galaxies. This is an important result since our galaxies where selected as ETGs: this finding provides further support for the existence of a  disc in at least NGC1349 and NGC3106. Similar results were obtained by \citet{catalantorrecilla17} using a larger sample of galaxies from the CALIFA survey.

%--------------------------------------------------------
%--------------------------------------------------------

%--------------------------------------------------------
%--------------------------------------------------------

The different properties of the ionised gas in our bulges and discs can also be noticed in Fig.~\ref{fig:emission}. We show the radial profiles for both the bulge and disc of the EW H$\alpha$, H$\alpha$ luminosity, [OIII]/H$\beta$, [NII]/H$\alpha$, and H$\alpha$/H$\beta$ line ratios. We already discussed the importance of the EW H$\alpha$ to identify the ionisation mechanisms in our bulges and discs. The H$\alpha$ luminosity shows generally a prominently-peaked profile for the bulge whereas a flatter trend with the galacto-centric radius is found for the discs. The distribution of the line ratios involved in the BPT diagram ([OIII]/H$\beta$, [NII]/H$\alpha$) show flat profiles. Finally, the H$\alpha$/H$\beta$ line ratio can be used as a proxy for the dust attenuation in the galaxy. We find that bulge values are higher for NGC1167 than those of the disc, therefore pointing towards the bulge of NGC1167 being more obscured by dust than its disc. Dust attenuation is similar for the bulges and discs of NGC1349 and NGC3106. Interestingly, a higher dust attenuation is generally associated with a higher amount of molecular gas \citep{scoville13,orellana17}. Thus, we can assume the bulge of NGC1167 has a larger content of molecular gas than the disc, whereas this amount is similar for the bulges and discs of NGC1349 and NGC3106. However, none of our bulges is forming stars. We will discuss this point further in the next section.

%--------------------------------------------------------
%--------------------------------------------------------

%----------------------------------------------
%----------------------------------------------
\section{New insights into the formation of bulges, discs, and ETGs}
\label{sec:discussion}

The analysis of both the stellar populations and ionised-gas presented in the previous section clearly shows the potential of our new methodology to shed new light on the formation and evolution of galaxies. The combination of {\sc c2d}+{\sc Pipe3D} produces a wealth of spatially resolved information for bulges and discs that needs to be interpreted in the framework of galaxy evolution scenarios. In fact, spatial variations on the stellar populations and ionised-gas properties are intimately connected with the dynamical processes that led to the formation of these structures, the degree of dissipation involved in the process, and the possible re-arrangement of material. We discuss here the possible formation mechanisms for the three galaxies selected as prototypes in this work. A statistically complete sample of CALIFA galaxies will be analysed in a forthcoming paper.

\subsection{Bulge formation and evolution}
A common characteristic of our three bulges is that they are old systems with most of their stars ($>90\%$) formed $>$ 6 Gyr ago. In fact, the age distribution of their stellar populations is quite homogeneous as a function of the galacto-centric radius (see Fig.~\ref{fig:SFH}). Regarding the stellar metallicity gradient of the bulge, NGC1349 and NGC3106 both show a steep negative gradient while this is almost flat for NGC1167. The spatially resolved analysis of the stellar populations in galaxy bulges has not been extensively studied in the literature. Most of the works performed so far have found luminosity-weighted negative gradients in metallicity and almost no gradients in age \citep{moorthy06,morelli08,morelli16,coccato18}. These results are similar to those found for ETGs when they are treated as a whole, i.e., without isolating the bulge regions \citep{mehlert03, sanchezblazquez06,rawle10,gonzalezdelgado14a}

Within the current observational paradigm, galactic bulges are divided into two different populations: classical and disc-like bulges, depending on their observational characteristics and formation mechanisms \citep[see][]{kormendykennicutt04,athanassoula05}. Several mechanisms have been proposed for the formation of classical bulges, all of them implying violent process at high redshift with some degree of dissipation. The most likely scenarios so far include that classical bulges can be created via dissipative collapse of protogalactic gas clouds \citep{eggen62,larson74}, by major mergers of galaxies \citep{kauffmann96,hopkins09}, or by the coalescence of giant SF clumps in primordial discs \citep{noguchi99,bournaud07}. From a theoretical point of view, chemical enrichment models based on a monolithic collapse scenario produce strong negative metallicity gradients \citep{eggen62, larson74, arimotoyoshii87}. Numerical simulations of galaxy mergers produce a diversity of stellar population gradients depending on the amount of gas, the recipe for the star formation feedback, and the gas/stellar accretion. Thus, some models predict that mergers tend to dilute metallicity gradients \citep{bekkishioya98, bekkishioya99} while others support that they maintain or enhance the previously existing gradients \citep{kawata01, kobayashi04,hirschmann15}. Unfortunately, the scenario involving coalescence of giant SF clumps has not been fully worked out to provide predictions against the spatially resolved stellar populations of bulges. Therefore, our results for NGC1349 and NGC3106 might be compatible with all formation scenarios of classical bulges, and the null gradient of NGC1167 may favour a merger origin. However, it seems clear from our analysis that our bulges have not suffered from the secular evolution processes which are thought to lead to the formation of disc-like bulges \citep{kormendykennicutt04}. This scenario generally involves some degree of SF in the bulges due to gas inflow towards the galaxy center promoted by either bars or spiral arms. Therefore, the lack of either young SF bursts in their SFH (Fig.~\ref{fig:integratedSFH}) or recent SF in the BPT diagram (Fig.~\ref{fig:BPT}) is not completely surprising since our galaxies are ETGs without bars or spiral arms. Therefore, the analysis of the stellar populations and ionised-gas properties for our bulges suggest an early formation of these structures, with an inside-out mass built up in the case of NGC1349 and NGC3106. Still, a complementary analysis including the stellar kinematics and the intrinsic shape of the structures would be needed to complete our view of these bulges \citep{mendezabreu18b, costantin18a, costantin18b}.

\subsection{Disc formation and evolution}
The disc component of our ETGs shows more complex stellar and ionised-gas properties consistent with previous results in the literature \citep[e.g.,][]{yoachimdalcanton08,sanchezblazquez14,morelli15}. We show in Table \ref{tab:massfrac} that discs are composed by younger stellar populations than bulges, with a significant amount of stars ($>20\%$) formed at 1 $<$ Age (Gyr) $<$ 6. In addition, the analysis of the ionisation source of the gas present in our ETGs reveals that all the current SF in these galaxies is happening in their disc components (see Fig. \ref{fig:BPT}). Therefore, this suggests that the mass built up of the discs took place after the bulge structure was already in place, neglecting the effects of radial migration or changes in the orbits. The spatially resolved distribution of the SFH and stellar metallicity imposes further constraints on this formation. NGC1349 and NGC3106 show a steep positive gradient in their outer disc stellar metallicity, reaching a minimum at about 1 disc effective radius, and keeping a lower value than the bulge within the central regions. On the other hand, the disc of NGC1167 shows a flat stellar metallicity gradient similar to its bulge, but with a slightly lower value. 

We discussed previously the possibility that our bulges could have been formed after a major merger event at high redshift. Since the discs of all our sample were formed after the bulge, it is possible that the disc was formed once the remnant gas after the merger settled up again into a cold configuration \citep{hopkins09}. This might explain the flat metallicity gradient of NGC1167 which could also shows a slightly lower value than the bulge due to the delay in the enrichment of the interstellar medium. Still, we cannot rule out that both component were formed at the same time, i.e., this galaxy is a pure elliptical. However, the metallicity radial profiles of NGC1349 and NGC3106 require different conditions. Regardless of whether the gas used to form the stars of the disc was a remnant from an early merger, or accreted from the cosmic web afterwards, it is clear that the mass growth of these discs continued for a longer time in the outer parts of the disc than in the center. Star formation in high mass discs was also found in \citet{catalantorrecilla17} using CALIFA data. This can be explained with either an outside-in star formation \citep{bedregal11,goddard17} or because the presence of the bulge stabilized the gaseous disc against star formation in the central parts of the galaxy \citep{martig09}. The analysis of the ionised gas in the BPT diagram (Fig.~\ref{fig:BPT}) shows that NGC1167 has no recent star formation in the disc, reinforcing the idea that it was formed from the remnant gas just after the merger leading to the bulge formation. However, the ionised-gas of NGC1349 and NGC3106 tells us again a different story. The current SF in those discs is located mainly in the outer part of the galaxies with some residual SF close to their centers. This is suggestive of a recent outside-in disc growth possibly driven by a central stabilisation of the disc due to the presence of the bulge \citep{martig09}.

%--------------------------------------------------------
\begin{figure}
\begin{center}
\includegraphics[width=0.45\textwidth]{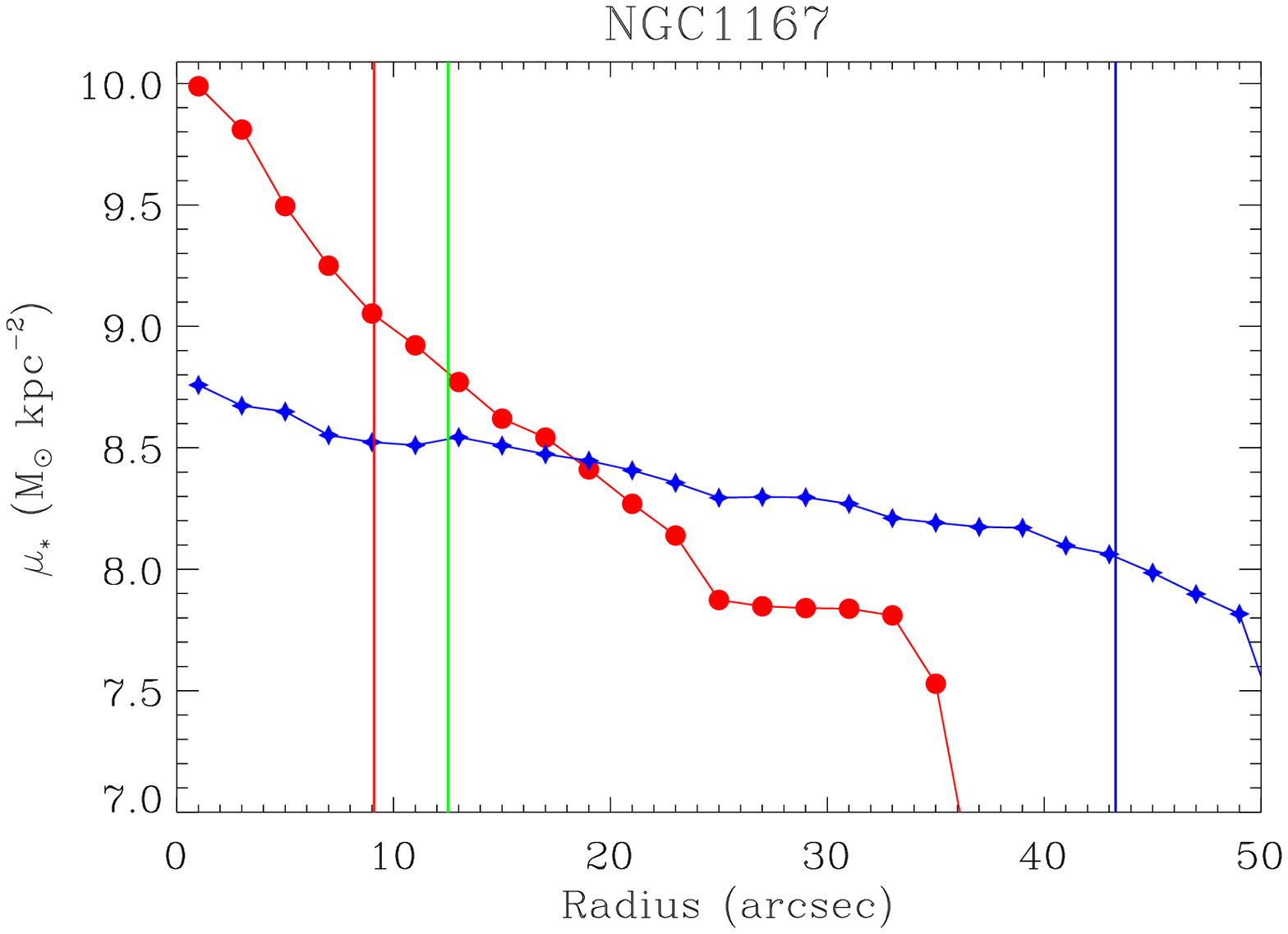}
\includegraphics[width=0.45\textwidth]{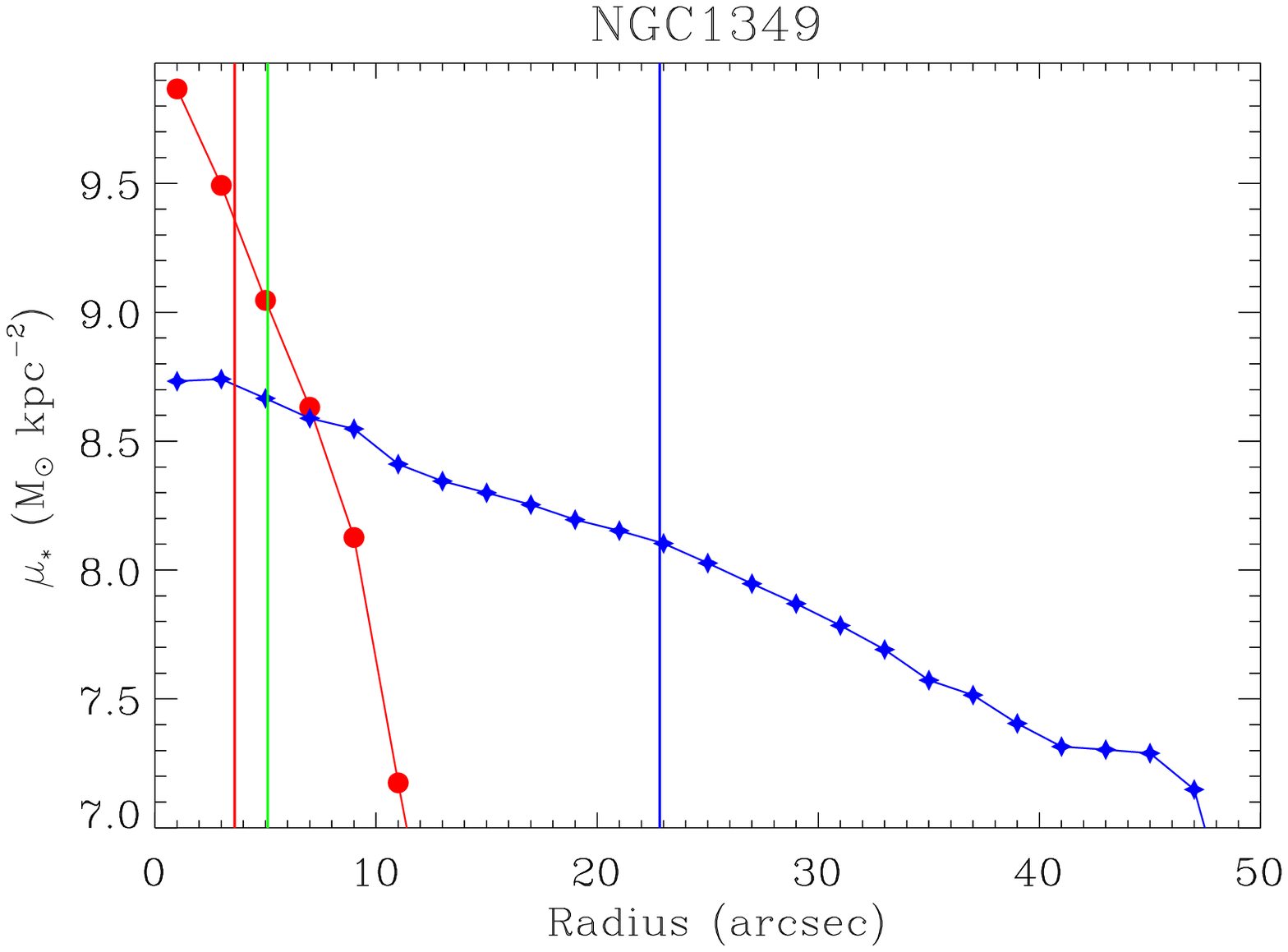}
\includegraphics[width=0.45\textwidth]{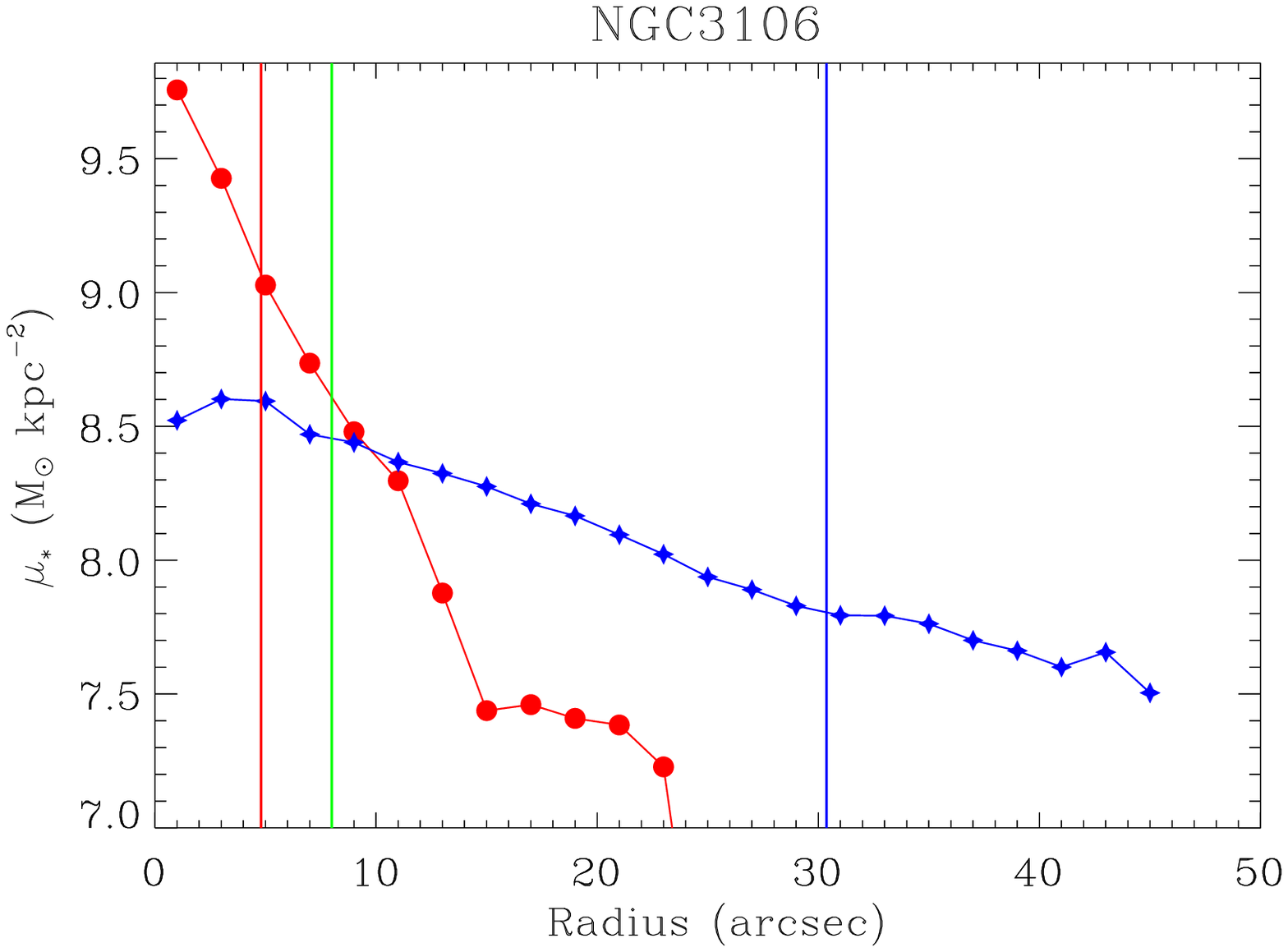}
\caption{Radial profiles of the stellar mass surface density for both the bulge (red) and disc (blue). The red, green, and blue vertical lines indicate the effective radius of the bulge, disc, and the radius where the surface brightness of the bulge and the disc are the same.}
\label{fig:SMD}
\end{center}
\end{figure}
%--------------------------------------------------------

\subsection{ETGs and the quenching mechanisms}

Another important point to understand the evolution of galactic structures involves understanding when and how their mass growth stopped. We cannot trace the possible dynamical re-arrangements of the stars between the bulge and disc, but we can study the mechanisms of fading or quenching of the SF in these components. We have demonstrated that our bulges were formed at relatively high-redshifts and that the current (residual) SF in these galaxies is happening (whenever it occurs) in discs. In addition, we provided some hints that the amount of molecular gas available in the bulge of NGC1167 might be higher than in its disc; for this galaxy, young stars are detected in the outer regions (see Fig.~\ref{fig:SFH}) but not in the bulge. Recently, \citet{utomo17} and \citet{colombo18} showed that the surface density of molecular gas is nearly constant for all Hubble types in the EDGE-CALIFA sample of galaxies \citep{bolatto17}, but the star formation surface density increases towards later Hubble types \citep[see also][]{gonzalezdelgado15}. Therefore they claim for a mechanism (morphological quenching) able to shutdown the SF in early-type galaxies rather than a shortage of molecular gas. Similar conclusions were reached by \citet{sanchez18}. In our sample, the lack of SF in the bulge of NGC1167 might be caused by the stabilisation against gravitational collapse induced by the massive bulge \citep{martig09}, nevertheless we cannot rule out other plausible explanations such as quenching due to AGN feedback \citep{cattaneo09}. NGC1349 and NGC3106 do not show signatures of an AGN present now at their galaxy centers, but we cannot discard an AGN episode happening in the past. Still, the surface mass densities shown in Fig.~\ref{fig:SMD} indicate that bulges might be massive enough to quench the SF by themselves \citep{franx08,bell12,whitaker17}.

The separated analysis of the bulges and discs in our sample of ETGs allows us to suggest a more complex formation scenario than in previous studies. \citet{gomes16} suggested that the presence of SF in the outskirts of these ETGs reflects their continuous process of inside-out mass growth. Similar conclusions were reached by \citet{gonzalezdelgado15} after analysing the age and metallicity gradients in the whole sample of ETGs present in the CALIFA survey. In general terms, our results are in agreement with those previously mentioned since our bulges were formed before the discs, and therefore the galaxy mass growth occurred inside-out. Nevertheless, by providing independent pieces of evidence for each structural component we have reached a more complex view of the life of these ETGs.

It is worth noting that, in the case of NGC1167, we cannot discard that this galaxy might be a single component (i.e., purely elliptical) galaxy. According to the photometric study carried out in \citet{mendezabreu17} this galaxy was classified as {\it unknown}, so it was not possible to decide whether it is an elliptical or lenticular. For the current analysis, we used the best bulge+disc photometric fit to the surface-brightness distribution, i.e., we assumed it is a two-component galaxy. The SFHs and stellar metallicity gradients of the bulge and disc components shown in this paper display some differences, however this galaxy shows a more similar behaviour between both components than the other two galaxies in the sample. A more extensive study on a statistically complete sample will help us determine whether our new methodology can be used to overcome the long-lasting difficulty of distinguishing elliptical and lenticular galaxies.

%----------------------------------------------
%----------------------------------------------
\section{Conclusions}
\label{sec:conclusions}

We present {\sc c2d}, a new spectro-photometric code to separate the spectral properties of the different galaxy components (bulge, disc and bar) using IFS datacubes. We demonstrate the robustness and accuracy of {\sc c2d} in the simplest case of galaxies composed by a central bulge and an outer disc. Our tests with mock galaxies demonstrate that we can recover the stellar population properties of bulges and discs for galaxies with $B/T>0.05$. As outputs, {\sc c2d} provides both the characteristic spectrum and the spatially resolved datacube of each structural component included in the fit. The resulting datacubes from {\sc c2d} are then analysed using the {\sc Pipe3D} algorithm to derive the stellar populations and ionised-gas properties of bulges and discs. The unique combination of {\sc c2d}+{\sc Pipe3D} produces a wealth of new spatially resolved information for each galaxy structural component that might be crucial to understand galaxy evolution processes.

Aiming at showing the potential of the {\sc c2d} methodology, we have applied our code to a sample of three ETGs observed within the CALIFA survey. This sample of ETGs was classified in \citet{gomes16} as i+, i.e., ETGs presenting an inner sector that displays low-ionization emission-line region (LINER) properties, and the outer one with H{\sc ii}-region characteristics. The CALIFA datacubes for the three ETGs were spectro-photometrically decomposed using {\sc c2d} into their bulges and discs, and then analysed using {\sc Pipe3D} in the same way as the original CALIFA datacubes. From the stellar population analysis we found that their bulges are old -most of their stars ($>90\%$) formed $>$6 Gyr ago- and generally present negative stellar metallicity gradients. Discs always show younger stellar populations than bulges ($>20\%$ formed at 1 $<$ Age (Gyr) $<$ 6) and positive (or flat) stellar metallicity gradients in their outer regions. Interestingly, our analysis of the ionised-gas reveals that the two different regions described in \citet{gomes16} are associated with the bulge and disc structures. We found that only the bulge datacube of NGC1167 has some regions compatible with Seyfert activity, while star-forming regions are unequivocally found in the galaxy discs. This is a very important result since we found star-forming ionised-gas even in the very central regions of the discs (see NGC3106 in Fig.~\ref{fig:BPT}).

We interpret our results in terms of an early formation of the three bulges in our sample involving some dissipation, either by monolithic collapse or major mergers. Discs show a more complex assembly process, but they were all formed after the central bulge. We suggest that NGC1167 might have formed its disc from the remnant gas after the merger in a radially homogeneous way. However, regardless of whether the gas used to form the stars of the discs was a remnant from an early merger, or accreted afterwards, it seems clear that the discs of NGC1349 and NGC3106 grew their mass outside-in. 

The analysis presented in the paper demonstrates the importance of combining photometric, stellar population, and ionised-gas information for the different structures shaping the galaxies to obtain a whole picture of galaxy evolution. The application of {\sc c2d}+{\sc Pipe3D} to more complex galaxies (including bars) and to the statistically significant sample observed with CALIFA will be explored in forthcoming papers.

%----------------------------------------------
%----------------------------------------------
\section*{Acknowledgements}
We thank the referee for a constructive report which helped to improve the manuscript. JMA acknowledge support from the Spanish Ministerio de Economia y Competitividad (MINECO) by the grant AYA2017-83204-P. AdLC acknowledges support from grant AYA2016- 77237-C3-1-P from the Spanish Ministry of Economy and Competitiveness (MINECO). SFS is grateful for the support of a CONACYT grant CB-285080 and funding from the PAPIIT-DGAPA-IA101217 and PAPIIT-DGAPA-IN100519(UNAM) projects.  This paper is based on data from the Calar Alto Legacy Integral Field Area Survey, CALIFA (http://califa.caha.es), funded by the Spanish Ministery of Science under grant ICTS-2009-10, and the Centro Astron\'omico Hispano-Alem\'an.

Based on observations collected at the Centro Astron\'omico Hispano Alem\'an (CAHA) at Calar Alto, operated jointly by the  Max-Planck Institut f\"ur Astronomie and the Instituto de Astrof\'isica de Andaluc\'ia (CSIC)

%%%%%%%%%%%%%%%%%%%%%%%%%%%%%%%%%%%%%%%%%%%%%%%%%%


\bibliographystyle{mnras}
\bibliography{reference} 

\begin{thebibliography}{}
\makeatletter
\relax
\def\mn@urlcharsother{\let\do\@makeother \do\$\do\&\do\#\do\^\do\_\do\%\do\~}
\def\mn@doi{\begingroup\mn@urlcharsother \@ifnextchar [ {\mn@doi@}
  {\mn@doi@[]}}
\def\mn@doi@[#1]#2{\def\@tempa{#1}\ifx\@tempa\@empty \href
  {http://dx.doi.org/#2} {doi:#2}\else \href {http://dx.doi.org/#2} {#1}\fi
  \endgroup}
\def\mn@eprint#1#2{\mn@eprint@#1:#2::\@nil}
\def\mn@eprint@arXiv#1{\href {http://arxiv.org/abs/#1} {{\tt arXiv:#1}}}
\def\mn@eprint@dblp#1{\href {http://dblp.uni-trier.de/rec/bibtex/#1.xml}
  {dblp:#1}}
\def\mn@eprint@#1:#2:#3:#4\@nil{\def\@tempa {#1}\def\@tempb {#2}\def\@tempc
  {#3}\ifx \@tempc \@empty \let \@tempc \@tempb \let \@tempb \@tempa \fi \ifx
  \@tempb \@empty \def\@tempb {arXiv}\fi \@ifundefined
  {mn@eprint@\@tempb}{\@tempb:\@tempc}{\expandafter \expandafter \csname
  mn@eprint@\@tempb\endcsname \expandafter{\@tempc}}}

\bibitem[\protect\citeauthoryear{{Abazajian} et~al.,}{{Abazajian}
  et~al.}{2009}]{abazajian09}
{Abazajian} K.~N.,  et~al., 2009, \mn@doi [\apjs]
  {10.1088/0067-0049/182/2/543}, \href
  {http://adsabs.harvard.edu/abs/2009ApJS..182..543A} {182, 543}

\bibitem[\protect\citeauthoryear{{Abraham}, {van den Bergh}, {Glazebrook},
  {Ellis}, {Santiago}, {Surma}  \& {Griffiths}}{{Abraham}
  et~al.}{1996}]{abraham96}
{Abraham} R.~G.,  {van den Bergh} S.,  {Glazebrook} K.,  {Ellis} R.~S.,
  {Santiago} B.~X.,  {Surma} P.,   {Griffiths} R.~E.,  1996, \mn@doi [\apjs]
  {10.1086/192352}, \href {http://adsabs.harvard.edu/abs/1996ApJS..107....1A}
  {107, 1}

\bibitem[\protect\citeauthoryear{{Argyle}, {M{\'e}ndez-Abreu}, {Wild}  \&
  {Mortlock}}{{Argyle} et~al.}{2018}]{argyle18}
{Argyle} J.~J.,  {M{\'e}ndez-Abreu} J.,  {Wild} V.,   {Mortlock} D.~J.,  2018,
  \mn@doi [\mnras] {10.1093/mnras/sty1691}, \href
  {http://adsabs.harvard.edu/abs/2018MNRAS.479.3076A} {479, 3076}

\bibitem[\protect\citeauthoryear{{Arimoto} \& {Yoshii}}{{Arimoto} \&
  {Yoshii}}{1987}]{arimotoyoshii87}
{Arimoto} N.,  {Yoshii} Y.,  1987, \aap, \href
  {http://adsabs.harvard.edu/abs/1987A%26A...173...23A} {173, 23}

\bibitem[\protect\citeauthoryear{{Athanassoula}}{{Athanassoula}}{2005}]{athanassoula05}
{Athanassoula} E.,  2005, \mn@doi [\mnras] {10.1111/j.1365-2966.2005.08872.x},
  \href {http://adsabs.harvard.edu/abs/2005MNRAS.358.1477A} {358, 1477}

\bibitem[\protect\citeauthoryear{{Baldwin}, {Phillips}  \&
  {Terlevich}}{{Baldwin} et~al.}{1981}]{baldwin81}
{Baldwin} J.~A.,  {Phillips} M.~M.,   {Terlevich} R.,  1981, \mn@doi [\pasp]
  {10.1086/130766}, \href {http://adsabs.harvard.edu/abs/1981PASP...93....5B}
  {93, 5}

\bibitem[\protect\citeauthoryear{{Bedregal}, {Cardiel}, {Arag{\'o}n-Salamanca}
  \& {Merrifield}}{{Bedregal} et~al.}{2011}]{bedregal11}
{Bedregal} A.~G.,  {Cardiel} N.,  {Arag{\'o}n-Salamanca} A.,   {Merrifield}
  M.~R.,  2011, \mn@doi [\mnras] {10.1111/j.1365-2966.2011.18752.x}, \href
  {http://adsabs.harvard.edu/abs/2011MNRAS.415.2063B} {415, 2063}

\bibitem[\protect\citeauthoryear{{Bekki} \& {Shioya}}{{Bekki} \&
  {Shioya}}{1998}]{bekkishioya98}
{Bekki} K.,  {Shioya} Y.,  1998, \mn@doi [\apj] {10.1086/305445}, \href
  {http://adsabs.harvard.edu/abs/1998ApJ...497..108B} {497, 108}

\bibitem[\protect\citeauthoryear{{Bekki} \& {Shioya}}{{Bekki} \&
  {Shioya}}{1999}]{bekkishioya99}
{Bekki} K.,  {Shioya} Y.,  1999, \mn@doi [\apj] {10.1086/306833}, \href
  {http://adsabs.harvard.edu/abs/1999ApJ...513..108B} {513, 108}

\bibitem[\protect\citeauthoryear{{Bell} et~al.,}{{Bell} et~al.}{2012}]{bell12}
{Bell} E.~F.,  et~al., 2012, \mn@doi [\apj] {10.1088/0004-637X/753/2/167},
  \href {http://adsabs.harvard.edu/abs/2012ApJ...753..167B} {753, 167}

\bibitem[\protect\citeauthoryear{{Ben{\'{\i}}tez} et~al.,}{{Ben{\'{\i}}tez}
  et~al.}{2013}]{benitez13}
{Ben{\'{\i}}tez} E.,  et~al., 2013, \mn@doi [\apj]
  {10.1088/0004-637X/763/2/136}, \href
  {http://adsabs.harvard.edu/abs/2013ApJ...763..136B} {763, 136}

\bibitem[\protect\citeauthoryear{{Birnboim} \& {Dekel}}{{Birnboim} \&
  {Dekel}}{2003}]{birnboimdekel03}
{Birnboim} Y.,  {Dekel} A.,  2003, \mn@doi [\mnras]
  {10.1046/j.1365-8711.2003.06955.x}, \href
  {http://adsabs.harvard.edu/abs/2003MNRAS.345..349B} {345, 349}

\bibitem[\protect\citeauthoryear{{Bolatto} et~al.,}{{Bolatto}
  et~al.}{2017}]{bolatto17}
{Bolatto} A.~D.,  et~al., 2017, \mn@doi [\apj] {10.3847/1538-4357/aa86aa},
  \href {http://adsabs.harvard.edu/abs/2017ApJ...846..159B} {846, 159}

\bibitem[\protect\citeauthoryear{{Bournaud}}{{Bournaud}}{2016}]{bournaud16}
{Bournaud} F.,  2016, {in Astrophysics and Space Science Library, Galactic
  Bulges, ed. E. Laurikainen, R. Peletier, D. Gadotti}.
 Vol. 418

\bibitem[\protect\citeauthoryear{{Bournaud}, {Elmegreen}  \&
  {Elmegreen}}{{Bournaud} et~al.}{2007}]{bournaud07}
{Bournaud} F.,  {Elmegreen} B.~G.,   {Elmegreen} D.~M.,  2007, \mn@doi [\apj]
  {10.1086/522077}, \href {http://adsabs.harvard.edu/abs/2007ApJ...670..237B}
  {670, 237}

\bibitem[\protect\citeauthoryear{{Caon}, {Capaccioli}  \& {D'Onofrio}}{{Caon}
  et~al.}{1993}]{caon93}
{Caon} N.,  {Capaccioli} M.,   {D'Onofrio} M.,  1993, \mnras, \href
  {http://adsabs.harvard.edu/abs/1993MNRAS.265.1013C} {265, 1013}

\bibitem[\protect\citeauthoryear{{Catal{\'a}n-Torrecilla}
  et~al.,}{{Catal{\'a}n-Torrecilla} et~al.}{2017}]{catalantorrecilla17}
{Catal{\'a}n-Torrecilla} C.,  et~al., 2017, \mn@doi [\apj]
  {10.3847/1538-4357/aa8a6d}, \href
  {http://adsabs.harvard.edu/abs/2017ApJ...848...87C} {848, 87}

\bibitem[\protect\citeauthoryear{{Cattaneo} et~al.,}{{Cattaneo}
  et~al.}{2009}]{cattaneo09}
{Cattaneo} A.,  et~al., 2009, \mn@doi [\nat] {10.1038/nature08135}, \href
  {http://adsabs.harvard.edu/abs/2009Natur.460..213C} {460, 213}

\bibitem[\protect\citeauthoryear{{Cid Fernandes} et~al.,}{{Cid Fernandes}
  et~al.}{2013}]{cidfernandes13}
{Cid Fernandes} R.,  et~al., 2013, \mn@doi [\aap]
  {10.1051/0004-6361/201220616}, \href
  {http://adsabs.harvard.edu/abs/2013A%26A...557A..86C} {557, A86}

\bibitem[\protect\citeauthoryear{{Cid Fernandes} et~al.,}{{Cid Fernandes}
  et~al.}{2014}]{cidfernandes14}
{Cid Fernandes} R.,  et~al., 2014, \mn@doi [\aap]
  {10.1051/0004-6361/201321692}, \href
  {http://cdsads.u-strasbg.fr/abs/2014A%26A...561A.130C} {561, A130}

\bibitem[\protect\citeauthoryear{{Coccato}, {Morelli}, {Corsini}, {Buson},
  {Pizzella}, {Vergani}  \& {Bertola}}{{Coccato} et~al.}{2011}]{coccato11}
{Coccato} L.,  {Morelli} L.,  {Corsini} E.~M.,  {Buson} L.,  {Pizzella} A.,
  {Vergani} D.,   {Bertola} F.,  2011, \mn@doi [\mnras]
  {10.1111/j.1745-3933.2011.01016.x}, \href
  {http://adsabs.harvard.edu/abs/2011MNRAS.412L.113C} {412, L113}

\bibitem[\protect\citeauthoryear{{Coccato}, {Fabricius}, {Saglia}, {Bender},
  {Erwin}, {Drory}  \& {Morelli}}{{Coccato} et~al.}{2018}]{coccato18}
{Coccato} L.,  {Fabricius} M.~H.,  {Saglia} R.~P.,  {Bender} R.,  {Erwin} P.,
  {Drory} N.,   {Morelli} L.,  2018, \mn@doi [\mnras] {10.1093/mnras/sty705},
  \href {http://adsabs.harvard.edu/abs/2018MNRAS.477.1958C} {477, 1958}

\bibitem[\protect\citeauthoryear{{Colombo} et~al.,}{{Colombo}
  et~al.}{2018}]{colombo18}
{Colombo} D.,  et~al., 2018, \mn@doi [\mnras] {10.1093/mnras/stx3233}, \href
  {http://adsabs.harvard.edu/abs/2018MNRAS.475.1791C} {475, 1791}

\bibitem[\protect\citeauthoryear{{Costantin}, {M{\'e}ndez-Abreu}, {Corsini},
  {Morelli}, {Aguerri}, {Dalla Bont{\`a}}  \& {Pizzella}}{{Costantin}
  et~al.}{2017}]{costantin17a}
{Costantin} L.,  {M{\'e}ndez-Abreu} J.,  {Corsini} E.~M.,  {Morelli} L.,
  {Aguerri} J.~A.~L.,  {Dalla Bont{\`a}} E.,   {Pizzella} A.,  2017, \mn@doi
  [\aap] {10.1051/0004-6361/201630302}, \href
  {http://adsabs.harvard.edu/abs/2017A%26A...601A..84C} {601, A84}

\bibitem[\protect\citeauthoryear{{Costantin}, {Corsini}, {M{\'e}ndez-Abreu},
  {Morelli}, {Dalla Bont{\`a}}  \& {Pizzella}}{{Costantin}
  et~al.}{2018a}]{costantin18b}
{Costantin} L.,  {Corsini} E.~M.,  {M{\'e}ndez-Abreu} J.,  {Morelli} L.,
  {Dalla Bont{\`a}} E.,   {Pizzella} A.,  2018a, \mn@doi [\mnras]
  {10.1093/mnras/sty1754}, \href
  {http://adsabs.harvard.edu/abs/2018MNRAS.481.3623C} {481, 3623}

\bibitem[\protect\citeauthoryear{{Costantin}, {M{\'e}ndez-Abreu}, {Corsini},
  {Eliche-Moral}, {Tapia}, {Morelli}, {Dalla Bont{\`a}}  \&
  {Pizzella}}{{Costantin} et~al.}{2018b}]{costantin18a}
{Costantin} L.,  {M{\'e}ndez-Abreu} J.,  {Corsini} E.~M.,  {Eliche-Moral}
  M.~C.,  {Tapia} T.,  {Morelli} L.,  {Dalla Bont{\`a}} E.,   {Pizzella} A.,
  2018b, \mn@doi [\aap] {10.1051/0004-6361/201731823}, \href
  {http://adsabs.harvard.edu/abs/2018A%26A...609A.132C} {609, A132}

\bibitem[\protect\citeauthoryear{{Di Matteo}, {Bournaud}, {Martig}, {Combes},
  {Melchior}  \& {Semelin}}{{Di Matteo} et~al.}{2008}]{dimatteo08}
{Di Matteo} P.,  {Bournaud} F.,  {Martig} M.,  {Combes} F.,  {Melchior} A.-L.,
   {Semelin} B.,  2008, \mn@doi [\aap] {10.1051/0004-6361:200809480}, \href
  {http://adsabs.harvard.edu/abs/2008A%26A...492...31D} {492, 31}

\bibitem[\protect\citeauthoryear{{Dimauro} et~al.,}{{Dimauro}
  et~al.}{2018}]{dimauro18}
{Dimauro} P.,  et~al., 2018, \mn@doi [\mnras] {10.1093/mnras/sty1379}, \href
  {http://adsabs.harvard.edu/abs/2018MNRAS.478.5410D} {478, 5410}

\bibitem[\protect\citeauthoryear{{Eggen}, {Lynden-Bell}  \& {Sandage}}{{Eggen}
  et~al.}{1962}]{eggen62}
{Eggen} O.~J.,  {Lynden-Bell} D.,   {Sandage} A.~R.,  1962, \mn@doi [\apj]
  {10.1086/147433}, \href {http://adsabs.harvard.edu/abs/1962ApJ...136..748E}
  {136, 748}

\bibitem[\protect\citeauthoryear{{Elmegreen}, {Elmegreen}, {Ravindranath}  \&
  {Coe}}{{Elmegreen} et~al.}{2007}]{elmegreen07}
{Elmegreen} D.~M.,  {Elmegreen} B.~G.,  {Ravindranath} S.,   {Coe} D.~A.,
  2007, \mn@doi [\apj] {10.1086/511667}, \href
  {http://adsabs.harvard.edu/abs/2007ApJ...658..763E} {658, 763}

\bibitem[\protect\citeauthoryear{{Erwin} et~al.,}{{Erwin}
  et~al.}{2015}]{erwin15}
{Erwin} P.,  et~al., 2015, \mn@doi [\mnras] {10.1093/mnras/stu2376}, \href
  {http://adsabs.harvard.edu/abs/2015MNRAS.446.4039E} {446, 4039}

\bibitem[\protect\citeauthoryear{{Feldmann}, {Carollo}, {Mayer}, {Renzini},
  {Lake}, {Quinn}, {Stinson}  \& {Yepes}}{{Feldmann} et~al.}{2010}]{feldmann10}
{Feldmann} R.,  {Carollo} C.~M.,  {Mayer} L.,  {Renzini} A.,  {Lake} G.,
  {Quinn} T.,  {Stinson} G.~S.,   {Yepes} G.,  2010, \mn@doi [\apj]
  {10.1088/0004-637X/709/1/218}, \href
  {http://adsabs.harvard.edu/abs/2010ApJ...709..218F} {709, 218}

\bibitem[\protect\citeauthoryear{{Franx}, {van Dokkum}, {F{\"o}rster
  Schreiber}, {Wuyts}, {Labb{\'e}}  \& {Toft}}{{Franx} et~al.}{2008}]{franx08}
{Franx} M.,  {van Dokkum} P.~G.,  {F{\"o}rster Schreiber} N.~M.,  {Wuyts} S.,
  {Labb{\'e}} I.,   {Toft} S.,  2008, \mn@doi [\apj] {10.1086/592431}, \href
  {http://adsabs.harvard.edu/abs/2008ApJ...688..770F} {688, 770}

\bibitem[\protect\citeauthoryear{{Freeman}}{{Freeman}}{1970}]{freeman70}
{Freeman} K.~C.,  1970, \mn@doi [\apj] {10.1086/150474}, \href
  {http://adsabs.harvard.edu/abs/1970ApJ...160..811F} {160, 811}

\bibitem[\protect\citeauthoryear{{Gadotti}}{{Gadotti}}{2009}]{gadotti09}
{Gadotti} D.~A.,  2009, \mn@doi [\mnras] {10.1111/j.1365-2966.2008.14257.x},
  \href {http://adsabs.harvard.edu/abs/2009MNRAS.393.1531G} {393, 1531}

\bibitem[\protect\citeauthoryear{{Galbany} et~al.,}{{Galbany}
  et~al.}{2016}]{galbany16}
{Galbany} L.,  et~al., 2016, \mn@doi [\mnras] {10.1093/mnras/stv2620}, \href
  {http://adsabs.harvard.edu/abs/2016MNRAS.455.4087G} {455, 4087}

\bibitem[\protect\citeauthoryear{{Galbany}, {Collett}, {M{\'e}ndez-Abreu},
  {S{\'a}nchez}, {Anderson}  \& {Kuncarayakti}}{{Galbany}
  et~al.}{2018}]{galbany18}
{Galbany} L.,  {Collett} T.~E.,  {M{\'e}ndez-Abreu} J.,  {S{\'a}nchez} S.~F.,
  {Anderson} J.~P.,   {Kuncarayakti} H.,  2018, \mn@doi [\mnras]
  {10.1093/mnras/sty1448}, \href
  {http://adsabs.harvard.edu/abs/2018MNRAS.479..262G} {479, 262}

\bibitem[\protect\citeauthoryear{{Garc{\'{\i}}a-Benito}
  et~al.,}{{Garc{\'{\i}}a-Benito} et~al.}{2015}]{garciabenito15}
{Garc{\'{\i}}a-Benito} R.,  et~al., 2015, \mn@doi [\aap]
  {10.1051/0004-6361/201425080}, \href
  {http://adsabs.harvard.edu/abs/2015A%26A...576A.135G} {576, A135}

\bibitem[\protect\citeauthoryear{{Garc{\'{\i}}a-Lorenzo}, {S{\'a}nchez},
  {Mediavilla}, {Gonz{\'a}lez-Serrano}  \&
  {Christensen}}{{Garc{\'{\i}}a-Lorenzo} et~al.}{2005}]{garcialorenzo05}
{Garc{\'{\i}}a-Lorenzo} B.,  {S{\'a}nchez} S.~F.,  {Mediavilla} E.,
  {Gonz{\'a}lez-Serrano} J.~I.,   {Christensen} L.,  2005, \mn@doi [\apj]
  {10.1086/427429}, \href {http://adsabs.harvard.edu/abs/2005ApJ...621..146G}
  {621, 146}

\bibitem[\protect\citeauthoryear{{Goddard} et~al.,}{{Goddard}
  et~al.}{2017}]{goddard17}
{Goddard} D.,  et~al., 2017, \mn@doi [\mnras] {10.1093/mnras/stw3371}, \href
  {http://adsabs.harvard.edu/abs/2017MNRAS.466.4731G} {466, 4731}

\bibitem[\protect\citeauthoryear{{Gomes} et~al.,}{{Gomes}
  et~al.}{2016a}]{gomes16}
{Gomes} J.~M.,  et~al., 2016a, \mn@doi [\aap] {10.1051/0004-6361/201525974},
  \href {http://adsabs.harvard.edu/abs/2016A%26A...585A..92G} {585, A92}

\bibitem[\protect\citeauthoryear{{Gomes} et~al.,}{{Gomes}
  et~al.}{2016b}]{gomes16b}
{Gomes} J.~M.,  et~al., 2016b, \mn@doi [\aap] {10.1051/0004-6361/201525976},
  \href {http://adsabs.harvard.edu/abs/2016A%26A...588A..68G} {588, A68}

\bibitem[\protect\citeauthoryear{{Gonz{\'a}lez Delgado} et~al.,}{{Gonz{\'a}lez
  Delgado} et~al.}{2014}]{gonzalezdelgado14a}
{Gonz{\'a}lez Delgado} R.~M.,  et~al., 2014, \mn@doi [\apjl]
  {10.1088/2041-8205/791/1/L16}, \href
  {http://adsabs.harvard.edu/abs/2014ApJ...791L..16G} {791, L16}

\bibitem[\protect\citeauthoryear{{Gonz{\'a}lez Delgado} et~al.,}{{Gonz{\'a}lez
  Delgado} et~al.}{2015}]{gonzalezdelgado15}
{Gonz{\'a}lez Delgado} R.~M.,  et~al., 2015, \mn@doi [\aap]
  {10.1051/0004-6361/201525938}, \href
  {http://adsabs.harvard.edu/abs/2015A%26A...581A.103G} {581, A103}

\bibitem[\protect\citeauthoryear{{H{\"a}u{\ss}ler} et~al.,}{{H{\"a}u{\ss}ler}
  et~al.}{2013}]{haussler13}
{H{\"a}u{\ss}ler} B.,  et~al., 2013, \mn@doi [\mnras] {10.1093/mnras/sts633},
  \href {http://adsabs.harvard.edu/abs/2013MNRAS.430..330H} {430, 330}

\bibitem[\protect\citeauthoryear{{Hinojosa-Go{\~n}i}, {Mu{\~n}oz-Tu{\~n}{\'o}n}
   \& {M{\'e}ndez-Abreu}}{{Hinojosa-Go{\~n}i} et~al.}{2016}]{hinojosagoni16}
{Hinojosa-Go{\~n}i} R.,  {Mu{\~n}oz-Tu{\~n}{\'o}n} C.,   {M{\'e}ndez-Abreu} J.,
   2016, \mn@doi [\aap] {10.1051/0004-6361/201527066}, \href
  {http://adsabs.harvard.edu/abs/2016A%26A...592A.122H} {592, A122}

\bibitem[\protect\citeauthoryear{{Hirschmann}, {Naab}, {Ostriker}, {Forbes},
  {Duc}, {Dav{\'e}}, {Oser}  \& {Karabal}}{{Hirschmann}
  et~al.}{2015}]{hirschmann15}
{Hirschmann} M.,  {Naab} T.,  {Ostriker} J.~P.,  {Forbes} D.~A.,  {Duc} P.-A.,
  {Dav{\'e}} R.,  {Oser} L.,   {Karabal} E.,  2015, \mn@doi [\mnras]
  {10.1093/mnras/stv274}, \href
  {http://adsabs.harvard.edu/abs/2015MNRAS.449..528H} {449, 528}

\bibitem[\protect\citeauthoryear{{Hopkins}, {Hernquist}, {Cox}, {Di Matteo},
  {Martini}, {Robertson}  \& {Springel}}{{Hopkins} et~al.}{2005}]{hopkins05}
{Hopkins} P.~F.,  {Hernquist} L.,  {Cox} T.~J.,  {Di Matteo} T.,  {Martini} P.,
   {Robertson} B.,   {Springel} V.,  2005, \mn@doi [\apj] {10.1086/432438},
  \href {http://adsabs.harvard.edu/abs/2005ApJ...630..705H} {630, 705}

\bibitem[\protect\citeauthoryear{{Hopkins}, {Cox}, {Dutta}, {Hernquist},
  {Kormendy}  \& {Lauer}}{{Hopkins} et~al.}{2009}]{hopkins09}
{Hopkins} P.~F.,  {Cox} T.~J.,  {Dutta} S.~N.,  {Hernquist} L.,  {Kormendy} J.,
    {Lauer} T.~R.,  2009, \mn@doi [\apjs] {10.1088/0067-0049/181/1/135}, \href
  {http://adsabs.harvard.edu/abs/2009ApJS..181..135H} {181, 135}

\bibitem[\protect\citeauthoryear{{Husemann}, {Wisotzki}, {S{\'a}nchez}  \&
  {Jahnke}}{{Husemann} et~al.}{2013}]{husemann13}
{Husemann} B.,  {Wisotzki} L.,  {S{\'a}nchez} S.~F.,   {Jahnke} K.,  2013,
  \mn@doi [\aap] {10.1051/0004-6361/201220076}, \href
  {http://adsabs.harvard.edu/abs/2013A%26A...549A..43H} {549, A43}

\bibitem[\protect\citeauthoryear{{Ibarra-Medel} et~al.,}{{Ibarra-Medel}
  et~al.}{2016}]{ibarramedel16}
{Ibarra-Medel} H.~J.,  et~al., 2016, \mn@doi [\mnras] {10.1093/mnras/stw2126},
  \href {http://adsabs.harvard.edu/abs/2016MNRAS.463.2799I} {463, 2799}

\bibitem[\protect\citeauthoryear{{Jahnke}}{{Jahnke}}{2002}]{jahnke02}
{Jahnke} K.,  2002, PhD thesis, University of Hamburg, Astrophysikalisches
  Institut Potsdam

\bibitem[\protect\citeauthoryear{{Jahnke}, {Wisotzki}, {S{\'a}nchez},
  {Christensen}, {Becker}, {Kelz}  \& {Roth}}{{Jahnke} et~al.}{2004}]{jahnke04}
{Jahnke} K.,  {Wisotzki} L.,  {S{\'a}nchez} S.~F.,  {Christensen} L.,  {Becker}
  T.,  {Kelz} A.,   {Roth} M.~M.,  2004, \mn@doi [Astronomische Nachrichten]
  {10.1002/asna.200310191}, \href
  {http://adsabs.harvard.edu/abs/2004AN....325..128J} {325, 128}

\bibitem[\protect\citeauthoryear{{Johnston}, {Arag{\'o}n-Salamanca},
  {Merrifield}  \& {Bedregal}}{{Johnston} et~al.}{2012}]{johnston12}
{Johnston} E.~J.,  {Arag{\'o}n-Salamanca} A.,  {Merrifield} M.~R.,   {Bedregal}
  A.~G.,  2012, \mn@doi [\mnras] {10.1111/j.1365-2966.2012.20813.x}, \href
  {http://adsabs.harvard.edu/abs/2012MNRAS.422.2590J} {422, 2590}

\bibitem[\protect\citeauthoryear{{Johnston}, {Arag{\'o}n-Salamanca}  \&
  {Merrifield}}{{Johnston} et~al.}{2014}]{johnston14}
{Johnston} E.~J.,  {Arag{\'o}n-Salamanca} A.,   {Merrifield} M.~R.,  2014,
  \mn@doi [\mnras] {10.1093/mnras/stu582}, \href
  {http://adsabs.harvard.edu/abs/2014MNRAS.441..333J} {441, 333}

\bibitem[\protect\citeauthoryear{{Johnston} et~al.,}{{Johnston}
  et~al.}{2017}]{johnston17}
{Johnston} E.~J.,  et~al., 2017, \mn@doi [\mnras] {10.1093/mnras/stw2823},
  \href {http://adsabs.harvard.edu/abs/2017MNRAS.465.2317J} {465, 2317}

\bibitem[\protect\citeauthoryear{{Kamann}, {Wisotzki}  \& {Roth}}{{Kamann}
  et~al.}{2013}]{kamann13}
{Kamann} S.,  {Wisotzki} L.,   {Roth} M.~M.,  2013, \mn@doi [\aap]
  {10.1051/0004-6361/201220476}, \href
  {http://adsabs.harvard.edu/abs/2013A%26A...549A..71K} {549, A71}

\bibitem[\protect\citeauthoryear{{Kauffmann}}{{Kauffmann}}{1996}]{kauffmann96}
{Kauffmann} G.,  1996, \mnras, \href
  {http://adsabs.harvard.edu/abs/1996MNRAS.281..487K} {281, 487}

\bibitem[\protect\citeauthoryear{{Kawata}}{{Kawata}}{2001}]{kawata01}
{Kawata} D.,  2001, \mn@doi [\apj] {10.1086/322309}, \href
  {http://adsabs.harvard.edu/abs/2001ApJ...558..598K} {558, 598}

\bibitem[\protect\citeauthoryear{{Kere{\v s}}, {Katz}, {Dav{\'e}}, {Fardal}  \&
  {Weinberg}}{{Kere{\v s}} et~al.}{2009}]{keres09}
{Kere{\v s}} D.,  {Katz} N.,  {Dav{\'e}} R.,  {Fardal} M.,   {Weinberg} D.~H.,
  2009, \mn@doi [\mnras] {10.1111/j.1365-2966.2009.14924.x}, \href
  {http://adsabs.harvard.edu/abs/2009MNRAS.396.2332K} {396, 2332}

\bibitem[\protect\citeauthoryear{{Kewley}, {Dopita}, {Sutherland}, {Heisler}
  \& {Trevena}}{{Kewley} et~al.}{2001}]{kewley01}
{Kewley} L.~J.,  {Dopita} M.~A.,  {Sutherland} R.~S.,  {Heisler} C.~A.,
  {Trevena} J.,  2001, \mn@doi [\apj] {10.1086/321545}, \href
  {http://adsabs.harvard.edu/abs/2001ApJ...556..121K} {556, 121}

\bibitem[\protect\citeauthoryear{{Kewley}, {Groves}, {Kauffmann}  \&
  {Heckman}}{{Kewley} et~al.}{2006}]{kewley06}
{Kewley} L.~J.,  {Groves} B.,  {Kauffmann} G.,   {Heckman} T.,  2006, \mn@doi
  [\mnras] {10.1111/j.1365-2966.2006.10859.x}, \href
  {http://adsabs.harvard.edu/abs/2006MNRAS.372..961K} {372, 961}

\bibitem[\protect\citeauthoryear{{Kobayashi}}{{Kobayashi}}{2004}]{kobayashi04}
{Kobayashi} C.,  2004, \mn@doi [\mnras] {10.1111/j.1365-2966.2004.07258.x},
  \href {http://adsabs.harvard.edu/abs/2004MNRAS.347..740K} {347, 740}

\bibitem[\protect\citeauthoryear{{Kormendy} \& {Kennicutt}}{{Kormendy} \&
  {Kennicutt}}{2004}]{kormendykennicutt04}
{Kormendy} J.,  {Kennicutt} Jr. R.~C.,  2004, \mn@doi [\araa]
  {10.1146/annurev.astro.42.053102.134024}, \href
  {http://adsabs.harvard.edu/abs/2004ARA%26A..42..603K} {42, 603}

\bibitem[\protect\citeauthoryear{{Lacerda} et~al.,}{{Lacerda}
  et~al.}{2018}]{lacerda18}
{Lacerda} E.~A.~D.,  et~al., 2018, \mn@doi [\mnras] {10.1093/mnras/stx3022},
  \href {http://adsabs.harvard.edu/abs/2018MNRAS.474.3727L} {474, 3727}

\bibitem[\protect\citeauthoryear{{Larson}}{{Larson}}{1974}]{larson74}
{Larson} R.~B.,  1974, \mn@doi [\mnras] {10.1093/mnras/166.3.585}, \href
  {http://adsabs.harvard.edu/abs/1974MNRAS.166..585L} {166, 585}

\bibitem[\protect\citeauthoryear{{Laurikainen}, {Salo}  \&
  {Buta}}{{Laurikainen} et~al.}{2005}]{laurikainen05}
{Laurikainen} E.,  {Salo} H.,   {Buta} R.,  2005, \mn@doi [\mnras]
  {10.1111/j.1365-2966.2005.09404.x}, \href
  {http://adsabs.harvard.edu/abs/2005MNRAS.362.1319L} {362, 1319}

\bibitem[\protect\citeauthoryear{{Martig}, {Bournaud}, {Teyssier}  \&
  {Dekel}}{{Martig} et~al.}{2009}]{martig09}
{Martig} M.,  {Bournaud} F.,  {Teyssier} R.,   {Dekel} A.,  2009, \mn@doi
  [\apj] {10.1088/0004-637X/707/1/250}, \href
  {http://adsabs.harvard.edu/abs/2009ApJ...707..250M} {707, 250}

\bibitem[\protect\citeauthoryear{{Meert}, {Vikram}  \& {Bernardi}}{{Meert}
  et~al.}{2016}]{meert16}
{Meert} A.,  {Vikram} V.,   {Bernardi} M.,  2016, \mn@doi [\mnras]
  {10.1093/mnras/stv2475}, \href
  {http://adsabs.harvard.edu/abs/2016MNRAS.455.2440M} {455, 2440}

\bibitem[\protect\citeauthoryear{{Mehlert}, {Thomas}, {Saglia}, {Bender}  \&
  {Wegner}}{{Mehlert} et~al.}{2003}]{mehlert03}
{Mehlert} D.,  {Thomas} D.,  {Saglia} R.~P.,  {Bender} R.,   {Wegner} G.,
  2003, \mn@doi [\aap] {10.1051/0004-6361:20030886}, \href
  {http://adsabs.harvard.edu/abs/2003A%26A...407..423M} {407, 423}

\bibitem[\protect\citeauthoryear{{M{\'e}ndez-Abreu}, {Aguerri}, {Corsini}  \&
  {Simonneau}}{{M{\'e}ndez-Abreu} et~al.}{2008}]{mendezabreu08a}
{M{\'e}ndez-Abreu} J.,  {Aguerri} J.~A.~L.,  {Corsini} E.~M.,   {Simonneau} E.,
   2008, \mn@doi [\aap] {10.1051/0004-6361:20078089}, \href
  {http://adsabs.harvard.edu/abs/2008A%26A...478..353M} {478, 353}

\bibitem[\protect\citeauthoryear{{M{\'e}ndez-Abreu} et~al.,}{{M{\'e}ndez-Abreu}
  et~al.}{2012}]{mendezabreu12}
{M{\'e}ndez-Abreu} J.,  et~al., 2012, \mn@doi [\aap]
  {10.1051/0004-6361/201117755}, \href
  {http://adsabs.harvard.edu/abs/2012A.26A...537A..25M} {537, A25}

\bibitem[\protect\citeauthoryear{{M{\'e}ndez-Abreu}, {Debattista}, {Corsini}
  \& {Aguerri}}{{M{\'e}ndez-Abreu} et~al.}{2014}]{mendezabreu14}
{M{\'e}ndez-Abreu} J.,  {Debattista} V.~P.,  {Corsini} E.~M.,   {Aguerri}
  J.~A.~L.,  2014, \mn@doi [\aap] {10.1051/0004-6361/201423955}, \href
  {http://adsabs.harvard.edu/abs/2014A%26A...572A..25M} {572, A25}

\bibitem[\protect\citeauthoryear{{M{\'e}ndez-Abreu} et~al.,}{{M{\'e}ndez-Abreu}
  et~al.}{2017}]{mendezabreu17}
{M{\'e}ndez-Abreu} J.,  et~al., 2017, \mn@doi [\aap]
  {10.1051/0004-6361/201629525}, \href
  {http://adsabs.harvard.edu/abs/2017A%26A...598A..32M} {598, A32}

\bibitem[\protect\citeauthoryear{{M{\'e}ndez-Abreu} et~al.,}{{M{\'e}ndez-Abreu}
  et~al.}{2018a}]{mendezabreu18}
{M{\'e}ndez-Abreu} J.,  et~al., 2018a, \mn@doi [\mnras]
  {10.1093/mnras/stx2804}, \href
  {http://adsabs.harvard.edu/abs/2018MNRAS.474.1307M} {474, 1307}

\bibitem[\protect\citeauthoryear{{M{\'e}ndez-Abreu}, {Costantin}, {Aguerri},
  {de Lorenzo-C{\'a}ceres}  \& {Corsini}}{{M{\'e}ndez-Abreu}
  et~al.}{2018b}]{mendezabreu18b}
{M{\'e}ndez-Abreu} J.,  {Costantin} L.,  {Aguerri} J.~A.~L.,  {de
  Lorenzo-C{\'a}ceres} A.,   {Corsini} E.~M.,  2018b, \mn@doi [\mnras]
  {10.1093/mnras/sty1694}, \href
  {http://adsabs.harvard.edu/abs/2018MNRAS.479.4172M} {479, 4172}

\bibitem[\protect\citeauthoryear{{Moorthy} \& {Holtzman}}{{Moorthy} \&
  {Holtzman}}{2006}]{moorthy06}
{Moorthy} B.~K.,  {Holtzman} J.~A.,  2006, \mn@doi [\mnras]
  {10.1111/j.1365-2966.2006.10722.x}, \href
  {http://adsabs.harvard.edu/abs/2006MNRAS.371..583M} {371, 583}

\bibitem[\protect\citeauthoryear{{Morelli} et~al.,}{{Morelli}
  et~al.}{2008}]{morelli08}
{Morelli} L.,  et~al., 2008, \mn@doi [\mnras]
  {10.1111/j.1365-2966.2008.13566.x}, \href
  {http://adsabs.harvard.edu/abs/2008MNRAS.389..341M} {389, 341}

\bibitem[\protect\citeauthoryear{{Morelli}, {Corsini}, {Pizzella}, {Dalla
  Bont{\`a}}, {Coccato}  \& {M{\'e}ndez-Abreu}}{{Morelli}
  et~al.}{2015}]{morelli15}
{Morelli} L.,  {Corsini} E.~M.,  {Pizzella} A.,  {Dalla Bont{\`a}} E.,
  {Coccato} L.,   {M{\'e}ndez-Abreu} J.,  2015, \mn@doi [\mnras]
  {10.1093/mnras/stv1357}, \href
  {http://adsabs.harvard.edu/abs/2015MNRAS.452.1128M} {452, 1128}

\bibitem[\protect\citeauthoryear{{Morelli}, {Parmiggiani}, {Corsini},
  {Costantin}, {Dalla Bont{\`a}}, {M{\'e}ndez-Abreu}  \& {Pizzella}}{{Morelli}
  et~al.}{2016}]{morelli16}
{Morelli} L.,  {Parmiggiani} M.,  {Corsini} E.~M.,  {Costantin} L.,  {Dalla
  Bont{\`a}} E.,  {M{\'e}ndez-Abreu} J.,   {Pizzella} A.,  2016, \mn@doi
  [\mnras] {10.1093/mnras/stw2285}, \href
  {http://adsabs.harvard.edu/abs/2016MNRAS.463.4396M} {463, 4396}

\bibitem[\protect\citeauthoryear{{Noguchi}}{{Noguchi}}{1999}]{noguchi99}
{Noguchi} M.,  1999, \mn@doi [\apj] {10.1086/306932}, \href
  {http://adsabs.harvard.edu/abs/1999ApJ...514...77N} {514, 77}

\bibitem[\protect\citeauthoryear{{Orellana} et~al.,}{{Orellana}
  et~al.}{2017}]{orellana17}
{Orellana} G.,  et~al., 2017, \mn@doi [\aap] {10.1051/0004-6361/201629009},
  \href {http://adsabs.harvard.edu/abs/2017A%26A...602A..68O} {602, A68}

\bibitem[\protect\citeauthoryear{{Peng}, {Ho}, {Impey}  \& {Rix}}{{Peng}
  et~al.}{2002}]{peng02}
{Peng} C.~Y.,  {Ho} L.~C.,  {Impey} C.~D.,   {Rix} H.,  2002, \mn@doi [\aj]
  {10.1086/340952}, \href {http://adsabs.harvard.edu/abs/2002AJ....124..266P}
  {124, 266}

\bibitem[\protect\citeauthoryear{{P{\'e}rez-Gonz{\'a}lez}
  et~al.,}{{P{\'e}rez-Gonz{\'a}lez} et~al.}{2008}]{perezgonzalez08}
{P{\'e}rez-Gonz{\'a}lez} P.~G.,  et~al., 2008, \mn@doi [\apj] {10.1086/523690},
  \href {http://adsabs.harvard.edu/abs/2008ApJ...675..234P} {675, 234}

\bibitem[\protect\citeauthoryear{{P{\'e}rez} et~al.,}{{P{\'e}rez}
  et~al.}{2013}]{perez13}
{P{\'e}rez} E.,  et~al., 2013, \mn@doi [\apjl] {10.1088/2041-8205/764/1/L1},
  \href {http://adsabs.harvard.edu/abs/2013ApJ...764L...1P} {764, L1}

\bibitem[\protect\citeauthoryear{{Rawle}, {Smith}  \& {Lucey}}{{Rawle}
  et~al.}{2010}]{rawle10}
{Rawle} T.~D.,  {Smith} R.~J.,   {Lucey} J.~R.,  2010, \mn@doi [\mnras]
  {10.1111/j.1365-2966.2009.15722.x}, \href
  {http://adsabs.harvard.edu/abs/2010MNRAS.401..852R} {401, 852}

\bibitem[\protect\citeauthoryear{{Salucci}, {Lapi}, {Tonini}, {Gentile},
  {Yegorova}  \& {Klein}}{{Salucci} et~al.}{2007}]{salucci07}
{Salucci} P.,  {Lapi} A.,  {Tonini} C.,  {Gentile} G.,  {Yegorova} I.,
  {Klein} U.,  2007, \mn@doi [\mnras] {10.1111/j.1365-2966.2007.11696.x}, \href
  {http://adsabs.harvard.edu/abs/2007MNRAS.378...41S} {378, 41}

\bibitem[\protect\citeauthoryear{{S{\'a}nchez-Bl{\'a}zquez}
  et~al.,}{{S{\'a}nchez-Bl{\'a}zquez} et~al.}{2006}]{sanchezblazquez06}
{S{\'a}nchez-Bl{\'a}zquez} P.,  et~al., 2006, \mn@doi [\mnras]
  {10.1111/j.1365-2966.2006.10699.x}, \href
  {http://adsabs.harvard.edu/abs/2006MNRAS.371..703S} {371, 703}

\bibitem[\protect\citeauthoryear{{S{\'a}nchez-Bl{\'a}zquez}
  et~al.,}{{S{\'a}nchez-Bl{\'a}zquez} et~al.}{2014}]{sanchezblazquez14}
{S{\'a}nchez-Bl{\'a}zquez} P.,  et~al., 2014, \mn@doi [\aap]
  {10.1051/0004-6361/201423635}, \href
  {http://adsabs.harvard.edu/abs/2014A%26A...570A...6S} {570, A6}

\bibitem[\protect\citeauthoryear{{S{\'a}nchez-Menguiano}
  et~al.,}{{S{\'a}nchez-Menguiano} et~al.}{2018}]{sanchezmenguiano18}
{S{\'a}nchez-Menguiano} L.,  et~al., 2018, \mn@doi [\aap]
  {10.1051/0004-6361/201731486}, \href
  {http://cdsads.u-strasbg.fr/abs/2018A%26A...609A.119S} {609, A119}

\bibitem[\protect\citeauthoryear{{S{\'a}nchez}, {Garcia-Lorenzo}, {Mediavilla},
  {Gonz{\'a}lez-Serrano}  \& {Christensen}}{{S{\'a}nchez}
  et~al.}{2004}]{sanchez04}
{S{\'a}nchez} S.~F.,  {Garcia-Lorenzo} B.,  {Mediavilla} E.,
  {Gonz{\'a}lez-Serrano} J.~I.,   {Christensen} L.,  2004, \mn@doi [\apj]
  {10.1086/423990}, \href {http://adsabs.harvard.edu/abs/2004ApJ...615..156S}
  {615, 156}

\bibitem[\protect\citeauthoryear{{S{\'a}nchez}, {Cardiel}, {Verheijen},
  {Pedraz}  \& {Covone}}{{S{\'a}nchez} et~al.}{2007}]{sanchez07}
{S{\'a}nchez} S.~F.,  {Cardiel} N.,  {Verheijen} M.~A.~W.,  {Pedraz} S.,
  {Covone} G.,  2007, \mn@doi [\mnras] {10.1111/j.1365-2966.2007.11335.x},
  \href {http://adsabs.harvard.edu/abs/2007MNRAS.376..125S} {376, 125}

\bibitem[\protect\citeauthoryear{{S{\'a}nchez} et~al.,}{{S{\'a}nchez}
  et~al.}{2012}]{sanchez12}
{S{\'a}nchez} S.~F.,  et~al., 2012, \mn@doi [\aap]
  {10.1051/0004-6361/201117353}, \href
  {http://adsabs.harvard.edu/abs/2012A%26A...538A...8S} {538, A8}

\bibitem[\protect\citeauthoryear{{S{\'a}nchez} et~al.,}{{S{\'a}nchez}
  et~al.}{2015}]{sanchez15}
{S{\'a}nchez} S.~F.,  et~al., 2015, \mn@doi [\aap]
  {10.1051/0004-6361/201424873}, \href
  {http://adsabs.harvard.edu/abs/2015A%26A...574A..47S} {574, A47}

\bibitem[\protect\citeauthoryear{{S{\'a}nchez} et~al.,}{{S{\'a}nchez}
  et~al.}{2016a}]{sanchez16b}
{S{\'a}nchez} S.~F.,  et~al., 2016a, \rmxaa, \href
  {http://adsabs.harvard.edu/abs/2016RMxAA..52...21S} {52, 21}

\bibitem[\protect\citeauthoryear{{S{\'a}nchez} et~al.,}{{S{\'a}nchez}
  et~al.}{2016b}]{sanchez16a}
{S{\'a}nchez} S.~F.,  et~al., 2016b, \rmxaa, \href
  {http://adsabs.harvard.edu/abs/2016RMxAA..52..171S} {52, 171}

\bibitem[\protect\citeauthoryear{{S{\'a}nchez} et~al.,}{{S{\'a}nchez}
  et~al.}{2016c}]{sanchez16}
{S{\'a}nchez} S.~F.,  et~al., 2016c, \mn@doi [\aap]
  {10.1051/0004-6361/201628661}, \href
  {http://adsabs.harvard.edu/abs/2016A%26A...594A..36S} {594, A36}

\bibitem[\protect\citeauthoryear{{S{\'a}nchez} et~al.,}{{S{\'a}nchez}
  et~al.}{2017}]{sanchez17}
{S{\'a}nchez} S.~F.,  et~al., 2017, \mn@doi [\mnras] {10.1093/mnras/stx808},
  \href {http://adsabs.harvard.edu/abs/2017MNRAS.469.2121S} {469, 2121}

\bibitem[\protect\citeauthoryear{{S{\'a}nchez} et~al.,}{{S{\'a}nchez}
  et~al.}{2018}]{sanchez18}
{S{\'a}nchez} S.~F.,  et~al., 2018, \rmxaa, \href
  {http://cdsads.u-strasbg.fr/abs/2018RMxAA..54..217S} {54, 217}

\bibitem[\protect\citeauthoryear{{S{\'a}nchez} et~al.,}{{S{\'a}nchez}
  et~al.}{2019}]{sanchez19}
{S{\'a}nchez} S.~F.,  et~al., 2019, \mn@doi [\mnras] {10.1093/mnras/sty2730},
  \href {http://cdsads.u-strasbg.fr/abs/2019MNRAS.482.1557S} {482, 1557}

\bibitem[\protect\citeauthoryear{{Schawinski}, {Thomas}, {Sarzi}, {Maraston},
  {Kaviraj}, {Joo}, {Yi}  \& {Silk}}{{Schawinski} et~al.}{2007}]{schawinski07}
{Schawinski} K.,  {Thomas} D.,  {Sarzi} M.,  {Maraston} C.,  {Kaviraj} S.,
  {Joo} S.-J.,  {Yi} S.~K.,   {Silk} J.,  2007, \mn@doi [\mnras]
  {10.1111/j.1365-2966.2007.12487.x}, \href
  {http://adsabs.harvard.edu/abs/2007MNRAS.382.1415S} {382, 1415}

\bibitem[\protect\citeauthoryear{{Scoville}}{{Scoville}}{2013}]{scoville13}
{Scoville} N.~Z.,  2013, {in Secular Evolution of Galaxies, J. Falc\'on-Barroso
  and J.~H. Knapen eds., Cambridge University Press, Cambridge}.
p.~491

\bibitem[\protect\citeauthoryear{{S\'ersic}}{{S\'ersic}}{1968}]{sersic68}
{S\'ersic} J.~L.,  1968, {Atlas de galaxias australes (Observatorio
  Astronomico, Universidad Nacional de Cordoba, Cordoba)}

\bibitem[\protect\citeauthoryear{{Simard} et~al.,}{{Simard}
  et~al.}{2002}]{simard02}
{Simard} L.,  et~al., 2002, \mn@doi [\apjs] {10.1086/341399}, \href
  {http://adsabs.harvard.edu/abs/2002ApJS..142....1S} {142, 1}

\bibitem[\protect\citeauthoryear{{Simard}, {Mendel}, {Patton}, {Ellison}  \&
  {McConnachie}}{{Simard} et~al.}{2011}]{simard11}
{Simard} L.,  {Mendel} J.~T.,  {Patton} D.~R.,  {Ellison} S.~L.,
  {McConnachie} A.~W.,  2011, \mn@doi [\apjs] {10.1088/0067-0049/196/1/11},
  \href {http://adsabs.harvard.edu/abs/2011ApJS..196...11S} {196, 11}

\bibitem[\protect\citeauthoryear{{Tabor}, {Merrifield}, {Arag{\'o}n-Salamanca},
  {Cappellari}, {Bamford}  \& {Johnston}}{{Tabor} et~al.}{2017}]{tabor17}
{Tabor} M.,  {Merrifield} M.,  {Arag{\'o}n-Salamanca} A.,  {Cappellari} M.,
  {Bamford} S.~P.,   {Johnston} E.,  2017, \mn@doi [\mnras]
  {10.1093/mnras/stw3183}, \href
  {http://adsabs.harvard.edu/abs/2017MNRAS.466.2024T} {466, 2024}

\bibitem[\protect\citeauthoryear{{Tonnesen} \& {Bryan}}{{Tonnesen} \&
  {Bryan}}{2009}]{tonnesenbryant09}
{Tonnesen} S.,  {Bryan} G.~L.,  2009, \mn@doi [\apj]
  {10.1088/0004-637X/694/2/789}, \href
  {http://adsabs.harvard.edu/abs/2009ApJ...694..789T} {694, 789}

\bibitem[\protect\citeauthoryear{{Utomo} et~al.,}{{Utomo}
  et~al.}{2017}]{utomo17}
{Utomo} D.,  et~al., 2017, \mn@doi [\apj] {10.3847/1538-4357/aa88c0}, \href
  {http://adsabs.harvard.edu/abs/2017ApJ...849...26U} {849, 26}

\bibitem[\protect\citeauthoryear{{Vazdekis} et~al.,}{{Vazdekis}
  et~al.}{2015}]{vazdekis15}
{Vazdekis} A.,  et~al., 2015, \mn@doi [\mnras] {10.1093/mnras/stv151}, \href
  {http://adsabs.harvard.edu/abs/2015MNRAS.449.1177V} {449, 1177}

\bibitem[\protect\citeauthoryear{{Vika}, {Bamford}, {H{\"a}u{\ss}ler}, {Rojas},
  {Borch}  \& {Nichol}}{{Vika} et~al.}{2013}]{vika13}
{Vika} M.,  {Bamford} S.~P.,  {H{\"a}u{\ss}ler} B.,  {Rojas} A.~L.,  {Borch}
  A.,   {Nichol} R.~C.,  2013, \mn@doi [\mnras] {10.1093/mnras/stt1320}, \href
  {http://adsabs.harvard.edu/abs/2013MNRAS.435..623V} {435, 623}

\bibitem[\protect\citeauthoryear{{Walcher} et~al.,}{{Walcher}
  et~al.}{2014}]{walcher14}
{Walcher} C.~J.,  et~al., 2014, \mn@doi [\aap] {10.1051/0004-6361/201424198},
  \href {http://adsabs.harvard.edu/abs/2014A%26A...569A...1W} {569, A1}

\bibitem[\protect\citeauthoryear{{Whitaker} et~al.,}{{Whitaker}
  et~al.}{2017}]{whitaker17}
{Whitaker} K.~E.,  et~al., 2017, \mn@doi [\apj] {10.3847/1538-4357/aa6258},
  \href {http://adsabs.harvard.edu/abs/2017ApJ...838...19W} {838, 19}

\bibitem[\protect\citeauthoryear{{Wisotzki}, {Becker}, {Christensen}, {Helms},
  {Jahnke}, {Kelz}, {Roth}  \& {Sanchez}}{{Wisotzki} et~al.}{2003}]{wisotzki03}
{Wisotzki} L.,  {Becker} T.,  {Christensen} L.,  {Helms} A.,  {Jahnke} K.,
  {Kelz} A.,  {Roth} M.~M.,   {Sanchez} S.~F.,  2003, \mn@doi [\aap]
  {10.1051/0004-6361:20031004}, \href
  {http://adsabs.harvard.edu/abs/2003A%26A...408..455W} {408, 455}

\bibitem[\protect\citeauthoryear{{Yoachim} \& {Dalcanton}}{{Yoachim} \&
  {Dalcanton}}{2008}]{yoachimdalcanton08}
{Yoachim} P.,  {Dalcanton} J.~J.,  2008, \mn@doi [\apj] {10.1086/590246}, \href
  {http://adsabs.harvard.edu/abs/2008ApJ...683..707Y} {683, 707}

\bibitem[\protect\citeauthoryear{{de Lorenzo-C{\'a}ceres} et~al.,}{{de
  Lorenzo-C{\'a}ceres} et~al.}{2019a}]{delorenzocaceres19b}
{de Lorenzo-C{\'a}ceres} A.,  et~al., 2019a, arXiv e-prints, \href
  {http://adsabs.harvard.edu/abs/2019arXiv190106394D} {}

\bibitem[\protect\citeauthoryear{{de Lorenzo-C{\'a}ceres}, {M{\'e}ndez-Abreu},
  {Thorne}  \& {Costantin}}{{de Lorenzo-C{\'a}ceres}
  et~al.}{2019b}]{delorenzocaceres19}
{de Lorenzo-C{\'a}ceres} A.,  {M{\'e}ndez-Abreu} J.,  {Thorne} B.,
  {Costantin} L.,  2019b, \mn@doi [\mnras] {10.1093/mnras/sty3520}, \href
  {http://adsabs.harvard.edu/abs/2019MNRAS.484..665D} {484, 665}

\bibitem[\protect\citeauthoryear{{de Souza}, {Gadotti}  \& {dos Anjos}}{{de
  Souza} et~al.}{2004}]{desouza04}
{de Souza} R.~E.,  {Gadotti} D.~A.,   {dos Anjos} S.,  2004, \mn@doi [\apjs]
  {10.1086/421554}, \href {http://adsabs.harvard.edu/abs/2004ApJS..153..411D}
  {153, 411}

\bibitem[\protect\citeauthoryear{{van den Bergh}, {Abraham}, {Ellis}, {Tanvir},
  {Santiago}  \& {Glazebrook}}{{van den Bergh} et~al.}{1996}]{vandenbergh96}
{van den Bergh} S.,  {Abraham} R.~G.,  {Ellis} R.~S.,  {Tanvir} N.~R.,
  {Santiago} B.~X.,   {Glazebrook} K.~G.,  1996, \mn@doi [\aj]
  {10.1086/118020}, \href {http://adsabs.harvard.edu/abs/1996AJ....112..359V}
  {112, 359}

\makeatother
\end{thebibliography}

%%%%%%%%%%%%%%%%%%%%%%%%%%%%%%%%%%%%%%%%%%%%%%%%%%
%%%%%%%%%%%%%%%%% APPENDICES %%%%%%%%%%%%%%%%%%%%%

%%%%%%%%%%%%%%%%%%%%%%%%%%%%%%%%%%%%%%%%%%%%%%%%%%

% Don't change these lines
\bsp	% typesetting comment
\label{lastpage}
\end{document}